\documentclass[fleqn,usenatbib]{mnras}

\usepackage{setspace,pdflscape,multirow}
\usepackage{newtxtext,newtxmath}
\usepackage{amsmath,gensymb,amssymb}
\usepackage{caption,array,booktabs,enumitem,siunitx}

\usepackage[T1]{fontenc}
\usepackage{ae,aecompl}

\usepackage{graphicx,xcolor}	% Including figure files

\usepackage{pifont}
\newcommand{\cmark}{\ding{51}}%
\newcommand{\xmark}{\ding{55}}%
\title[Mg~II Variability]{Behaviour of the Mg~II 2798{\AA} Line Over the Full Range of AGN Variability}

\author[Homan et al.]
{David Homan,$^{1}$\thanks{E-mail: dhoman@roe.ac.uk}
Chelsea L. MacLeod$^{2}$,
Andy Lawrence$^{1}$, 
Nicholas P. Ross$^{1}$  
\newauthor and Alastair Bruce$^{1}$
\\
$^{1}$Institute for Astronomy, University of Edinburgh, Royal Observatory, Edinburgh, EH9 3HJ, United Kingdom\\ 
$^{2}$ Center for Astrophysics, Harvard \& Smithsonian, 60 Garden St, Cambridge, MA 02138, U.S.A.\\
}

\date{Accepted XXX. Received YYY; in original form ZZZ}

\pubyear{2020}

\begin{document}
\label{firstpage}
\pagerange{\pageref{firstpage}--\pageref{lastpage}}
\maketitle

\begin{abstract}
We investigate the responsiveness of the 2798{\AA} Mg~II broad emission line in AGN on timescales of several years. Our study is based on a sample of extremely variable AGN as well as a broad population sample. The observed response of the line in previous studies has been mixed. By focussing on extreme variability ($|\Delta g|> 1$) we find that Mg~II clearly does respond to the continuum. However, the degree of responsiveness varies strikingly from one object to another: we see cases of Mg~II changing by as much as the continuum, more than the continuum, or very little at all. In 74\% of the highly variable sample the behaviour of Mg~II corresponds with that of H$\beta$, with 30\% of the objects showing large variations in both lines. We do not detect any change in the line width that would correspond to Broad Line Region `breathing', in accordance with results from the literature. Some of the objects in our highly variable sample show a clear asymmetry in the Mg~II profile. This skewness can be both to the blue and the red of the line centre. Results from our broad population sample show that highly variable quasars have lower Eddington ratios. This result holds for the variability of the  continuum, but the correlation is significantly reduced for the variability of the Mg~II line. For the first time, we present an overview of the value of the intrinsic Baldwin Effect for Mg~II in a large sample.
\end{abstract}


%%%%%%%%%%%%%%%%%%%%%%%%%%%%%%%%%%%%%%%%%%%%%%%%%%

%%%%%%%%%%%%%%%%% BODY OF PAPER %%%%%%%%%%%%%%%%%%

\section{Introduction}
\label{sec:intro}
\noindent Variability is inherent to AGN emission and affects all parts of the electromagnetic spectrum. The timescales and amplitude of fluctuations differ across the spectral range. Changes in the UV-optical continuum output are accompanied by changes in broad emission lines (BELs), often at a smaller amplitude than the continuum \citep[][]{PET08,BEN13,SUN15}. Average optical continuum variability on a timescale of months-years is $\sim$10-20\% \citep[e.g.][]{MCL10}. However a subset of Extremely Variable Quasars (EVQs) can, on the same timescales, show variability in the order of several \citep{LAW16,RUM18,GRAH19}. Extreme continuum variability is closely linked to the concept of changing-look (CL) AGN \citep{TOH76,SHAPP14,LAM15,MCL16}, in which the change in broad H$\beta$ is such that the AGN changes between a type 1 and a type 2 classification \citep{KHA71,OST81}. The extreme variability of continuum and BELs in these objects provides the opportunity for greater insight into AGN evolution and structure.\\
\indent The focus of this work is on the response of the broad Mg~II$\lambda2798$ emission line. This line differs from most other emission lines in its formation mechanism, because it is collisional \citep{MCA72,NET80,GUO19a}\footnote{Recombination of Mg~III is expected to occur at a negligible rate \citep[e.g.][]{GUO19a}}. Results from previous studies with regard to the response of Mg~II to the continuum have been mixed. Some studies have found a response \citep{CLA91,REI94,MET06,HRY14}, although the variable fraction of the flux is usually only a few percent. An exception to this is the quasar LBQS 2113-4538, studied by \citet{HRY14}, which shows a fluctuation in Mg~II Equivalent Width (EW) of $\sim$25\% over several hundred days. A similar number of studies finds no correlation between continuum changes and the variability of Mg~II \citep{TRE07,CAC15}.\\
\indent This complex behaviour has also been studied using larger samples consisting of survey data. \citet{WOO08} finds peak-to-peak changes of up to 25\% on a timescale of approximately a year, indicating the potential for large flux changes. In a study of $\sim$9,000 Stripe 82 quasars \citet{KOK14} find that both Mg~II and the Fe~II complex in the same wavelength range vary only slightly, when compared to the Balmer lines. If Mg~II does respond to the continuum, this line could be a promising candidate for Reverberation Mapping (RM) studies at high $z$. Early results of the SDSS-IV RM programme showed a credible Mg~II lag in six quasars (0.3$<z<$0.8) out of a sample of $\sim$100 objects \citep{SHE16b}. A different study, also based on the SDSS-RM data, finds that Mg~II is only slightly less variable than H$\beta$ \citep{SUN15}.\\
\indent In comparisons from object to object, the line flux of Mg~II is found to correlate with the optical continuum at 5100 {\AA} \citep{SHE12} and the width of Mg~II is found to correlate strongly with the widths of H$\alpha$ and H$\beta$ \citep{SHE12,WAN19}. These correlations suggest that Mg~II and the Balmer lines are formed in the same, or a similar, region in the BLR \citep{MCLU04}. This would broadly agree with photoionisation models \citep{KOR04,BAS14a}. \citet{DON09} find that the Equivalent Width (EW) of Mg~II is strongly anti-correlated with the Eddington ratio, but not with the continuum luminosity at 3000{\AA}. These authors argue that the lack of correlation with the luminosity underlies the Mg~II ensemble Baldwin Effect.\\
\indent An investigation into epoch to epoch Mg~II variability in a sample of SDSS quasars is presented in \citet{ZHU17}, who find that the Mg~II line tracks the continuum and that the responsivity of the line decreases with a higher initial continuum state and with a higher Eddington ratio. Both conclusions are supported by the results of \citet{SUN15}. \cite{YAN19} report a weak correlation between the change in the 3000{\AA} continuum and the change in Mg~II luminosity, based on spectroscopic follow up of 16 objects. The variability of Mg~II is of smaller amplitude than that of the UV continuum, but does appear to track it. A suggested explanation is that only part of the line responds to the ionising flux (cf. \citealt{KOR04}).\\
\indent In contrast to the flux changes, the Full Width at Half Maximum (FWHM) of Mg~II does not change with the fluctuating continuum. The lack of change in line width suggests that the radius to luminosity relationship (on which single epoch BH mass estimates are based) does not hold for Mg~II. This marks a particular difference with broad H$\beta$. Interestingly, \citet{ROI13} found a class of quasars characterised by a strongly suppressed broad H$\alpha$ and H$\beta$ emission and a very prominent Mg~II line, without signs of reddening. This population fits well in the evolutionary sequence presented in \citet{GUO19b}: Based on a photo-ionisation model \citet{GUO19b} find that Mg~II responds consistently to continuum changes, but at a much diminished level compared to the Balmer lines.\\
\indent In this paper we aim to show that the Mg~II line does respond to continuum variability. The response is complex and we detect a range of behaviours of the Mg~II line, both in flux and in line profile. We consider the variability of the broad Mg~II line in two samples. The first sample consists of 43 highly variable quasars. 40 of these are part of the CLQ candidate observations described in \citet[][hereafter referred to as MCL19]{MCL19}. This paper aims to present a complementary analysis of this data-set. The remaining three objects are part of the SDSS CLQ search presented in \cite{MCL16}. The second sample is a broad population sample, based on the SDSS quasar catalogue.\\
\indent The structure of this paper is as follows. In Section~\ref{sec:data} we describe our observations and the selection of our data-sets. Mg~II line characteristics are calculated from new fits to the spectra in both samples. Section~\ref{sec:method} presents the fitting method used, as well as several methods we will employ to quantify the changes in the spectra. In Section~\ref{sec:mgii_clq} we discuss the connection between Mg~II and H$\beta$ variability. The changes in the flux of the Mg~II line and its relation to the continuum flux are discussed in Section~\ref{sec:flux}. Following the discussion of the flux, we will consider the line profile and the possibility of extracting information about the kinematics of the line forming regions in Section~\ref{sec:profile}. The connection to physical parameters is discussed in Section~\ref{sec:phys}. We discuss the implications of our findings in Section~\ref{sec:interpretation} and summarise our conclusions in Section~\ref{sec:conclusion}. In Appendix~\ref{app:tests} we provide additional details on our statistical methods. Where necessary we assume a standard flat cosmology with $\Omega_\Lambda=0.7$, $\Omega_m=0.3$, and $H_0 = 70$ km s$^{-1}$ Mpc$^{-1}$.

\section{Mg~II Samples}
\label{sec:data}
\noindent This study is based on two samples: a set of 43 highly variable objects, 40 of which have new observations (the \textbf{Supervariable sample}), and the set of SDSS DR14 quasars for which we have repeat spectra available that cover Mg~II (the \textbf{Full Population sample}). The bases of the samples are the quasar catalogues for DR7 \citep{SCH10} and DR14 \citep{PAR14} respectively, referred to here as DR7Q and DR14Q.
\subsection{Observations}
\label{subsec:data_obs}
All observations for SDSS were made using the 2.5m Sloan telescope at Apache Point \citep{GUN}, as part of the SDSS-I/II, SDSS-III/BOSS, and SDSS-IV/eBOSS observation campaigns. The spectroscopic reduction pipelines for SDSS I/II and BOSS respectively are presented in \cite{SDSSI} and \cite{SDSSIII}. Of particular importance to this study is that the DR14 reduction pipeline for quasar spectra includes a correction for an error in the flux calibration due to atmospheric differential refraction which is present in SDSS III/BOSS spectra \citep{MAR,HAR}.\\
\indent The new observations were all part of the study discussed in MCL19. Here we provide a brief overview, for a full description please see MCL19. The observations were made using the Intermediate Spectrograph and Imaging System (ISIS) on the the 4.2m WHT in La Palma, with the Blue Channel Spectrograph on the MMT on the 6.5m telesecope (Mt. Hopkins, Arizona) and the Low Dispersion Survey Spectrograph 3 (LDSS3)-C on the 6.5m Magellan telescope. All observations were reduced using standard long slit spectroscopy methods. The flux calibration of the MMT and Magellan spectra has been scaled to that of SDSS by matching the [OIII]4959 and [OIII]5007 fluxes. For the WHT spectra we deemed the flux calibration of our spectra to be of sufficient quality not to require this additional grey scaling.
\subsection{Sample Selection}
\label{sec:data_sample}
\subsubsection{The Supervariable Sample}
\label{sec:data_samplevar}
The Supervariable sample is a combination of two data-sets. The first is the CLQ sample presented in \citet{MCL16} (MCL16) and the second is the CLQ candidate sample presented in MCL19. The CLQs discussed in MCL16 were found by using Pan-STARRS (PS1) and SDSS light curve data and subsequently comparing BOSS and SDSS-I/II spectra. From both data-sets the objects that have Mg~II in their spectra are included. From the MCL16 sample 3 objects qualified.\\
\indent The selection criteria, discussed in detail in MCL19, are listed below:
\begin{enumerate}
	\itemsep0em
	\item Quasars included in the DR7 quasar catalogue \citep{SCH10}
	\item And with no BOSS spectrum following the initial SDSS observation
	\item And a large magnitude change in PS1 and SDSS photometry: max $|\Delta g|>1$ and $|\Delta r|>0.5$ 
	\item $z < 0.83$, so that H$\beta$ is in the spectral range (for WHT, MMT and Magellan)
	\item There is no associated radio source (this excludes e.g. blazars)
	\item The most recent photometry (2013) indicates a change in $g$ of at least 1 mag compared to the SDSS epoch
\end{enumerate}
Table~\ref{tab:clq_candidate_sel} summarises these selections. A total of 130 candidates were observed and 40 of these objects had Mg~II in the spectrum, qualifying for the Supervariable sample. In combination with the three MCL16 objects, this gives a \emph{total of 43 objects}. It should be emphasised the Supervariable sample consists of both confirmed CLQs and observed CLQ \emph{candidates}. All objects are characterised by strong variation in the optical flux and can be considered EVQs. The sample is listed in Table~\ref{tab:var_results}.\\
\begin{table}
\centering
\caption{Overview of the selection of CLQ candidates for observations, as presented in MCL19. This table is based on Table 1 in MCL19. The Supervariable sample consist of these 40 objects plus 3 known CLQs from previous studies.}
\label{tab:clq_candidate_sel}
\begin{tabular}{l | r}
    \toprule
	\toprule
	Criterion & N$_{obj}$ \\
	\hline
	DR7 catalogue & 105,783 \\
	\hspace{3mm}  No BOSS spectrum & 79,838 \\
	\hspace{6mm} max $|\Delta g |> 1$, $|\Delta r| > 0.5$ \& $z<0.83$ & 1,727 \\
	\hspace{9mm} No radio detection & 1,403 \\
	\hspace{12mm} Recent $|\Delta g| > 1$  & 262\\
	\hspace{15mm} Observed & 130\\
	\hspace{18mm} Mg~II in the spectrum & \textbf{40}\\
	\bottomrule
	\bottomrule
\end{tabular}
\end{table}
\subsubsection{The Full Population Sample}
\label{sec:data_samplesdss}
\begin{figure}
\includegraphics[width=\columnwidth]{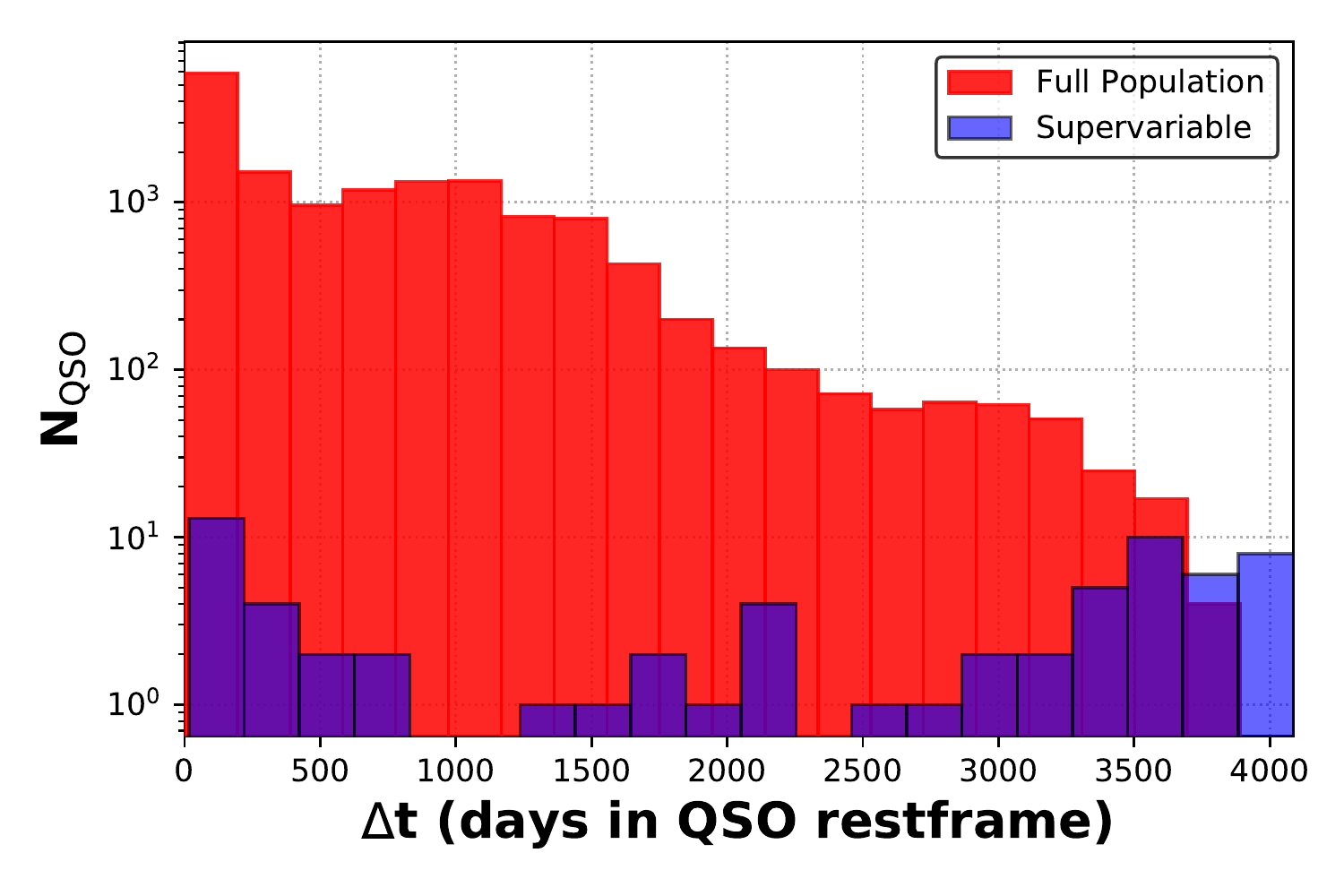}
\caption{Histogram of the elapsed time between spectral epochs for the Supervariable sample (\emph{blue}) and the Full Population sample (\emph{red}). Both samples cover a large range in $\Delta t$ and could therefore contain changes in Mg~II responsivity as a result of two different underlying effects: structural changes in the BLR, possible only for $\Delta t \sim$years, and large fluctuations of the continuum. $\Delta$t is defined as the time between two sequential spectra.}
\label{fig:mgii_dt_hist}
\end{figure}

For each object in DR14Q with more than one spectrum available the pair with the \emph{largest difference} between observation dates was selected, with a required minimum of 30 days. For the included objects, only the pair of spectra with the largest time difference is used from here on. The motivation for selecting the objects with the largest temporal baseline was to select for fitting the objects most likely to show large flux changes. This is based on the assumption that large continuum changes are more likely the longer we observe a quasar.\\
\indent We exclude BALQs from the sample as these objects could represent an irregular subset of quasars, with strong outflow signatures in their spectra \citep[e.g.][]{HAM19}. Further checks remove observations with an SDSS \texttt{z\_warning} (indicative of a redshift problem, \citet{BOL07}) and observations for which the exposure time is not registered in the file. The Mg~II$\lambda$2798 line is required to be in the spectral range for both the SDSS I/II and BOSS spectrographs. Assuming that for a proper fit of the Mg~II line it is necessary to have flux data at $\pm$100 {\AA} about the Mg~II line centre, and using the smaller $\lambda$-range available to the SDSS spectrograph (3800--9200 {\AA}), we are able to include quasars with redshift $0.41<z<2.17$. The Full Population sample consists of 15,824 objects; see Table~\ref{tab:sdss_sel}.\\
\begin{table}
\caption{The selection criteria for the Full Population sample, where the number in bold is the total number of quasars included in this sample. The table also shows the number of objects in a subsample of objects with high quality spectra available.}
\label{tab:sdss_sel}
\centering
\begin{tabular}{l|r}
	\toprule
	\toprule
	Criterion & N$_{obj}$ \\
	\hline
	DR14 catalogue & 523,356 \\
	\hspace{3mm} N$_{spec}\ge 2$ & 72,747 \\
	\hspace{6mm} $\Delta t > 30$days & 61,147 \\
	\hspace{9mm} not BALQ & 56,724 \\
	\hspace{12mm} z-warning/exposure/coordinates & 55,028\\
	\hspace{15mm} Mg~II in range & 15,824 \\
	\hspace{18mm} Successful pipeline fit & \textbf{15,101}\\
	\bottomrule
	\bottomrule
\end{tabular}
\end{table}
\noindent The distributions of the rest-frame timespans between epochs are displayed in Figure~\ref{fig:mgii_dt_hist}. For QSOs in the Supervariable sample with more than two spectra available, $\Delta t$ is calculated between spectra 1--2, 2--3, up to (N-1)--N, in chronological order, giving a total of 108 spectral pairs for the 43 objects. The distributions clearly indicate the wide range in intervals covered by our study, extending into a decade-long baseline for assessing spectral evolution.

\section{Tracking the Changes in Mg~II}
\label{sec:method}
\begin{figure}
\centering
\includegraphics[width=\columnwidth]{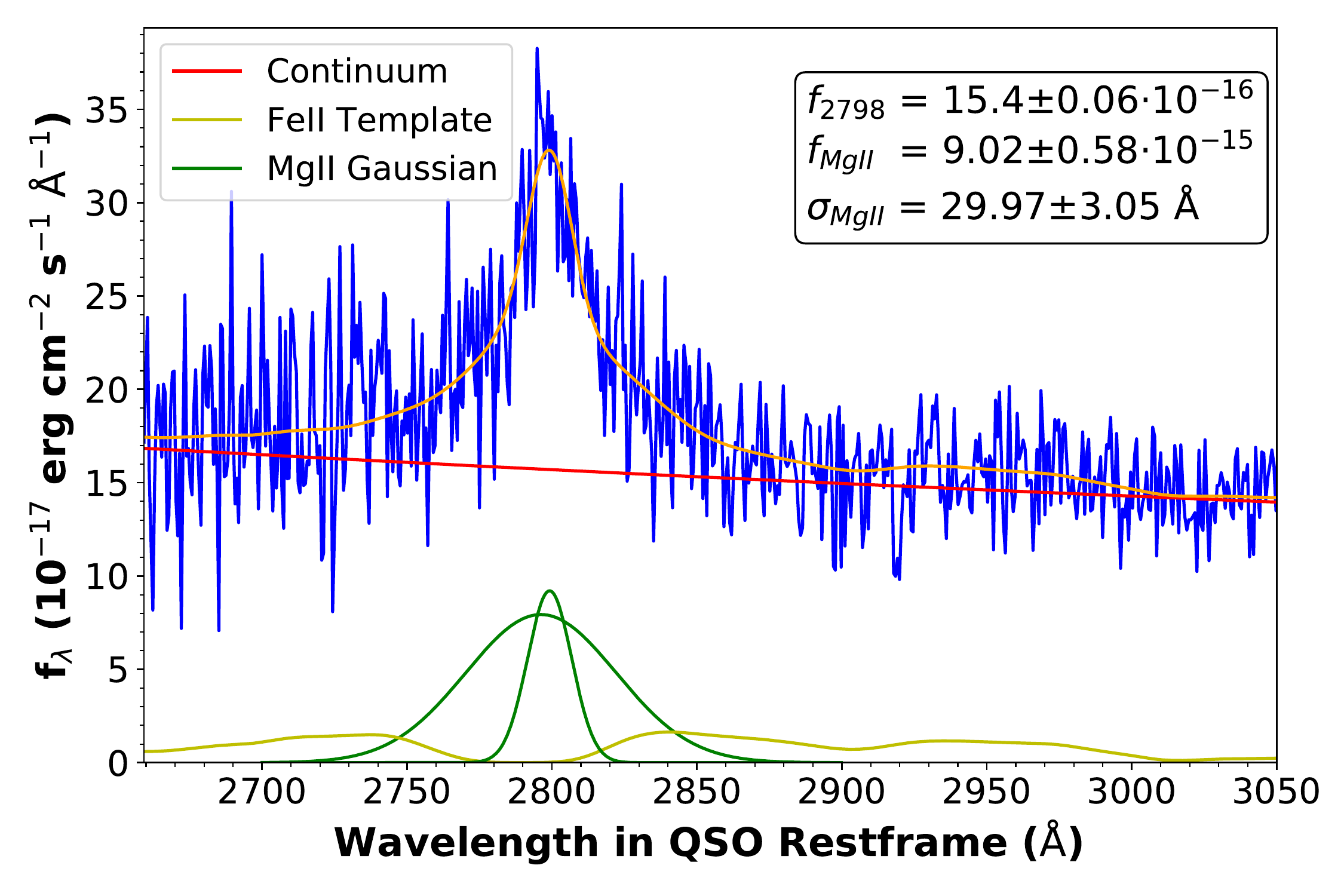}
\caption{Example of the result of a fit by the pipeline, for the SDSS spectrum of J212436 taken on MJD 52200. The spectrum is shown in blue and the fit components are: \emph{red}: power law continuum; \emph{green}: Gaussian fit to Mg~II (this can be one or two Gaussians, both of which constitute broad emission lines); \emph{yellow}: Fe~II template by Vestergaard et al., smoothed with a Gaussian filter; \emph{orange}: the combined fit to the spectrum. Note in particular the presence of Fe~II emission in both the red and the blue wing of the Mg~II line. The most important fitted parameters are listed in the top right corner.}
\label{fig:ex_fit}
\end{figure}

%1%
The spectra of the Supervariable and the Full Population samples are fitted using the \texttt{lmfit} package. The pipeline written for this purpose uses an iterative process to improve the fit quality. In brief overview the stages of the pipeline are:
\begin{enumerate}
	\itemsep0em
	\item Load spectrum and create mask for low quality flux bins;
	\item Correct for Galactic dust extinction;
	\item Trim spectrum to fitting range;
	\item Fit the relevant spectral component;
	\item Use fitting results for the next iteration of fitting;
\end{enumerate}
The steps (iv)-(v) are iterated for each spectral component, to improve the quality of the fit. In step (v) we use the fitting results as the new initial values and for sigma clipping.\\
%2%
\indent The components of the model used to fit the Mg~II line are the continuum power law, the Fe~II template, and either one or two Gaussian profiles for the Mg~II line itself. The number of iterations and the sigma-clipping limit are free parameters in the procedure, which were calibrated using the Supervariable sample. The power law continuum is defined as $A\nu^{-\alpha}$. The wavelength range for fitting this component is selected to contain a good representation the near-UV continuum, ranging from 2300{\AA} to 3088{\AA}.  This range is also used to jointly fit the power law and the Fe~II template. The Fe~II template used in this step is that presented by \cite{VES}. The best-fit continuum+Fe~II model is subtracted from the spectra.\\
\indent The final component of the fit is the Mg~II line itself, which is calculated in the window 2700{\AA}$<$$\lambda$$<$2900{\AA}. The line is represented with one or two Gaussian profiles. We limited the number of possible Gaussians to two, as the primary interest of this study is in flux changes: a large number of Gaussians increases the chance of spurious flux measurements, in particular due to remnant Fe~II contamination in the line wings. The second Gaussian, although narrower than the other component, does not constitute a `narrow' line in the same sense as e.g. the [OIII] narrow lines. The Akaike information criterion is calculated for a single-Gaussian and a two-Gaussian fit, the line model with the lowest AIC is selected. A table with an overview of all fitting parameters is included in the Online Supplement and the pipeline can be found here\footnote{https://github.com/dshoman/MgII}. An example fit to a spectrum is shown in Figure~\ref{fig:ex_fit}. \\
\indent For each epoch the spectral fitting provides: the continuum flux level at the line centre ($f_{2798}$), a Mg~II line flux ($f_{\mathrm{MgII}}$), the Mg~II line width ($\sigma_{\mathrm{MgII}}$), and the central wavelength of the Mg~II line. In the case where two Gaussians are fit, their combined flux constitutes the broad Mg~II emission. From the Mg~II line centre we derive the QSO redshift. This value of $z$, rather than the SDSS value, will be used for all results presented here. In the case of a two-Gaussian line model, $\sigma_{\mathrm{MgII}}$ is derived from the broadest Gaussian, as this most accurately represents the full width of the line (see e.g. Figure~\ref{fig:ex_fit}). In cases of low S/N, not all spectra can be fit successfully by the pipeline. Objects with one or both spectra for which no fit could be found are excluded. This issue only applies to a small fraction of the Full Population sample: 1446 spectra ($\sim$5\%). In the results presented further on, the number of objects in the Full Population sample is 15,101 (Table~\ref{tab:sdss_sel}).\\
%3%
\indent The measure of emission strength for the main results is a flux rather than the EW. This is similar to the method used in e.g. \citet{ZHU17} and \citet{YAN19}. Although a study of EW provides important information about the behaviour of broad emission lines \citep{SHE11,DON09,KOR04}, considering the line and continuum flux as separate parameters can allow for more insight into the behaviour of each component. For the purpose of comparison to other studies, the EW is used in Section~\ref{sec:flux_ibe} to calculate the intrinsic Baldwin Effect.\\
%4%
\indent We use several measures to quantify Mg~II variability. Different methods of normalisation highlight different characteristics of the data. The first category of normalisations pertains to the fluxes only, and is with respect to one spectral epoch. We consider both a normalisation to the high and the low state. For normalisation to the high state, the spectral epoch with the highest \emph{continuum} flux is identified for each object. The values $f_{\mathrm{MgII}}$ and $f_{2798}$ in all other epochs are then divided by $f_{\mathrm{MgII}}$ and $f_{2798}$ of the high state spectrum.  The emphasis in this normalisation is on small flux changes, and the associated changes in line flux. The case of normalisation to the low state, using the epoch of lowest $f_{2798}$, allows for a clearer view of whether the line flux is able to track an increase in the continuum. The second category is a normalisation of the changes in $f_{2798}$, $f_{\mathrm{MgII}}$, and $\sigma_{\mathrm{MgII}}$. The quantities $\Delta f_{2798}$ and $\Delta f_{\mathrm{MgII}}$ are defined as the change in flux, from epoch to epoch, divided by the flux of the second epoch.\\
%5%
\indent To quantify the responsivity in the normalised flux samples, as well as any differences in behaviour, we fit two empirical functions to the normalised data. The functions are fit to the epoch-normalised flux data. The first function is a simple linear response of the line to changes in the continuum, and the second function has two linear components. The fitting is iterative, which improves the assessment of the uncertainties associated with the fit. The measurement errors on the fluxes (propagated from the errors on the fits) appear to underestimate the uncertainties in the data. The latter are indicated by the scatter. Our uncertainties are therefore estimated using an initial fit to the data, after which the standard deviation of the residuals is set to be the uniform error. The next iteration of the fit provides the parameters presented in this study.\\
%6%
\indent We use a number of methods to test the line responsivity (see Appendix~\ref{app:tests} for more detail). In addition to well known statistical tests, we define a \emph{Sequence Test}. This is a test to discern which of the two fitting functions is most appropriate to describe the normalised data sets. The method measures the distribution of data points above or below the fitted functions and compares it with the binomial distribution. We also define a \emph{Responsivity Measure}. This metric is based on a simple criterion: the perpendicular distance in the normalised flux plane from the line that would indicate a 1:1 correspondence between line and continuum flux. The responsivity measure will referred to as $\alpha_{rm}$.

\section{Mg~II and Changing-Look AGN}
\label{sec:mgii_clq}
\begin{figure}
\centering
\includegraphics[width=.5\textwidth]{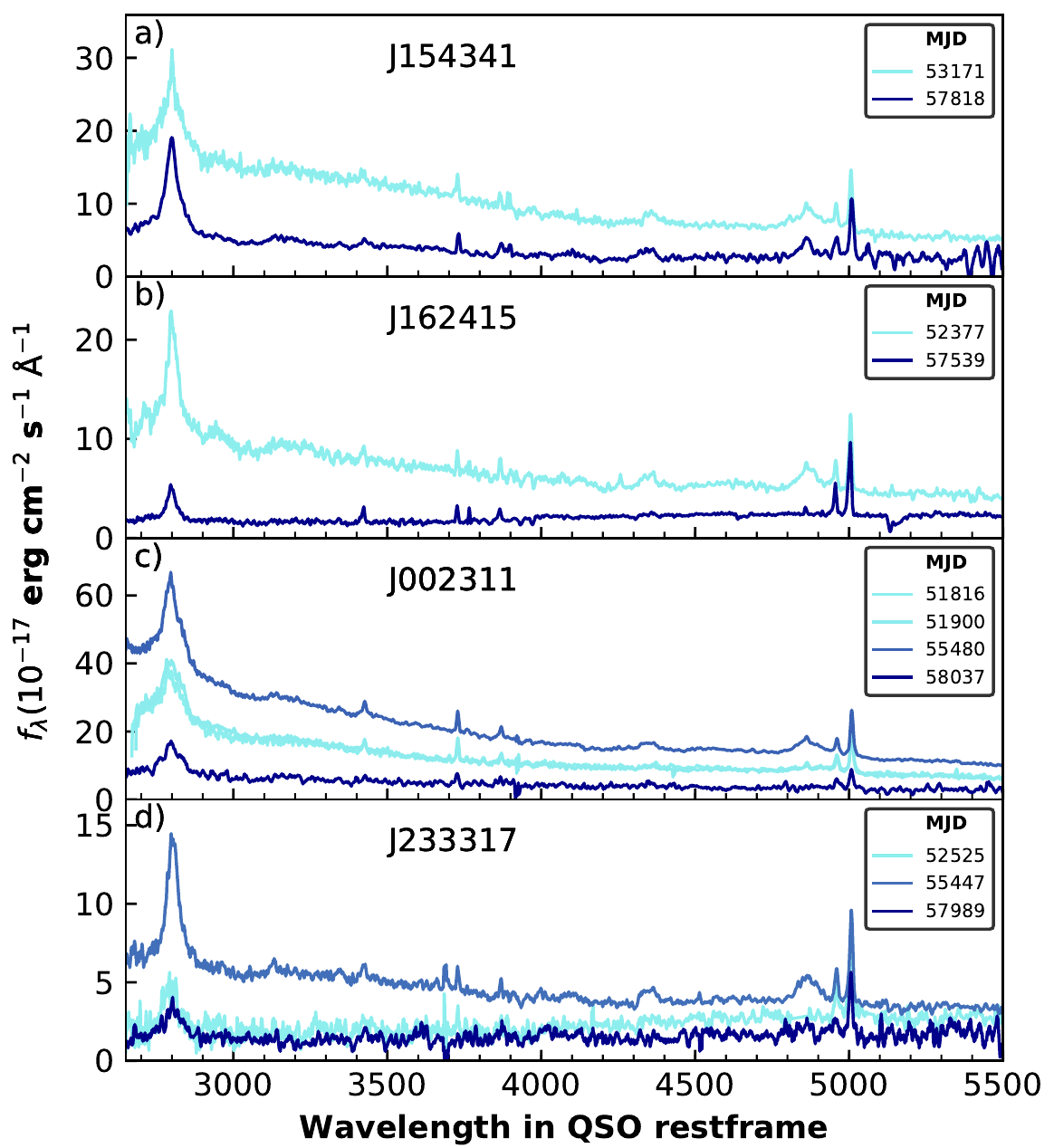}
\caption{Examples of the wide range of behaviours detected in the Supervariable sample. \emph{a)}: J154341 displays a diminishing continuum, with the Mg~II line flux remaining stable. \emph{b)}: J162415, the Mg~II line flux tracks the lowering continuum quite closely. Note that this is also a CLQ. \emph{c)}: J002311, where we see a rising continuum, without a strong response in the Mg~II flux. We have enough spectra to see the spectrum both rise and fall: after the lack of response to a rising continuum, the Mg~II tracks the continuum more closely after it dims. \emph{d)}: J233317, which shows the Mg~II flux responding to a rising continuum, and returning to a lower state subsequently.}
\label{fig:mgii_var_behaviour}
\end{figure}

\noindent There is a wide range of variability among the objects in the Supervariable sample, see e.g. Table~\ref{tab:var_results} column 7 and the spectra shown in Figure~\ref{fig:mgii_var_behaviour}.
The four objects included in this figure were selected as examples of four categories of Mg~II variability: the top two panels show objects which both have a dimming continuum: the Mg~II line changes only in one of the two cases. The bottom two panels show a similar difference in response, but for a rising continuum. Of particular interest is J002311 (panel $c$), for which the range of observations covers the object brightening and re-dimming. The response from Mg~II as the continuum rises (between MJD 51900 and 55480) is very limited, but once the continuum diminishes again (MJD 58037) the reaction is more pronounced. The responsivity of the line therefore not only differs from object to object, \emph{but also from epoch to epoch}.\\
\indent The bottom three objects in Figure~\ref{fig:mgii_var_behaviour} can be classified as CLQs, as only the narrow line component of H$\beta$ remains in the low state spectra. The classification is summarised in Table~\ref{tab:var_avgtime}. The categories in Table~\ref{tab:var_avgtime} are based on two classifications: variability in H$\beta$ is based on whether the object was classified as a CLQ in MCL19 and the classification of Mg~II variability requires $|\Delta f_{\mathrm{MgII}}|$$\ge$$0.5$. The category `no Mg~II' covers a stable line flux, as the continuum changes. Although the relation between variability in Mg~II and in H$\beta$ is not one-to-one, there does appear to be a connection: objects with changing H$\beta$ are more likely to show variability in Mg~II than not.\\
\begin{table}
\centering
\caption{Four categories of variability displayed in the Supervariable sample based on H$\beta$ and Mg~II line flux changes. The H$\beta$ variability is based on a visual identification as a CLQ in MCL19 and the Mg~II variability requires $|\Delta f_{\mathrm{MgII}}|$$\ge$$0.5$. The second column lists the number of objects that fall in each category and the third column the average of time spans of the spectral epochs for the included objects. The value in parentheses is the standard deviation of $\Delta t$.}
\label{tab:var_avgtime}
\begin{tabular}{l | c c}
\toprule
\toprule
Type of Variability & N$_{\mathrm{obj}}$ & avg. $\Delta t_{\mathrm{spec}}$ in days\\
\hline
H$\beta$, Mg~II & 13 & 3,335 (772) \\
H$\beta$, no Mg~II & 3 & 3,871 (198) \\
no H$\beta$, Mg~II & 8 & 3,375 (527) \\
no H$\beta$, no Mg~II & 19 & 3,379 (769) \\
\bottomrule
\bottomrule
\end{tabular}
\end{table}
\indent The categorisation in Table~\ref{tab:var_avgtime} indicates that there is no direct correlation between the elapsed time between spectral epochs and the type of variability detected. There is a large scatter in the range of $\Delta t_{\mathrm{spec}}$ (see also Figure~\ref{fig:mgii_dt_hist}) and this scatter is present in all four categories. This potentially indicates that the change in ionising flux, rather than the elapsed time is the driving force behind the variability of Mg~II as well as of H$\beta$.

\section{How Variable Can Mg~II Flux Be?}
\label{sec:flux}
Using the methods outlined in section~\ref{sec:method} we fit and normalise the data in our two samples. The results of the spectral fitting of the Supervariable sample are summarised in Table~\ref{tab:var_results}. The parameters calculated for the Supervariable sample represent the behaviour of the Mg~II line in a sample with a large range in time-spans between observations.
\subsection{Luminosities}
\begin{figure}
\centering
\includegraphics[scale=0.45]{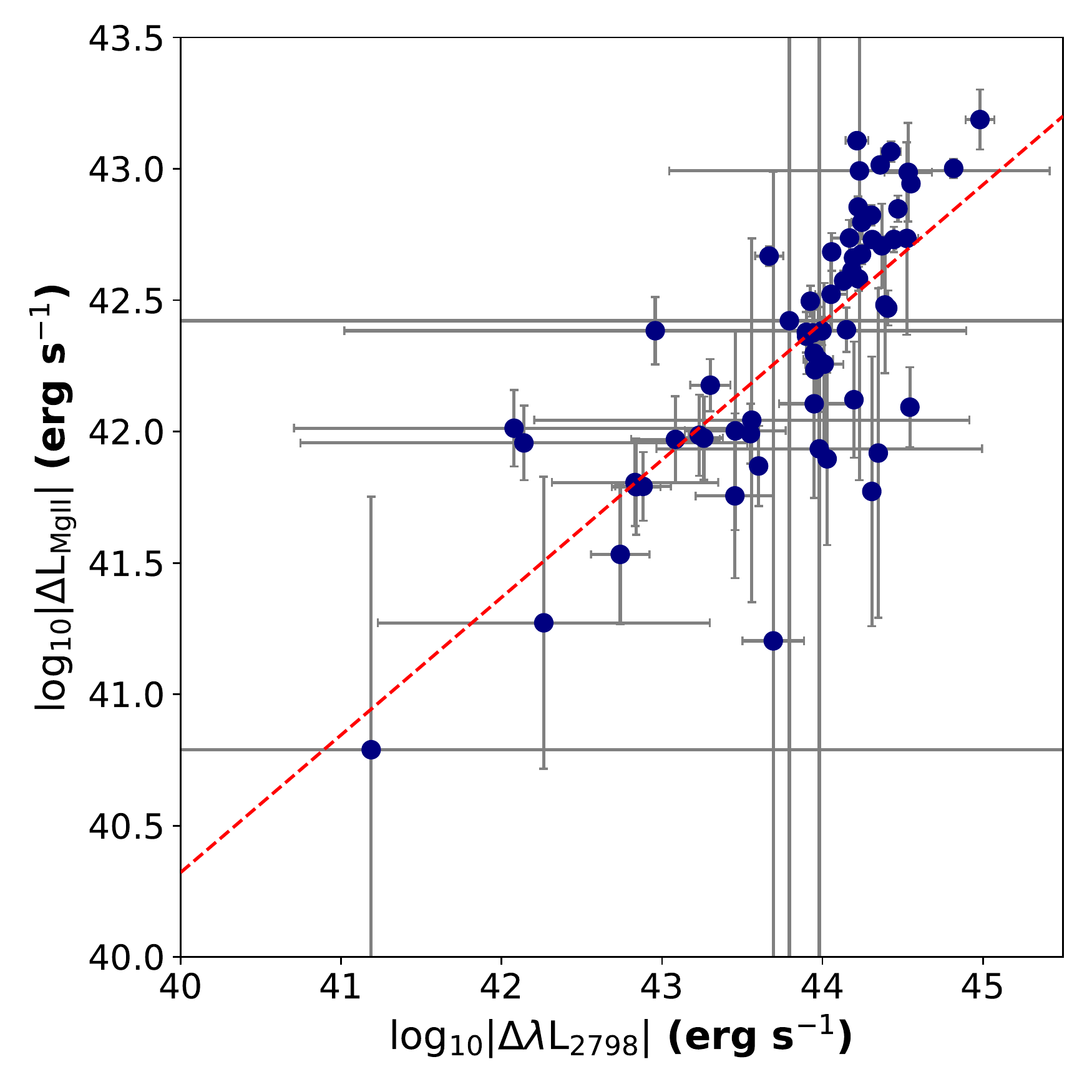}
\caption{There is a positive correlation between the changes in line (L$_{\mathrm{MgII}}$) and continuum (L$_{2798}$) luminosity for the Supervariable sample: a linear regression of log L$_{\mathrm{MgII}}$ on log L$_{2798}$ (\emph{red}) gives a positive slope of 0.52. The plot displays the changes in luminosity from epoch to epoch, including the information from all spectra available for every objects. The horizontal axis represents the absolute value of the change in continuum luminosity at 2798{\AA} and the vertical axis the absolute value of the change in the Mg~II line luminosity.}
\label{fig:mgii_lum_var}
\end{figure}

\indent An overview of the change in line and continuum luminosities of the Supervariable sample, using the absolute values of the changes, is shown in Figure~\ref{fig:mgii_lum_var}. The data in this figure concern the change from epoch to epoch in total energy emitted per second. A linear regression of the change in line flux on the change in continuum gives a slope of 0.52. This indicates a positive but non-linear correlation between the two parameters. The plot also illustrates the large range in luminosities covered by the Supervariable sample.
\begin{landscape}
\begin{table}
\caption{Overview of the results of the spectral fitting for the Supervariable sample, listed by SDSS in column 1. Column 2 indicates whether $|\Delta f_{\mathrm{MgII}}|$$\ge$$0.5$ (\cmark), is smaller than 0.5 (\xmark), or the continuum is stable (-). Column 3 lists whether the object was identified as a CLQ in MCL19. Column 4 is the time span between the spectral epochs, in the AGN restframe. Columns 5 and 6 list the changes in line and continuum luminosity, columns 7 and 8 are the fractional changes in line flux, and column 9 the change in line width. For objects with more than two spectra available the values in columns 5--9 are between the two spectra with the largest $|\Delta\mathrm{L}_{2798}|$. Columns 10 list the measured QSO redshift, and columns 11--13 in turn give log$_{10}$ of the the bolometric luminosity, the black hole mass and the Eddington ratio, as provided by \citet{SHE11}. Column 14 provides the maximum absolute change in g-band photometry over the period between spectral observations, based on SDSS, PS1 \citep{HEI} and Catalina Real Time Survey \citep[CRTS;][]{DRA09} data. The final column lists the telescope with which the latest spectral observation was made. A machine-readable version of this table is available in the Online Supplement.}
\label{tab:var_results}
\begin{tabular}{l l | c c | c | c c | c c | c | c c c c c c }
\toprule
\toprule
   &  SDSS name &  Mg~II & CLQ & max $\Delta t_{\mathrm{spec}}$ & {$\Delta\mathrm{L}_{2798}$} & {$\Delta\mathrm{L}_{\mathrm{MgII}}$} & {$\Delta{f_{2798}}$} & {$\Delta{f_{\mathrm{MgII}}}$} & {$\Delta\sigma$} & {$z$} & {$L_{Bol}$} &  {M$_{BH}$} &  {$\frac{L_{bol}}{L_{Edd}}$} & max $|\Delta g|$ & Telescope\\
  &   &   &   &  (RF days) &  {($10^{42}$ erg s$^{-1}$)} &  {($10^{42}$ erg s$^{-1}$)} &   &   &  {({\AA})} & & & & & & \\
\hline
 1 & 001113.46-110023.5 & \cmark & \cmark & 3906 & 39.5 $\pm$ 3.7 & 0.9 $\pm$ 0.2 & 0.69 & 0.5 & 4.2 $\pm$ 3.6 & 0.495 & 45.2 & 8.2 & -1.1 & 1.2 & MMT \\
 2 & 001206.25-094536.3 & \cmark & \cmark & 3484 & 135.6 $\pm$ 21.9 & 2.2 $\pm$ 1.8 & 0.87 & 0.54 & 17.1 $\pm$ 11.1 & 0.566 & 45.5 & 8.5 & -1.1 & 2.0 & Magellan \\
 3 & 002311.06+003517.5 & \cmark & \cmark & 4378 & 319.9 $\pm$ 10.8 & 4.9 $\pm$ 0.4 & 0.91 & 0.55 & -4.9 $\pm$ 13.2 & 0.422 & 45.5 & 9.2 & -1.8 & 1.0 & MMT \\
 4 & 002450.50+003447.7 & \xmark & \xmark & 3795 & 110.4 $\pm$ 9.2 & 1.3 $\pm$ 0.2 & 0.69 & 0.33 & -10.7 $\pm$ 6.7 & 0.524 & 45.5 & 8.7 & -1.3 & 1.1 & MMT \\
 5 & 002714.21+001203.7 & - & \xmark & 4086 & -17.2 $\pm$ 53.6 & -0.5 $\pm$ 0.8 & -0.08 & -0.17 & 0.5 $\pm$ 1.4 & 0.454 & 45.5 & 8.2 & -0.8 & 0.5 & MMT \\
 6 & 010529.61-001104.0 & \xmark & \xmark & 3353 & 29.4 $\pm$ 15.0 & 0.4 $\pm$ 0.4 & 0.52 & 0.12 & -3.6 $\pm$ 6.8 & 0.738 & 45.2 & 8.3 & -1.2 & 0.8 & MMT \\
 7 & 013458.36-091435.4 & \xmark & \cmark & 4041 & 168.0 $\pm$ 7.6 & 0.6 $\pm$ 0.2 & 0.59 & 0.1 & -8.4 $\pm$ 7.1 & 0.443 & 45.6 & 8.5 & -1.0 & 1.4 & MMT \\
 8 & 022556.07+003026.7 & \cmark & \cmark & 2920 & 68.9 $\pm$ 2.3 & 2.0 $\pm$ 0.1 & 0.88 & 0.82 & -9.9 $\pm$ 11.3 & 0.504 & 45.4 & 8.4 & -1.1 & 1.7 & SDSS \\
 9 & 022652.24-003916.5 & \cmark & \cmark & 2206 & 30.0 $\pm$ 2.5 & 0.9 $\pm$ 0.3 & 0.75 & 0.5 & 3.0 $\pm$ 16.6 & 0.625 & 45.3 & 7.5 & -0.3 & 1.4 & SDSS \\
 10 & 034144.72-002854.3 & \cmark & \xmark & 3419 & -100.8 $\pm$ 13.9 & -4.4 $\pm$ 0.4 & -1.98 & -1.8 & 5.8 $\pm$ 11.5 & 0.623 & 45.4 & 9.9 & -2.6 & 1.2 & MMT \\
 11 & 034330.48-073703.1 & \xmark & \xmark & 2404 & 97.0 $\pm$ 10.7 & 2.3 $\pm$ 0.3 & 0.71 & 0.46 & 20.1 $\pm$ 5.6 & 0.742 & 45.5 & 8.7 & -1.3 & 1.0 & Magellan \\
 12 & 074502.90+374947.0 & \cmark & \xmark & 3895 & 66.2 $\pm$ 3.5 & 1.5 $\pm$ 0.2 & 0.69 & 0.77 & 21.3 $\pm$ 8.2 & 0.593 & 45.3 & 8.4 & -1.2 & 1.3 & MMT \\
 13 & 074719.65+215245.9 & \xmark & \xmark & 3674 & 39.7 $\pm$ 4.2 & 1.5 $\pm$ 0.2 & 0.62 & 0.36 & -4.9 $\pm$ 8.5  & 0.462 & 45.2 & 8.7 & -1.6 & 1.2 & MMT \\
 14 & 075228.65+405931.7 & \xmark & \xmark & 4349 & 48.9 $\pm$ 4.9 & 1.2 $\pm$ 0.1 & 0.34 & 0.19 & 1.1 $\pm$ 1.2 & 0.423 & 45.4 & 8.2 & -0.9 & 1.3 & MMT \\
 15 & 081916.15+345050.3 & \xmark & \xmark & 767 & 19.7 $\pm$ 2.1 & 0.4 $\pm$ 0.1 & 0.43 & 0.17 & 7.0 $\pm$ 5.4 & 1.500 & 45.9 & 9.3 & -1.4 & 2.1 & WHT \\
 16 & 090104.75+345524.2 & \cmark & \cmark & 2855 & 140.4 $\pm$ 48.0 & 4.0 $\pm$ 1.7 & 0.78 & 0.73 & -0.3 $\pm$ 21.2 & 0.564 & 45.7 & 9.2 & -1.6 & 1.8 & MMT \\
 17 & 092223.17+444629.0 & \cmark & \xmark & 3322 & 74.5 $\pm$ 419.0 & 4.3 $\pm$ 24.0 & 0.76 & 0.76 & 11.5 $\pm$ 8.6 & 0.536 & 45.5 & 8.9 & -1.5 & 1.4 & MMT \\
 18 & 094132.89+000731.1 & \xmark & \xmark & 3911 & 28.2 $\pm$ 325.2 & 1.2 $\pm$ 13.2 & 0.37 & 0.37 & -0.2 $\pm$ 4.8 & 0.489 & 45.5 & 8.0 & -0.7 & 1.2 & MMT \\
 19 & 102016.04+014231.7 & \xmark & \xmark & 3746 & -54.2 $\pm$ 10.9 & -1.6 $\pm$ 0.4 & -0.79 & -0.44 & -5.6 $\pm$ 8.0 & 0.448 & 45.3 & 8.1 & -0.9 & 1.4 & MMT \\
 20 & 111348.64+494522.4 & \xmark & \xmark & 3071 & 88.3 $\pm$ 10.0 & 1.1 $\pm$ 0.7 & 0.44 & 0.12 & -2.8 $\pm$ 5.6 & 0.659 & 45.7 & 9.2 & -1.6 & 1.5 & WHT \\
 21 & 123228.08+141558.7 & \xmark & \xmark & 3084 & 53.1 $\pm$ 5.2 & -0.4 $\pm$ 0.3 & 0.62 & -0.2 & -2.5 $\pm$ 9.0 & 0.427 & 45.3 & 8.6 & -1.4 & 1.2 & MMT \\
 22 & 132815.50+223823.8 & \cmark & \xmark & 2149 & 78.6 $\pm$ 5.8 & 2.1 $\pm$ 0.2 & 0.76 & 0.63 & -2.8 $\pm$ 7.7 & 0.486 & 45.4 & 8.8 & -1.5 & 1.4 & WHT \\
 23 & 145519.88+364800.4 & - & \xmark & 3043 & 3.9 $\pm$ 19.2 & -1.0 $\pm$ 0.4 & 0.03 & -0.32 & 9.8 $\pm$ 7.8 & 0.525 & 45.5 & 8.6 & -1.1 & 1.3 & WHT \\
 24 & 152044.63+394930.3 & - & \xmark & 3581 & 23.9 $\pm$ 10.4 & -0.1 $\pm$ 0.3 & 0.38 & -0.04 & 4.7 $\pm$ 7.0 & 0.439 & 45.2 & 8.5 & -1.4 & 1.2 & MMT \\
 25 & 153734.06+461358.9 & \cmark & \cmark & 3704 & 92.2 $\pm$ 17.6 & 3.3 $\pm$ 0.6 & 0.76 & 0.79 & -24.2 $\pm$ 3.2 & 0.378 & 45.3 & 8.5 & -1.3 & 1.4 & MMT \\
 26 & 153912.76+524540.2 & \cmark & \cmark & 3896 & 43.1 $\pm$ 8.9 & 1.2 $\pm$ 0.3 & 0.66 & 0.78 & -10.7 $\pm$ 4.8 & 0.415 & 45.3 & 8.3 & -1.2 & 1.1 & MMT \\
 27 & 154341.64+385319.3 & \xmark & \xmark & 3241 & 109.3 $\pm$ 14.6 & -0.4 $\pm$ 1.1 & 0.67 & -0.11 & -3.4 $\pm$ 7.4 & 0.432 & 45.4 & 9.1 & -1.8 & 1.2 & MMT \\
 28 & 160226.01+323019.1 & \xmark & \xmark & 3350 & 83.6 $\pm$ 8.8 & 0.2 $\pm$ 0.3 & 0.4 & 0.07 & 7.0 $\pm$ 6.9 & 0.557 & 45.7 & 8.8 & -1.1 & 1.1 & MMT \\
 29 & 162415.02+455130.0 & \cmark & \cmark & 3479 & 93.9 $\pm$ 5.7 & 2.5 $\pm$ 0.1 & 0.83 & 0.73 & 6.4 $\pm$ 5.9 & 0.481 & 45.4 & 8.5 & -1.2 & 1.2 & WHT \\
 30 & 163031.47+410145.8 & \xmark & \xmark & 3603 & 88.2 $\pm$ 5.9 & 2.9 $\pm$ 0.3 & 0.51 & 0.49 & -17.5 $\pm$ 10.5 & 0.531 & 45.5 & 8.8 & -1.4 & 1.3 & MMT \\
 31 & 164053.05+451033.7 & \xmark & \xmark & 3766 & 47.7 $\pm$ 197.6 & 0.4 $\pm$ 11.3 & 0.63 & 0.23 & -7.2 $\pm$ 4.3 & 0.422 & 45.4 & 8.9 & -1.6 & 2.6 & MMT \\
 32 & 212436.64-065722.1 & \xmark & \xmark & 3734 & 76.7 $\pm$ 6.5 & 0.6 $\pm$ 0.3 & 0.51 & 0.2 & -8.1 $\pm$ 4.2 & 0.430 & 45.4 & 8.3 & -1.0 & 2.1 & MMT \\
 33 & 214613.31+000930.8 & \cmark & \cmark & 1547 & -43.5 $\pm$ 4.2 & -1.8 $\pm$ 0.3 & -4.1 & -1.44 & 13.5 $\pm$ 24.2 & 0.621 & 45.2 & 8.9 & -1.8 & 0.5 & SDSS \\
 34 & 223133.89+003312.7 & \xmark & \cmark & 3917 & 41.4 $\pm$ 4.3 & 0.8 $\pm$ 0.2 & 0.45 & 0.29 & -12.3 $\pm$ 6.1 & 0.476 & 45.4 & 8.6 & -1.3 & 1.2 & MMT \\
 35 & 223500.50-004940.7 & \cmark & \xmark & 3559 & -86.4 $\pm$ 36.7 & -1.9 $\pm$ 1.3 & -1.45 & -0.71 & -4.5 $\pm$ 9.0 & 0.642 & 45.1 & 8.5 & -1.5 & 0.5 & MMT \\
 36 & 224017.28-011442.8 & \cmark & \xmark & 3396 & -62.2 $\pm$ 6.1 & -1.1 $\pm$ 0.2 & -1.23 & -0.96 & -20.1 $\pm$ 4.9 & 0.502 & 45.1 & 8.3 & -1.3 & 1.2 & MMT \\
 37 & 224829.47+144418.0 & \cmark & \cmark & 3634 & 173.2 $\pm$ 4.4 & 4.3 $\pm$ 0.3 & 0.87 & 0.66 & 25.6 $\pm$ 7.7 & 0.424 & 45.5 & 8.9 & -1.5 & 1.5 & MMT \\
 38 & 225240.37+010958.7 & \cmark & \cmark & 3529 & 98.7 $\pm$ 5.5 & 4.5 $\pm$ 0.3 & 0.8 & 0.63 & 5.7 $\pm$ 7.5 & 0.534 & 45.3 & 8.9 & -1.7 & 1.7 & MMT \\
 39 & 230427.86-000803.2 & \cmark & \xmark & 3632 & -63.3 $\pm$ 16.1 & -2.3 $\pm$ 0.4 & -1.07 & -1.46 & -8.0 $\pm$ 6.4 & 0.526 & 45.2 & 8.7 & -1.6 & 1.1 & MMT \\
 40 & 231552.95+011406.1 & \xmark & \xmark & 3643 & -17.9 $\pm$ 14.0 & -0.1 $\pm$ 0.3 & -0.51 & -0.12 & -5.3 $\pm$ 9.1 & 0.521 & 45.1 & 8.5 & -1.6 & 1.1 & MMT \\
 41 & 231742.68+011425.2 & \xmark & \cmark & 3654 & 40.6 $\pm$ 11.9 & 0.8 $\pm$ 0.5 & 0.53 & 0.43 & -1.9 $\pm$ 17.7 & 0.518 & 45.3 & 9.0 & -1.8 & 1.6 & MMT \\
 42 & 233317.38-002303.4 & \cmark & \cmark & 3821 & 66.8 $\pm$ 2.1 & 1.8 $\pm$ 0.1 & 0.79 & 0.64 & -0.2 $\pm$ 7.9 & 0.513 & 45.3 & 10.2 & -3.0 & 1.7 & Magellan \\
 43 & 234623.42+010918.1 & \cmark & \xmark & 3635 & -71.5 $\pm$ 11.7 & -5.6 $\pm$ 0.2 & -2.9 & -3.64 & -7.0 $\pm$ 9.7 & 0.509 & 45.3 & 8.9 & -1.7 & 2.0 & MMT \\
\bottomrule
\bottomrule
\end{tabular}
\end{table}
\end{landscape}
\subsection{Flux Normalisations}
\begin{figure*}
\hspace*{-1.1cm}\begin{minipage}{.55\textwidth}
\includegraphics[width=\textwidth]{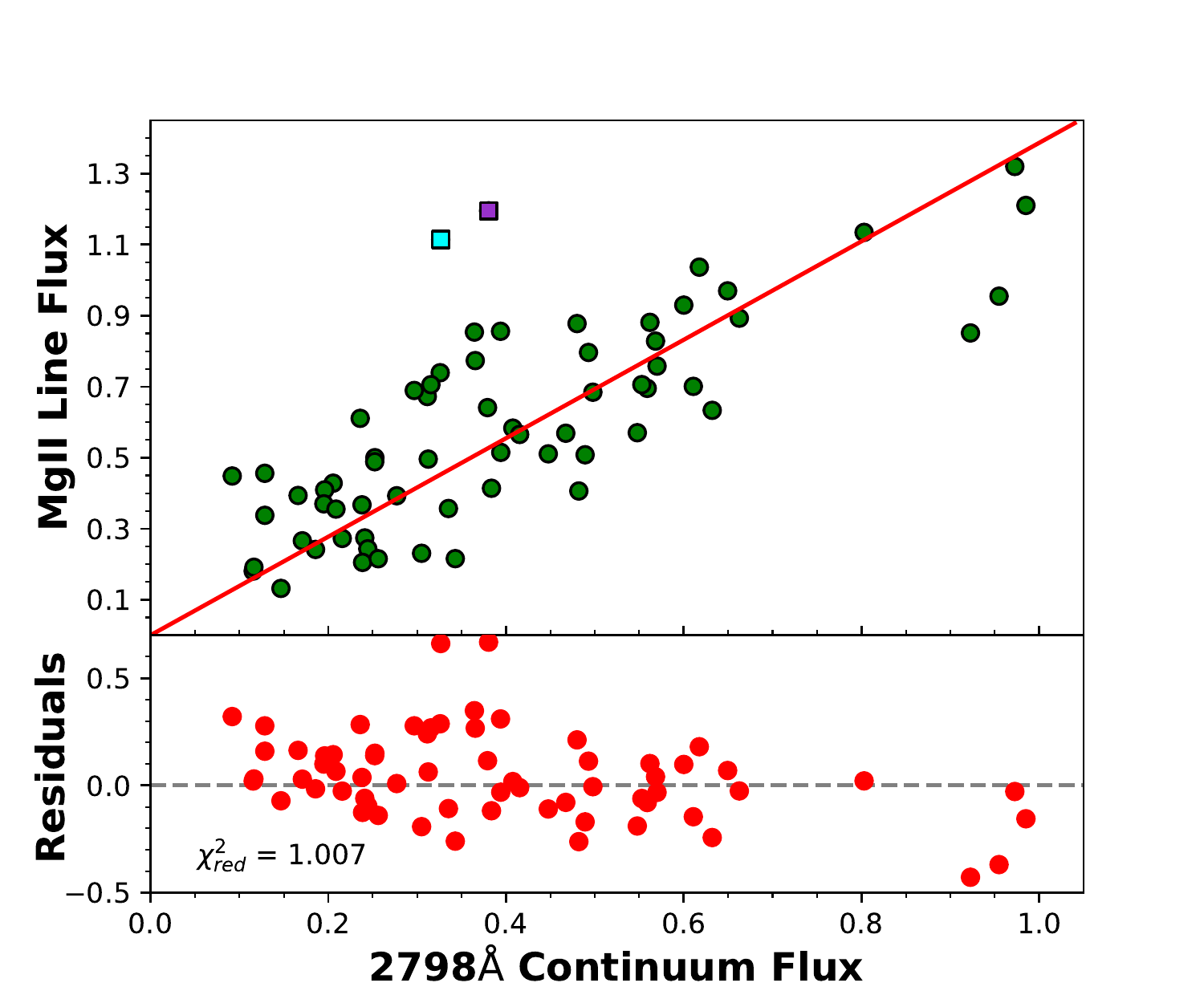}
\end{minipage}%
\begin{minipage}{.56\textwidth}
\includegraphics[scale=.55]{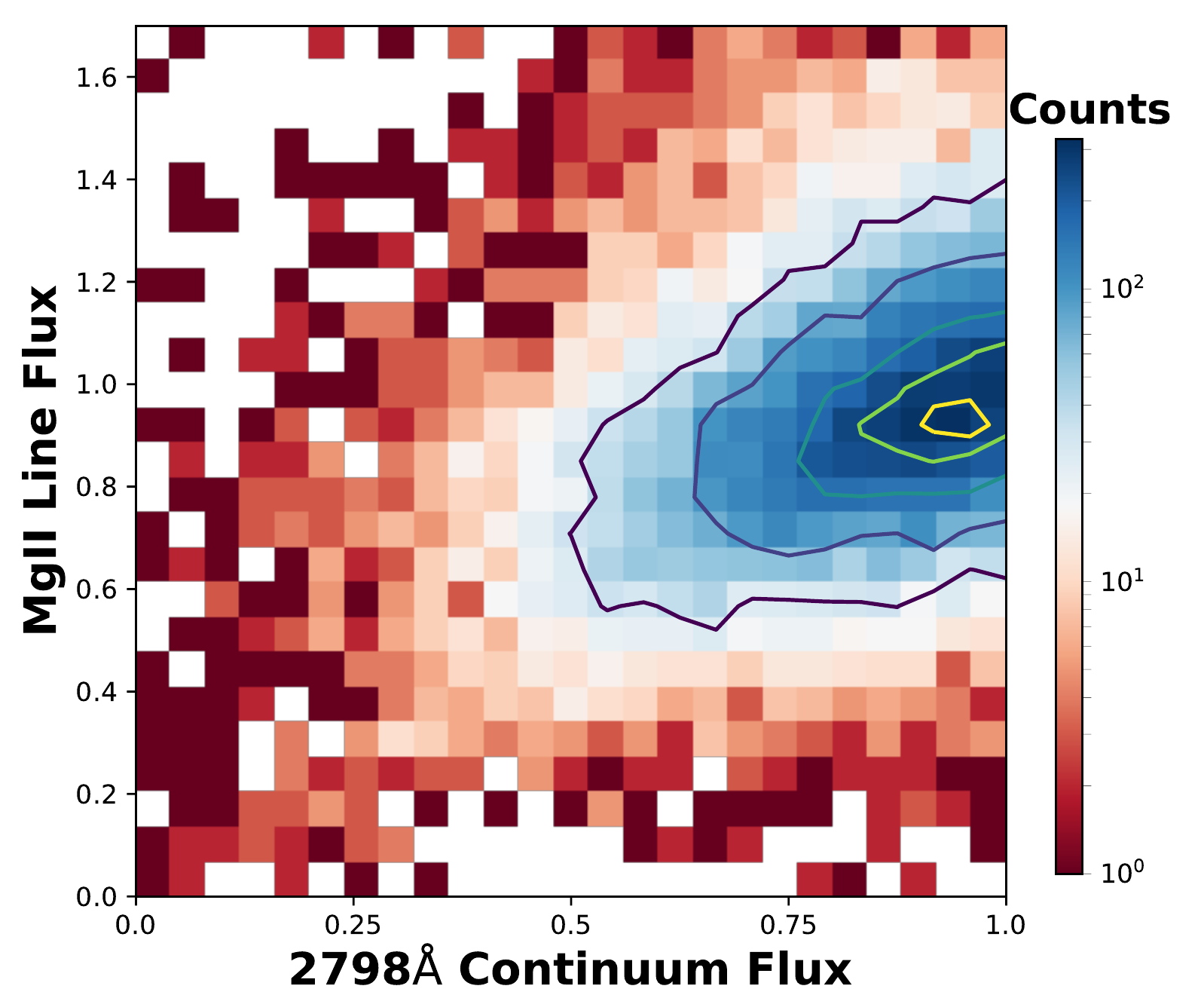}
\end{minipage}
\caption{\emph{left,top}: Fluxes in the Supervariable sample, normalised to the epoch of $maximum$ continuum. There is a correlation between large continuum changes and large line flux changes. The plot shows the line flux plotted against the continuum as well as the fitted linear responsivity function. Two outlying data points have been marked with squares: J154341 on MJD 57818 (cyan) and J123228 on MJD 57845 (purple). \emph{left,bottom}: The residuals and $\chi^2_{red}$ for the response function. \emph{right}: The Mg~II line and 2798{\AA} continuum fluxes for the Full Population sample, normalised to the epoch of maximum continuum. The data show a clear correlation. Most objects in the sample are concentrated on the right side of the plot, exhibiting only small changes in both line and continuum. The colours in this 2D histogram correspond to the number of objects. The contours are placed at 10, 25, 50, 75, and 90\% of the maximum count number for the sample.}
\label{fig:mgii_linfit_max}
\end{figure*}

\begin{figure*}
\centering
\hspace*{-1.1cm}\begin{minipage}{.55\textwidth}
\includegraphics[width=\textwidth]{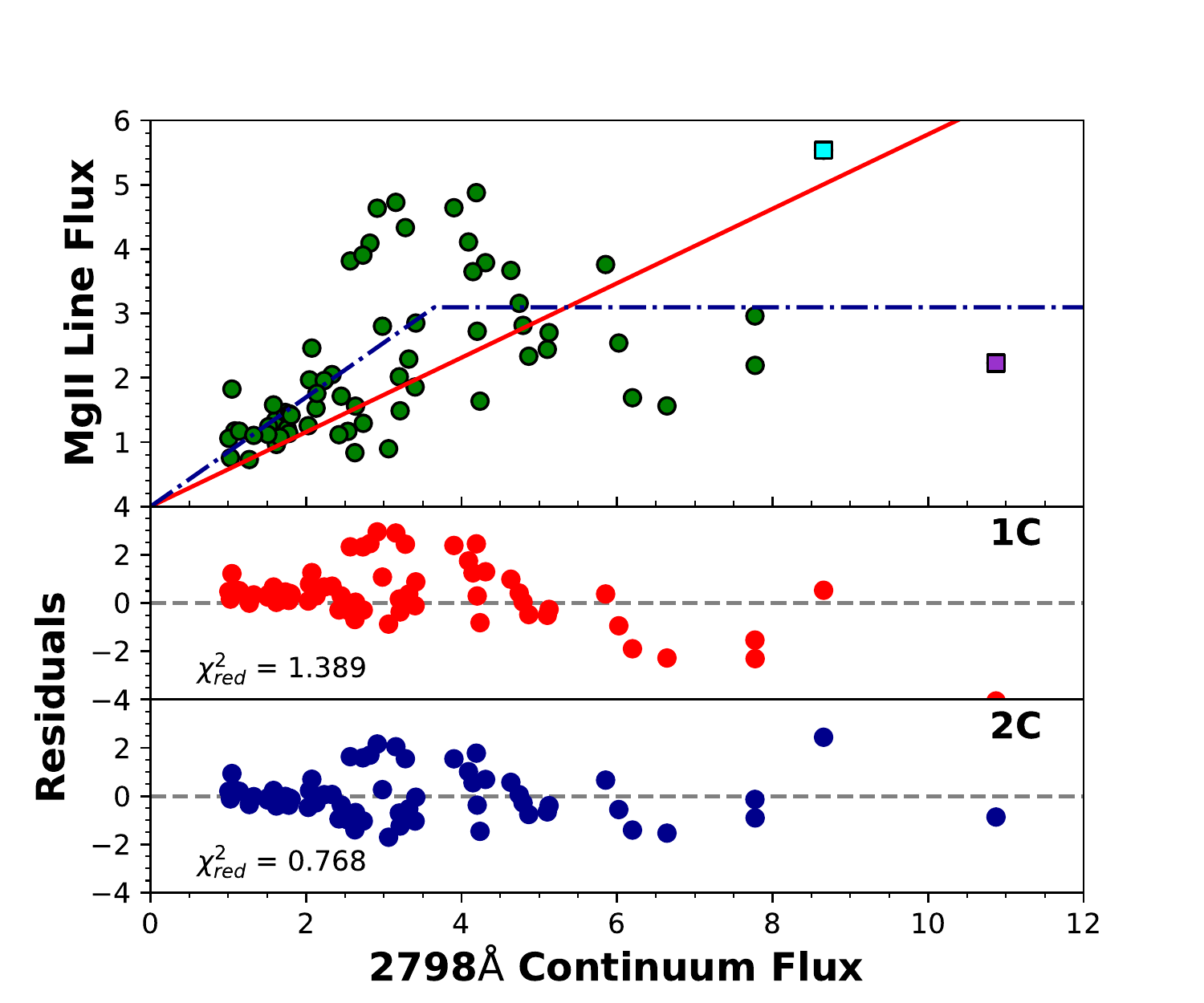}
\end{minipage}%
\begin{minipage}{.56\textwidth}
\includegraphics[scale=.55]{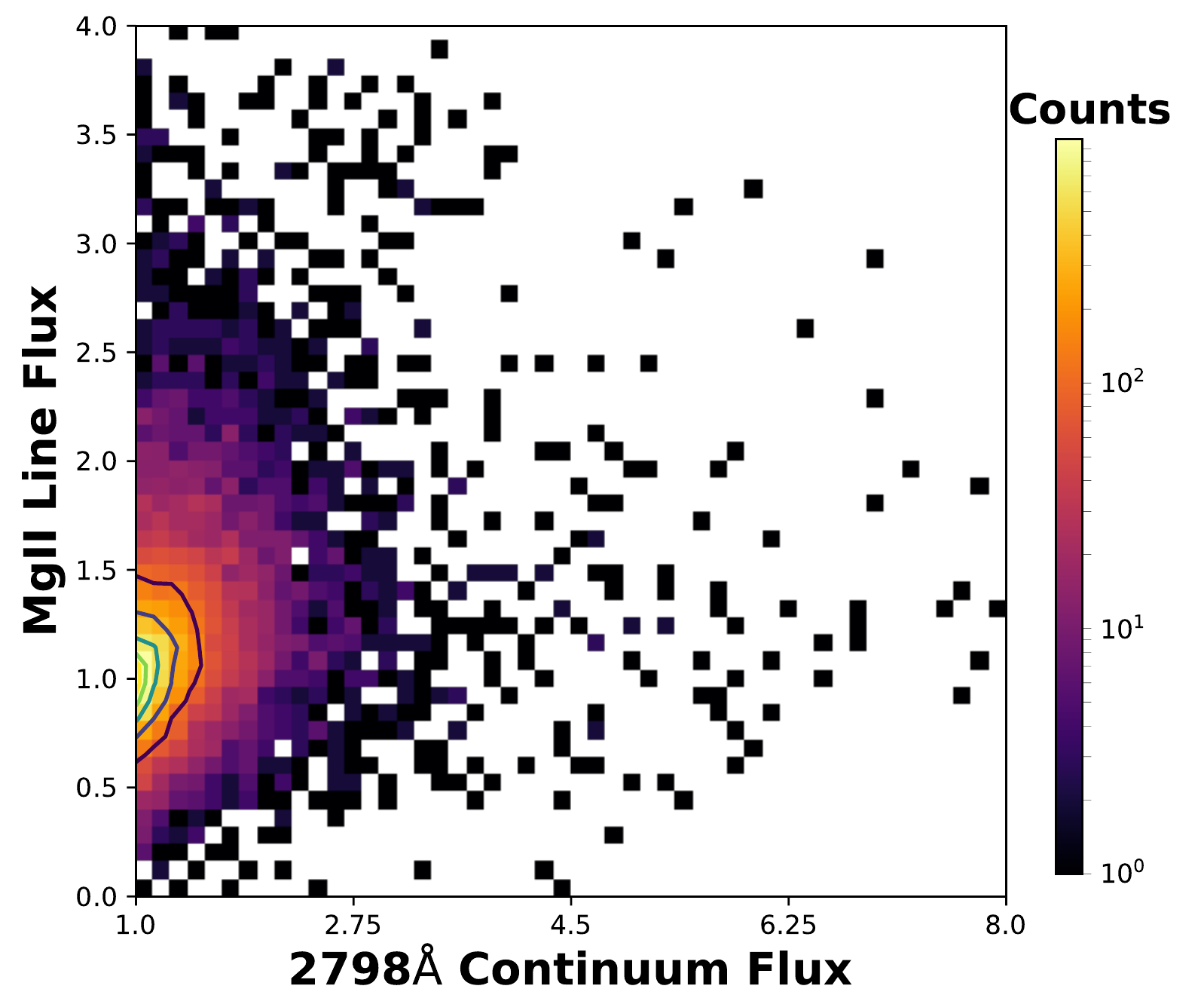}
\end{minipage}
\caption{\emph{left,top}: Normalised to the epoch of $minimum$ continuum level, the data for the Supervariable quasar sample show a significant scatter. This is particularly clear for the largest fractional flux changes. The panel also shows the two fitted responsivity functions. Two outliers are marked with squares: J022556 on MJD (cyan) and J002311 on MJD 55480 (purple). These objects are discussed in more detail at the end of section~\ref{sec:flux}. \emph{right,bottom}: The residuals from the two fitted functions as well as the $\chi^2_{red}$ for the fits. The two-component fit (lower) results in a smaller reduced $\chi^2$ value. As $\chi^2_{red} < 1$, this could be a result over overfitting. \emph{left}: The distribution in the $f_{2798}$-$f_{\mathrm{MgII}}$ parameter space of the Full Population sample, when normalised to the epoch of minimum continuum. The distribution shows a linear correlation. The layout of this figure is the same as that of the right panel in Figure~\ref{fig:mgii_linfit_max}.}
\label{fig:mgii_linfit_min}
\end{figure*}

\noindent The results of the normalisation to the epoch of the highest continuum are presented in Figure~\ref{fig:mgii_linfit_max}. All fluxes in this figure are fractional changes in flux: the further away from unity the normalised value is, the greater the relative change. The data for both samples show a strong scatter, but there is a pattern to the relation between line and continuum fluxes. Large $relative$ changes in continuum are matched by large relative changes in line emission. For smaller changes in continuum flux the scatter in line fluxes becomes greater. The distribution of fluxes in the Full Population sample is concentrated in the region where normalised line and continuum fluxes are $\sim$1. These are spectral pairs where the overall flux change has been small, likely relating to quasars that were not varying significantly over the period of observation. The population does show a clear linear correlation, however.\\
\indent For the Full Population sample we have visually examined the quality of the data and fits for 236 of the strongest outliers in the normalised sample. In a small fraction of the cases ($\sim$5\%) large changes were associated with changes in data quality between observations, such as missing data in the spectrum that partially covered the Mg~II range in one of the epochs. A relatively low S/N in both spectra may have affected the reliability of the fit in $\sim$30\% of the spectra. However, for most objects the measured flux change is visible in the spectrum. Given the small fraction of outliers overall, we are therefore confident in the identification of the trend visible in the right panel of Figure~\ref{fig:mgii_linfit_max}.\\
\indent The data in Figure~\ref{fig:mgii_linfit_max} are fit with the response functions defined in Section~\ref{sec:method}. A summary of results of the relevant statistical tests is given in Table~\ref{tab:stat_norm_both}. For the Supervariable sample the results of the one-component fit are included in red in the plot. The right bottom panel shows the associated residuals, as well as the $\chi^2_{red}$. The sequence test statistic, S (section~\ref{sec:method}), for this fit is 31 corresponding to a p-value ($p_s$) of 0.1. Fitting the same functions to the Full Population sample, we find a slight preference for the two-component function (Table~\ref{tab:stat_norm_both}).\\
\indent For the normalisation to the minimum epoch (Figure~\ref{fig:mgii_linfit_min}) the scatter in both samples is clearly visible. The distribution of the data of the Supervariable sample suggests a pattern different from simple linear correspondence. There is a sharp initial rise, followed by a possible flattening of the responsivity. Note that the scatter for $f_c \gtrsim 6$ is particularly large. This is reflected in a lower Spearman correlation coefficient than for the normalisation to the maximum epoch, 0.67 versus 0.78 (Table~\ref{tab:stat_norm_both}). The tentative pattern in the data can be tested using the two responsivity functions. The results of both fits are included in Figure~\ref{fig:mgii_linfit_min}. The residuals are shown in the bottom two panels. The reduced $\chi^2$ value is lower for the two component fit, but the fact that the value is less than one indicates this function is likely overfitting the data. The results of the sequence test (listed in Table~\ref{tab:stat_norm_both}) indicate that neither fit describes the data particularly well.\\
\indent For the Full Population sample (right panel in Figure~\ref{fig:mgii_linfit_min}), the coefficients (Table~\ref{tab:stat_norm_both}) clearly indicate a correlation. Evidence of levelling off is not as clear as it was for the Supervariable sample. This is possibly due to the strong concentration of the distribution around objects with relatively small flux and continuum changes. A firm conclusion is that the Mg~II line and UV continuum flux are clearly correlated and scale linearly. This is an ensemble effect: the large scatter suggests that individual objects can show a different correspondence between line and continuum. The difference from object to object will be a combination of intrinsic differences between the objects, and an artefact of the cadence of observations. The latter is caused by the fact that we only have two snapshots of the full line response, which means we will observe the line in different stages of its response for every target (in agreement with the evolutionary `sequence' proposed in \cite{GUO19b}).\\
\begin{table}
\centering
\caption{Correlation coefficients and the fitting statistics for the two normalisations, for the Supervariable and Full Population samples. The top row lists the Spearman ($r_s$) and Pearson ($r$) correlation coefficients. The bottom rows list the $\chi^2_{red}$ and the p-value associated with the sequence test statistic S, for the one and two component linear fits.}
\label{tab:stat_norm_both}
\begin{tabular}{l | c  c | c  c}
\toprule
\toprule
& \multicolumn{2}{c}{Maximum} & \multicolumn{2}{c}{Minimum}\\
& Supervar. & Full Pop. & Supervar. & Full Pop. \\
\hline
$r_s$ & 0.78 & 0.32 & 0.67 & 0.30 \\
$r$ & 0.76 & 0.30 & 0.46 & 0.32 \\
\hline
\begin{tabular}{@{}c c@{}} 1C: & $\chi^2_{red}$ \\ & $p_s$ \end{tabular} & \begin{tabular}{@{}c@{}} 1.01 \\ 0.10 \end{tabular} & \begin{tabular}{@{}c@{}} 1.15 \\ $\sim$1 \end{tabular} & \begin{tabular}{@{}c@{}} 1.39 \\ 0.94 \end{tabular} & \begin{tabular}{@{}c@{}} 1.28 \\ $\sim$1 \end{tabular}\\
\hline
\begin{tabular}{@{}c c@{}} 2C: & $\chi^2_{red}$ \\ & $p_s$ \end{tabular} & \begin{tabular}{@{}c@{}} 0.92 \\ 0.47 \end{tabular} & \begin{tabular}{@{}c@{}} 1.02 \\ $\sim$1 \end{tabular} & \begin{tabular}{@{}c@{}} 0.77 \\ 0.97 \end{tabular} & \begin{tabular}{@{}c@{}} 0.93 \\ 0.98 \end{tabular}\\
\bottomrule
\bottomrule
\end{tabular}
\end{table}
\indent The metrics of correlation and responsivity for the two samples are summarised in Table~\ref{tab:test_varsdss}. In all cases the Pearson and Spearman coefficients show a strong correlation between the line and the continuum flux. The columns labelled `slope' list the gradient of the one-component linear fit to the data. The error on the slope was calculated in a Monte Carlo simulation (N=10,000) by removing a random selection of data points each iteration, 10 for the Supervariable sample and 3,000 for the Full Population Sample, and re-fitting. The slopes for the two data-sets differ significantly. For both normalisations the gradient for the Supervariable sample indicates a stronger response of the line to the continuum than in the Full Population sample. This means that for a sample dominated by large continuum flux changes the line responsivity differs from a sample dominated by quasars with a more tranquil continuum.\\
\begin{table}
\centering
\caption{A comparison of various metrics of responsivity between the Supervariable and the Full Population sample for both normalisations. The included values are the slope of a linear fit to the data, the Pearson coefficient ($r$) and the Spearman coefficient ($r_S$).}
\label{tab:test_varsdss}
\begin{tabular}{l | c c c | c c c}
\toprule
& \multicolumn{3}{c}{Maximum} & \multicolumn{3}{c}{Minimum}\\
& Slope  & $r$ & $r_S$ & Slope  & $r$ & $r_S$\\
\hline \hline
Supervariable & $1.39 \pm 0.03$ & 0.76 & 0.78 & $0.58 \pm 0.03$ & 0.46 & 0.67\\
Full Population & $1.13 \pm 0.00$ & 0.30 & 0.32 & $0.74 \pm 0.01$ & 0.32 & 0.30\\
\bottomrule
\end{tabular}
\end{table}

\subsubsection{The Responsivity Measure}
\begin{figure}
\centering
\includegraphics[width=\columnwidth]{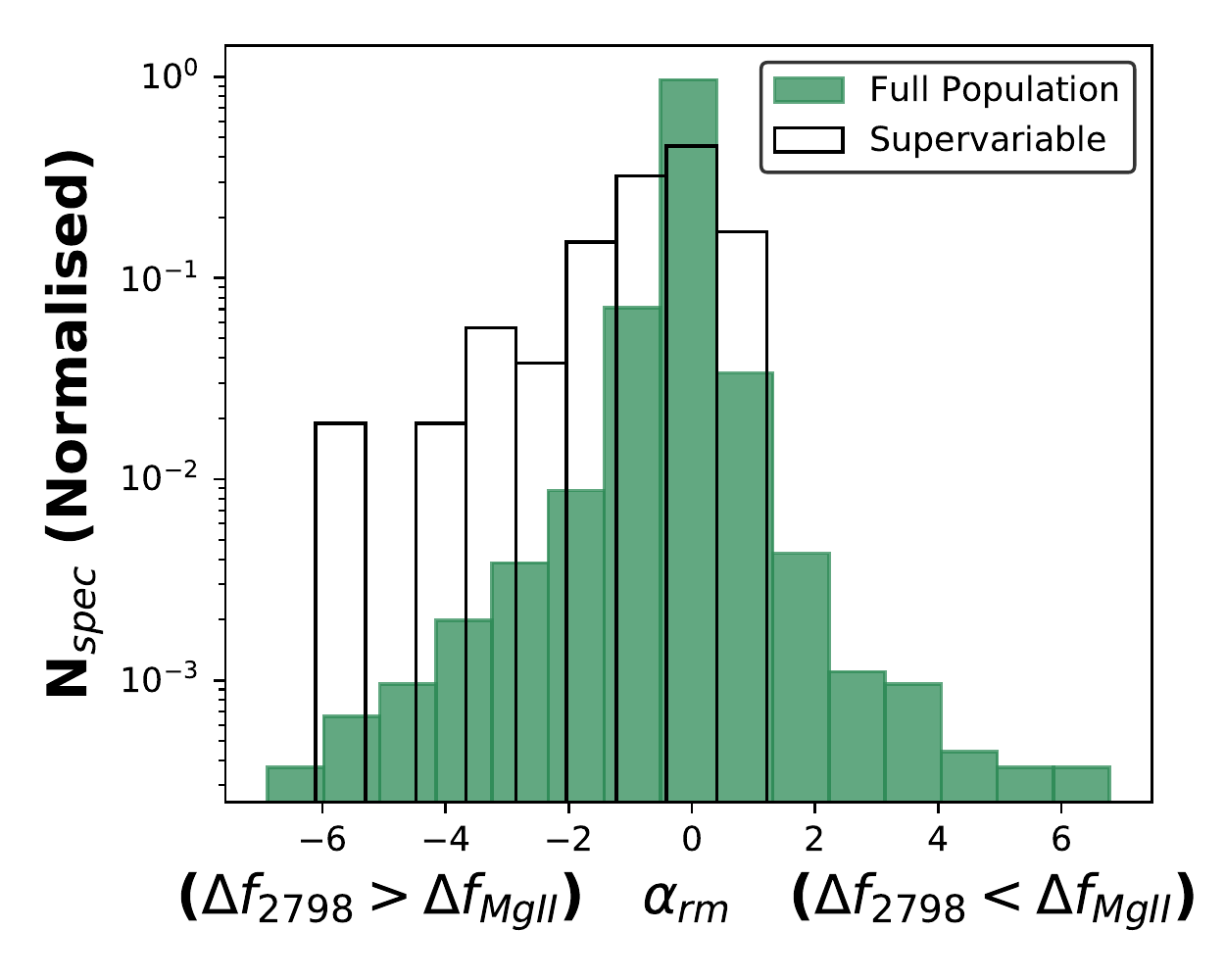}
\caption{A comparison of the distribution of $\alpha_{rm}$ in the two Mg~II samples. Both the Supervariable and the Full Population sample appear skewed to negative values of $\alpha_{rm}$, although this trend is more prominent for the Supervariable sample. Negative values are associated with an under-responsive Mg~II line flux and positive values with over-responsivity.}
\label{fig:mgii_rm_hist_sdssvar}
\end{figure}

\begin{figure}
\centering
\includegraphics[scale=.9]{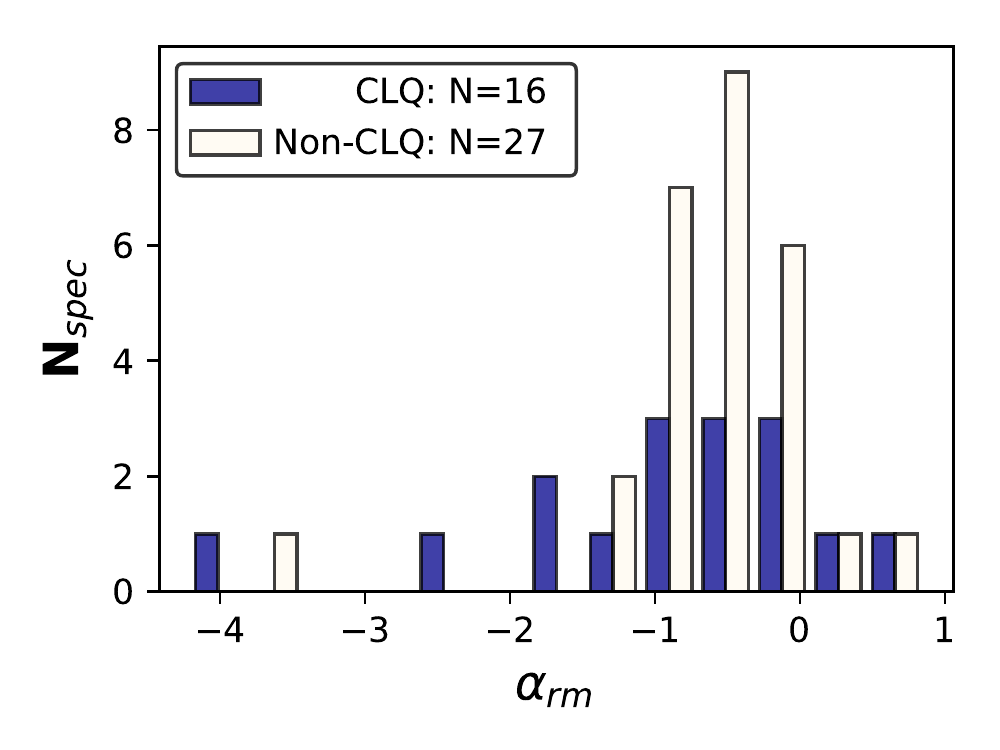}
\caption{The distribution of the responsivity measure among the CLQs and non-CLQs in the Supervariable sample. The average value of $\alpha_{rm}$ is used for objects with more than two spectra available.}
\label{fig:mgii_rm_clq}
\end{figure}

\noindent The distribution of the responsivity measure, $\alpha_{rm}$ (section~\ref{sec:method}), in the two samples is shown in Figure~\ref{fig:mgii_rm_hist_sdssvar}. The responsivity measure was chosen to easily identify over-responsive and under-responsive objects in the data set. Objects with $\alpha_{rm}>0$ have a larger relative change in line flux than in the continuum and can be considered over-responsive. Conversely, $\alpha_{rm}<0$ indicates an under-responsive object. The distribution of the Supervariable sample is clearly skewed to the negative values. 
The same skewness can be seen in the distribution of $\alpha_{rm}$ for the Full Population sample, although in this case the distribution is strongly dominated by objects with $\alpha_{rm}$$\sim$0. The $\alpha_{rm}$ distribution therefore shows the same pattern as observed in the normalisation plots. Most objects show only limited variability of both line and continuum, resulting in a responsivity measure close to zero.\\
\indent There are a number of spectra on the positive side of the distribution, representing strong responses in the Mg~II line. For the objects that show a response greater than 1:1, the extremes in $\alpha_{rm}$ are not as large as for the low-responsivity cases. This implies that the difference between fractional flux changes in continuum and line flux for objects with $\alpha_{rm}>0$ is overall smaller than for the less responsive objects.  It should be noted, however, that if the average response function is shallower than a 1:1 response the objects with $\alpha_{rm}>0$ are truly exceptionally responsive.\\
\indent The responsivity parameter can also be used to quantify the link between Mg~II and H$\beta$ variability. As seen in Table~\ref{tab:var_avgtime}, the correspondence between CLQ behaviour and Mg~II variability varies. Figure~\ref{fig:mgii_rm_clq} shows the distribution of $\alpha_{rm}$ for CLQ and non-CLQ objects in the Supervariable sample. For objects with more than one spectrum the average of the responsivity measures is used. The distribution for the CLQs is clearly broader than for the non-CLQ objects and stretches to more negative values. For both distributions the centre lies slightly below 0, corresponding to objects where Mg~II only slightly tracks the continuum. We therefore again see a range of behaviours: some CLQs also show strong variability in Mg~II, whereas for some CLQs Mg~II barely responds at all to the changing continuum.

\subsubsection{Interesting Outliers}
\begin{figure*}
\includegraphics[width=\textwidth]{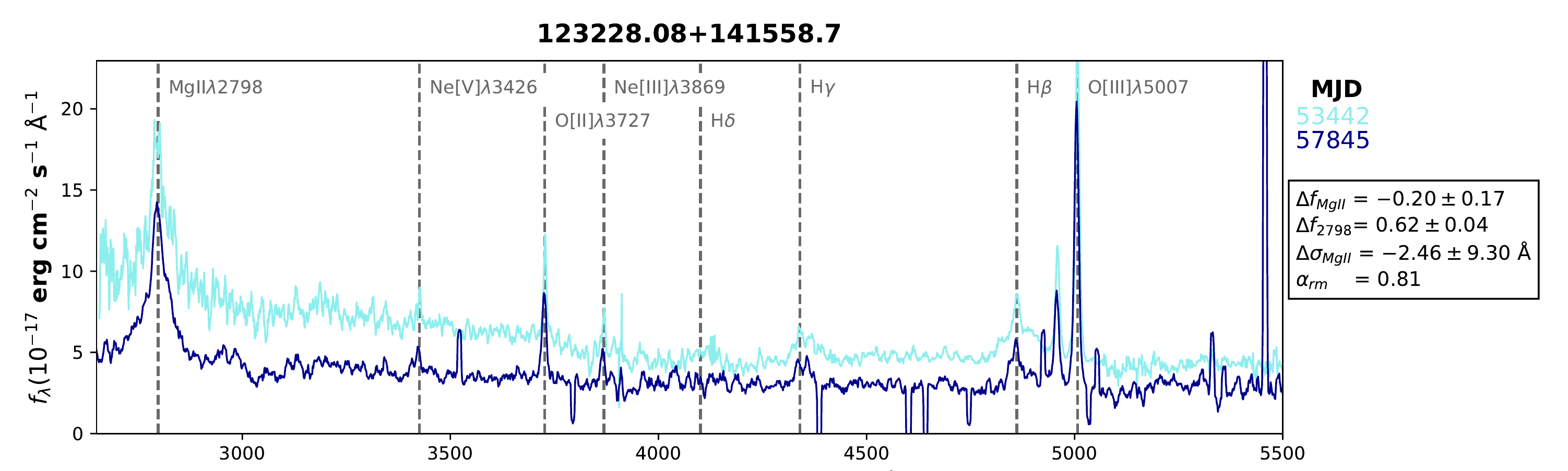}
\includegraphics[width=\textwidth]{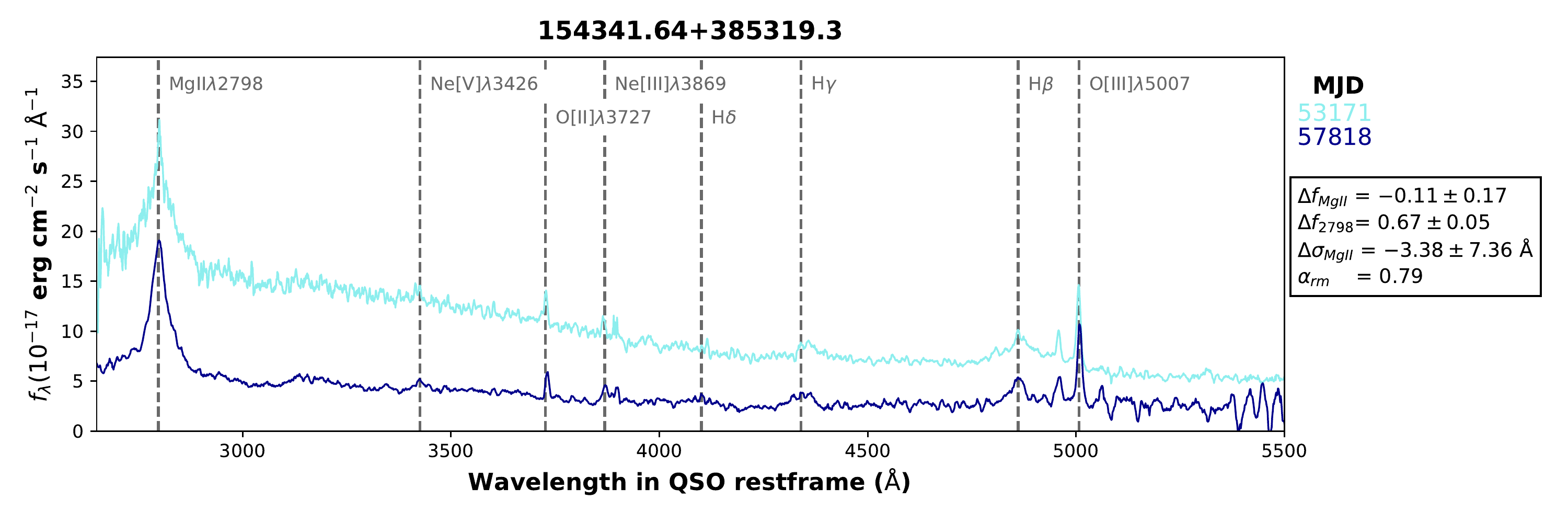}
\caption{The spectra for J123228 (\emph{top}) and J154341 (\emph{bottom}), the two outliers marked in the left panel of Figure~\ref{fig:mgii_linfit_max}. Both objects show large continuum changes, but a limited response of the Mg~II flux. The change in H$\beta$ is also limited for both objects.}
\label{fig:spec_outlier_12_15}
\end{figure*}

\begin{figure*}
\centering
\begin{minipage}{.49\textwidth}
\includegraphics[width=\textwidth]{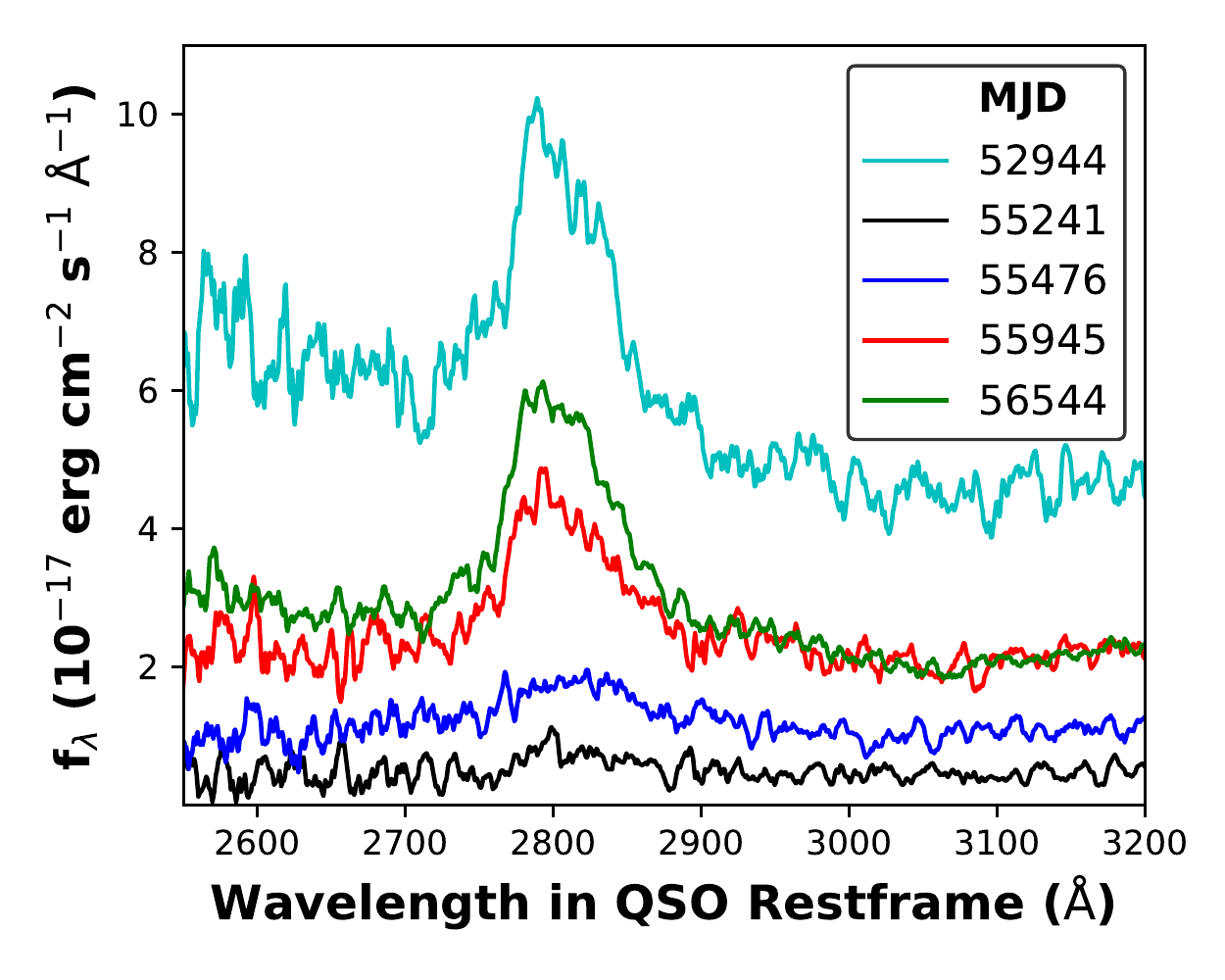}
\end{minipage}%
\begin{minipage}{.49\textwidth}
\includegraphics[width=.95\textwidth]{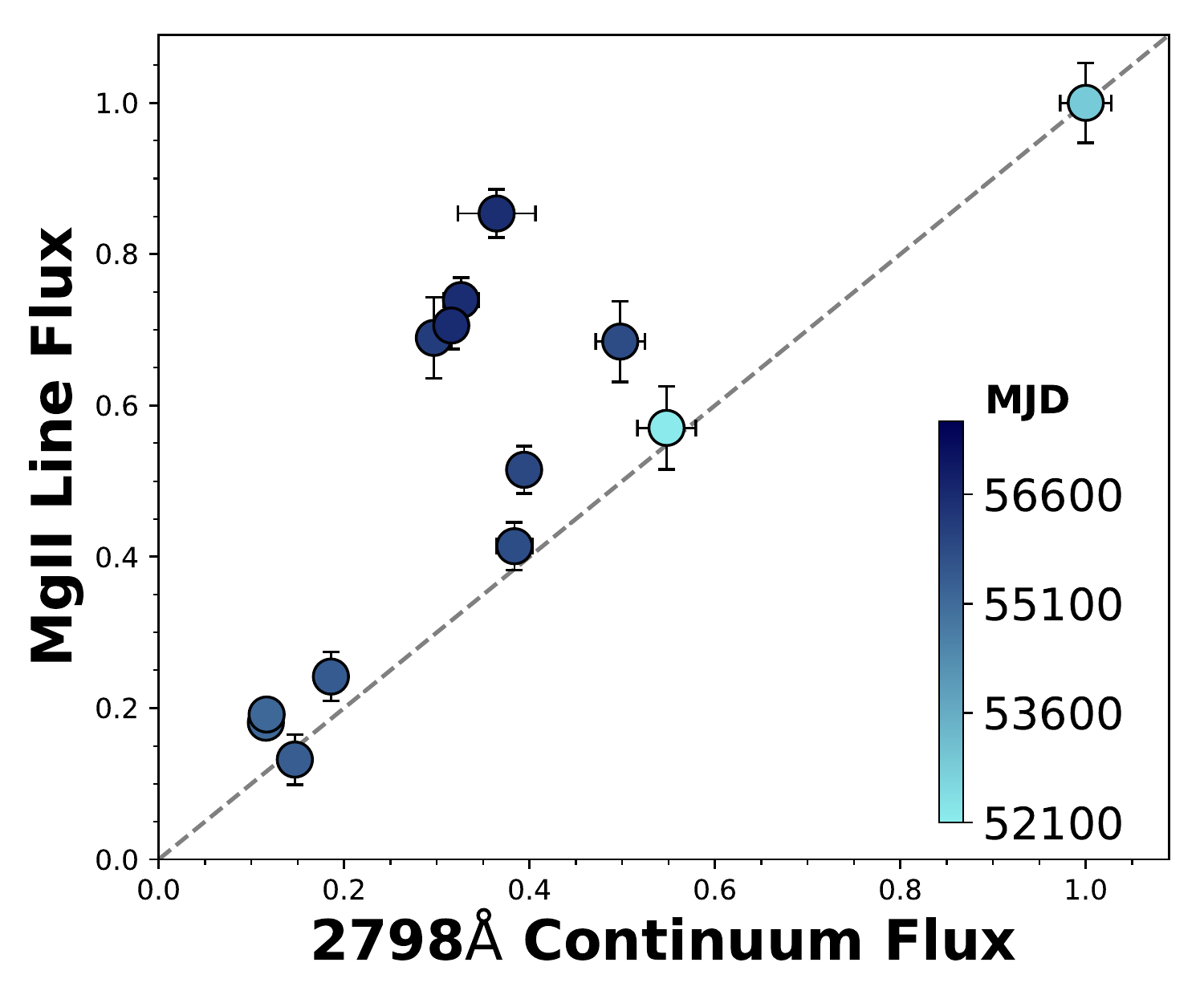}
\end{minipage}
\caption{\emph{left}: The changing  Mg~II line flux in J022556. The high cadence of observations for this object allows us to track the line flux as the continuum falls and rises over the course of $\sim$8 years in the QSO restframe. \emph{right}: The Mg~II line fluxes normalised to the maximum continuum. The epochs of the spectra are indicated with the colour, increasing chronologically from light to dark. This plot is based on the normalisation of the fluxes to the epoch of maximum continuum. A 1:1 response is indicated with the dashed grey line. The initial increase in line flux is matched approximately 1:1 by the Mg~II line. Following a collapse in line and continuum, however, the gradient of the line responsivity is considerably steeper as the continuum recovers.}
\label{fig:mgii_022556}
\end{figure*}

The responsivity functions and $\alpha_{rm}$ are used to define the response of the Supervariable sample. Compared to these measures there are individual objects that stand out. The first of these are identified in Figure~\ref{fig:mgii_linfit_max}, using the normalisation to the epoch of maximum continuum. The data points are found at the top left of the plot. These objects showed a large decrease of the continuum associated with an increase in the line flux. The spectra for the two objects, J123228 and J154341, are shown in Figure~\ref{fig:spec_outlier_12_15}. Neither of the objects lose their broad H$\beta$ flux. This could indicate that the response of the BLR to the drop in continuum has not occurred yet. The elapsed rest-frame time in both cases is close to two decades. An alternative hypothesis is that the observed line fluxes are in response to a flare in the continuum that occurred in the intervening years. The increasing line flux in J154341 seems coupled to a narrowing of the line, indicating a change in the physical conditions in the line forming region.\\
\indent The clearest outlier with respect to the normalisation to the lowest continuum (Figure~\ref{fig:mgii_linfit_min}; purple square), shows a continuum change of a factor $\sim$10 coupled to a line flux change of a factor $\sim$2. The object associated with this spectrum is J002311. The spectra for this object are shown in panel $c$ of Figure~\ref{fig:mgii_var_behaviour}. The spectrum marked in the right panel of Figure~\ref{fig:mgii_linfit_min} was taken on MJD 55480. The high state continuum was associated with a strong broad H$\beta$ line, which disappears in the lower state spectra. The cadence of observations for J002311 allows for a more detailed view of the evolution. The object is varying rapidly: it doubles its continuum flux on a restframe timescale of $\sim$14 years, and drops it by a factor 10 in the following decade. It is plausible the observation was made before the line had had enough time to respond in full to the continuum changes.\\
\indent Another interesting spectrum is marked in cyan in Figure~\ref{fig:mgii_linfit_min}. This epoch shows a line flux change greater than a factor 5. The spectrum belongs to J022556, for which we have a high number of spectra available because the object was part of extra deep BOSS plates\footnote{https://www.sdss.org/dr12/spectro/special\_plates/}. This QSO was identified as a CLQ in MCL16, where its strong Mg~II variability was also noted. The high cadence of observations allows us to see there is evidence of varying responsivity for this object.\\
\indent Figure~\ref{fig:mgii_022556} shows a close-up of the changing Mg~II line for J022556 (\emph{left}) as well as the line and continuum fluxes using the normalisation to the maximum continuum (\emph{right}). The spectra cover $\sim$8 years in the QSO restframe. The shading of the markers indicates the progression of time. The first epoch available (SDSS, MJD 52200) is represented by the light blue marker in the centre of the plot. The initial rise in the continuum is to the epoch of maximum continuum (top right). The gradient of this response is approximately unity. In the next spectrum the continuum has collapsed, and the line has also almost disappeared (bottom left corner). When the continuum rises again, the gradient is considerably steeper.\\
\indent The epoch dependent responsivity visible in J022556 (and J002311) indicates that the scatter around the responsivity functions (in Figures~\ref{fig:mgii_linfit_max} and \ref{fig:mgii_linfit_min}) could be the result of sampling the varying objects at different stages during the BLR response to the continuum. The response of the line to the continuum does not follow a single gradient for \emph{the same} object, therefore we can expect considerable scatter around any linear response function fitted to a large ensemble.\\
\subsection{The Intrinsic Baldwin Effect}
\label{sec:flux_ibe}
\begin{figure}
\centering
\includegraphics[scale=0.5]{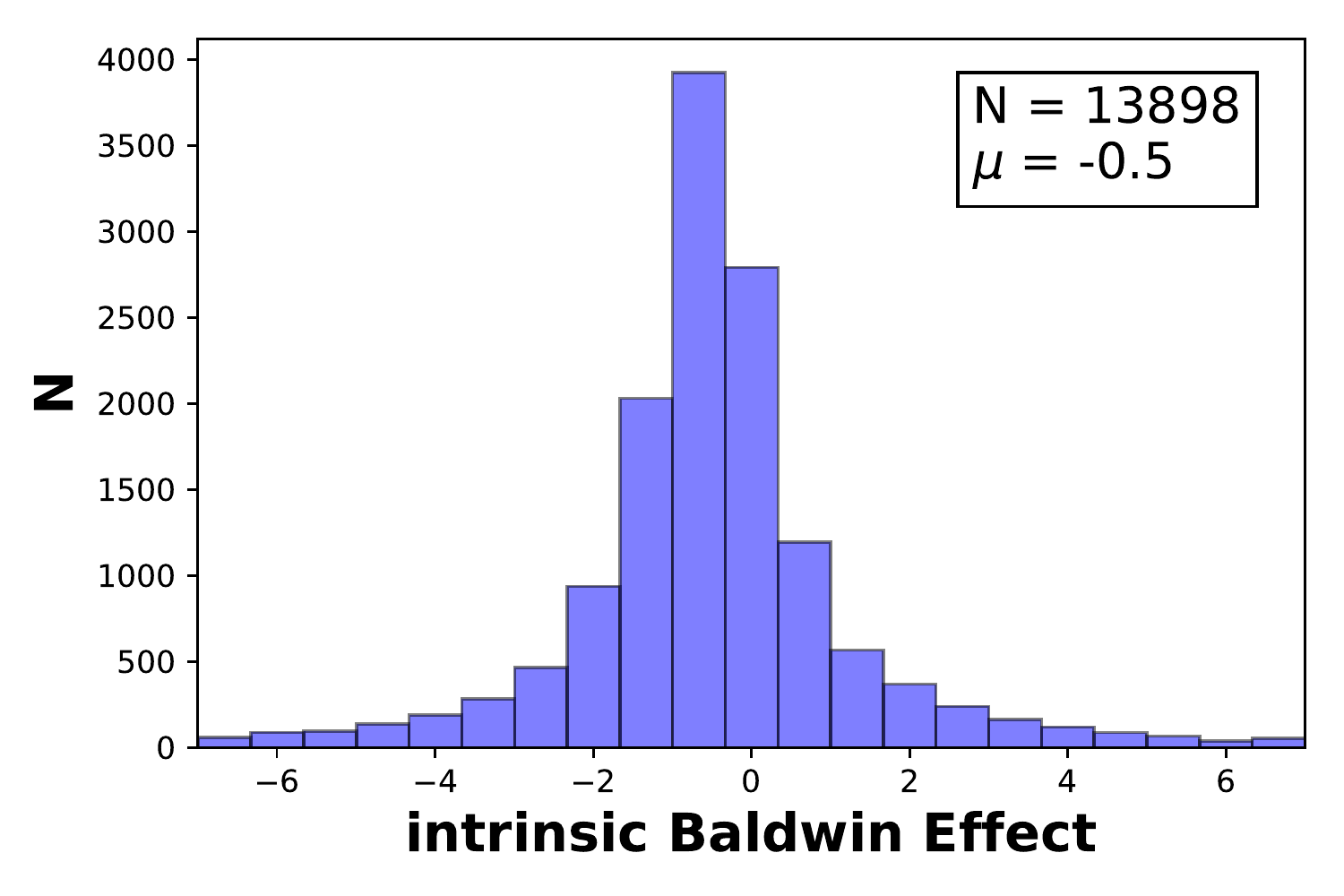}
\caption{A histogram of the intrinsic Baldwin Effect for the Full Population sample. The number of spectral pairs, N, and the sample average, $\mu$, are listed in the top right corner. The iBE is calculated as $\Delta$EW/$\Delta$L$_{2798}$ for each pair. Although there are some strong outliers, the distribution is clearly centred around $-0.5$. This is significantly lower than the slope of the ensemble BE of $-0.16$ calculated for the data from \citet{SHE11}.}
\label{fig:mgii_ibe}
\end{figure}

A final measure for the responsivity is the intrinsic Baldwin Effect (iBE), which measures the change in EW compared to the change in continuum luminosity for individual objects. For the Full Population sample the iBE is calculated as $\Delta$EW/$\Delta$L$_{2798}$ for each spectral pair. For the Supervariable sample the same is done for each chronologically sequential pair. The results for the Full Population sample are shown in Figure~\ref{fig:mgii_ibe}.\\
\indent The distribution is largely symmetric and centred around a slope of $-0.5$. For the Supervariable sample the average slope is $-0.54$. Unfortunately there are no literature values of the Mg~II iBE to compare these results to. For other UV lines such as CIV$\lambda$1549 and for the Balmer lines the slope of the intrinsic BE is significantly steeper than for the ensemble BE \citep{OSM99,RAK17}. We find that the same holds for Mg~II: the eBE of the DR7 sample has a slope of $-0.16$, shallower than the slope of the iBE.

\subsection{Line Responsivity over Time}
\label{sec:mgii_sdss_dt}
\begin{figure}
\centering
\includegraphics[width=\columnwidth]{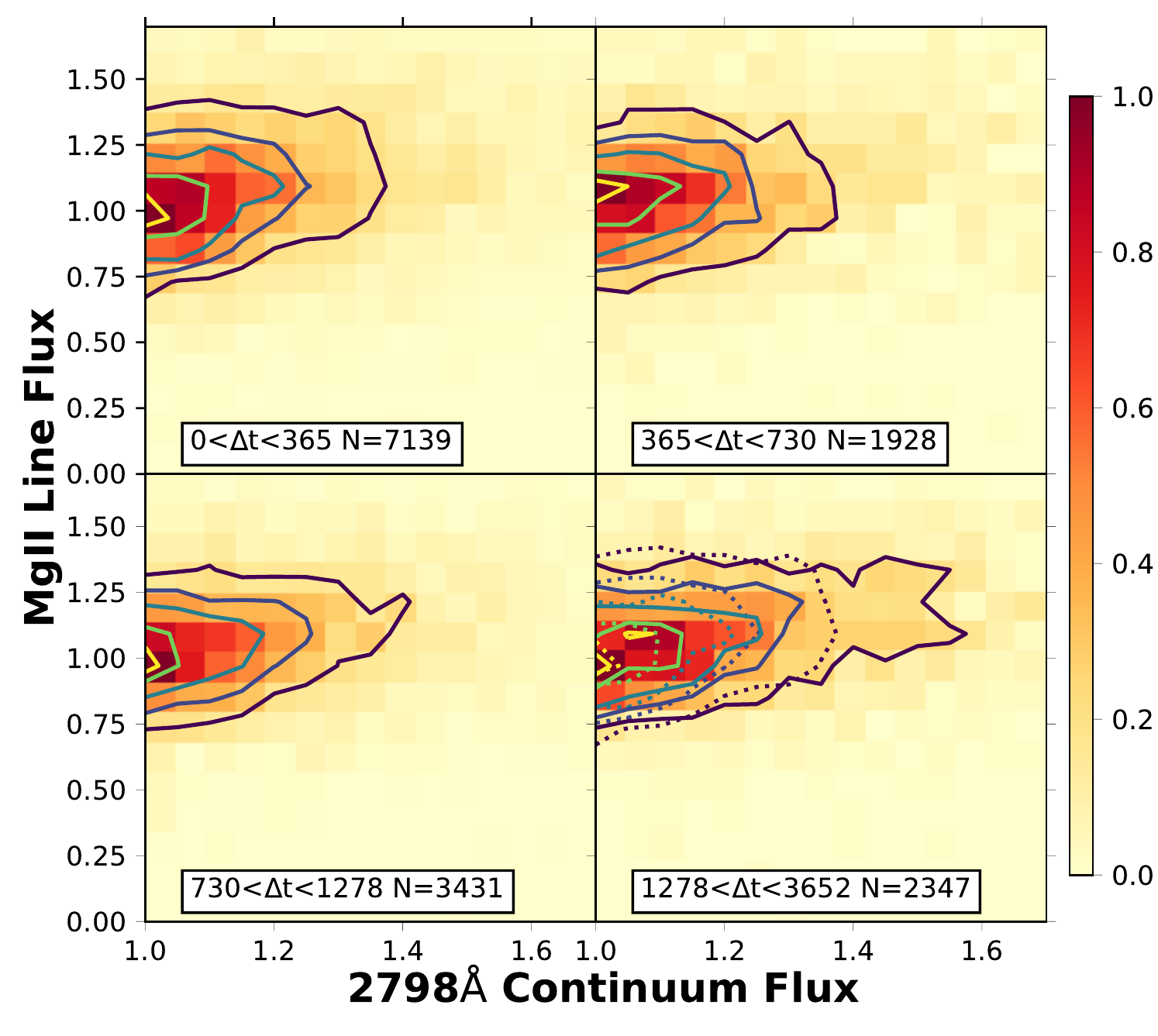}
\caption{The Full Population sample normalised to the minimum continuum state, binned by $\Delta$t between observations, in days in the QSO restframe. The contours are set to 10, 25, 50, 75, and 90\% of the maximum value. To observe any difference in response over time the contours from the shortest elapsed time bin (\emph{top left}) are included in the panel for the longest elapsed time (\emph{bottom right}). The maximum number of counts per bin in each panel has been normalised to one, to allow for comparison between the four subsets.}
\label{fig:mgii_dtbin_sdss}
\end{figure}

If the line responsivity changes over time, this should manifest itself as a dependency on $\Delta t$, the elapsed time between the spectral measurements for each pair. The Full Population sample is large enough to investigate the effect of $\Delta t$ in more detail. The first test is to divide the data-set into $\Delta t$ bins and calculate the normalised fluxes in the subsets as in the previous section. The results are shown in Figure~\ref{fig:mgii_dtbin_sdss}. To emphasise the visualisation of the outliers, the normalisation to the epoch of minimum continuum is used. To ensure a good sample size in each $\Delta t$ bin, the bin size increases for longer elapsed times. The time bins represent re-observations of the object within one year, between one and two years, between two and three-and-a-half years, and three-and-a-half up to 10 years respectively. All counts have been normalised by dividing by the maximum per plot.\\
\indent Two changes take place in the $\Delta t$ bins in Figure~\ref{fig:mgii_dtbin_sdss}: the shape of the contours changes and the gradients of the line responsivity appear to decrease. The first effect is visible as an elongation of the contours as $\Delta t$ increases. This implies that large changes in the continuum are more common for large $\Delta t$.  Changes in flux that are very unlikely to occur within the first year become more likely as $\Delta t$ increases. A longer waiting time means a greater chance of observing a large change in UV continuum. This agrees with established results on quasar continuum variability \citep{MCL10,WEL11}. The change in shape also implies that for greater $\Delta t$ the distribution has a larger spread.\\
\indent The second change visible in Figure~\ref{fig:mgii_dtbin_sdss} is in the slope of the linear correlation and can be seen in the plots as a gradual tilting of the contours. The contrast is most pronounced between the shortest and longest time differences: as a visual aid the contours from the top left plot have been included as dotted lines in the bottom right plot. Although the effect is subtle, the slope decreases over time. This effect can be further quantified by a linear fit to the data. This is the same one-component fitting function used in the previous section. The fitted slopes and the errors are listed in Table~\ref{tab:dt_slopes}. The results show a steady decline with $\Delta t$ in the slope of the responsivity. For a larger elapsed time between observations the response of the line to continuum changes therefore appears to be smaller.\\
\begin{table}
\centering
\caption{Slopes of the linear correlation between normalised $f_{\mathrm{MgII}}$ and $f_{2798}$ for the $\Delta t$ subsets shown in Figure~\ref{fig:mgii_dtbin_sdss}. The slope shows a slight but steady evolution with $\Delta t$: the line responsivity decreases for longer time intervals between measurements. The error on the slopes is calculated from the covariance matrix produced by the \texttt{scipy} fitting routine. The error on the measurements was estimated based on the scatter, as described in section~\ref{sec:method}. $\Delta t$ is given in the QSO restframe.}
\label{tab:dt_slopes}
\begin{tabular}{l| c c}
\toprule
\toprule
RF $\Delta t$ (days)& Slope & Uncertainty\\
\hline
$0<\Delta t<365$ & 0.746 & $0.006$\\
$365<\Delta t<730$ & 0.745 & $0.011$\\
$730<\Delta t<1278$ & 0.738 & $0.007$\\
$1278<\Delta t<3652$ & 0.724 & $0.007$\\
\bottomrule
\bottomrule
\end{tabular}
\end{table}

\begin{figure}
\centering
\includegraphics[scale=.5]{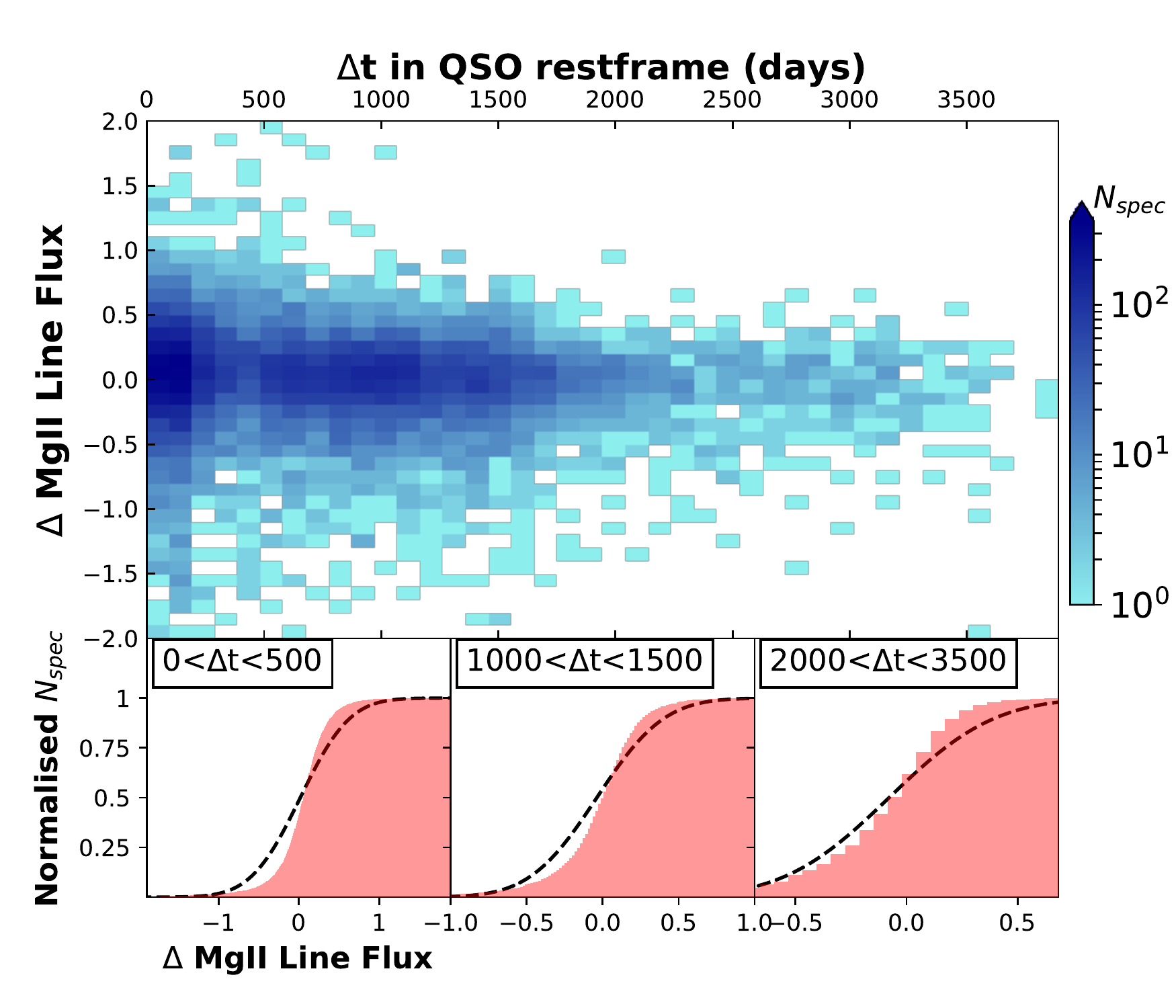}
\caption{\emph{top}: The normalised change in line fluxes ($\Delta f_{\mathrm{MgII}}$) plotted against the elapsed time between observations, in days in the QSO restframe. \emph{bottom}: Testing for normalcy on three different $\Delta t$ cuts in the data shows a significant deviation from a normal distribution. The three cumulative histograms represent the subsamples, marginalised over $\Delta t$. The normal CDF, calculated with the sample average and sample standard deviation, is plotted as the black dashed line as a visual aid for the comparison.}
\label{fig:mgii_ldt_sdss}
\end{figure}

\noindent A different approach to investigating the effect of elapsed time on the line responsivity is illustrated in Figure~\ref{fig:mgii_ldt_sdss}. Here we compare the distribution of the normalised change in line flux ($\Delta f_{\mathrm{MgII}}$) for several $\Delta t$ subsets. The top panel of Figure~\ref{fig:mgii_ldt_sdss} shows the distribution of $\Delta f_{\mathrm{MgII}}$ against $\Delta t$. The Full Population sample is skewed toward objects with a relatively small $\Delta t$, as is also evident in Figure~\ref{fig:mgii_dt_hist}. The sections in $\Delta t$ have been chosen to cover both the full range of elapsed times and include enough objects in each subsample to allow for meaningful statistical analysis. The $\Delta t$ bins are listed in the bottom three panels and the associated sample sizes are included in Table~\ref{tab:dt_moments}.\\
\indent If there is no pattern to the flux changes, we expect the data in each bin to follow a normal distribution. For each of the three subsets the distribution of $\Delta f_{\mathrm{MgII}}$ is therefore compared with a normal distribution using the Lilliefors test. The bottom three panels of Figure~\ref{fig:mgii_ldt_sdss} show the cumulative distributions of the three subsets, as well as the CDF of the normal distribution for a visual comparison. The results of the test are quite clear, rejecting the hypothesis the data are drawn from the normal distribution. The Lilliefors test provides a p-value$<10^{-10}$ in each case.\\
\noindent The distributions of $\Delta f_{\mathrm{MgII}}$ in the $\Delta t$ bins indicated in Figure~\ref{fig:mgii_ldt_sdss} and their deviation from a normal distribution can be quantified by calculating the sample moments. These are presented in Table~\ref{tab:dt_moments}. The moments of the distribution do change with time. The mean shifts to negative values as $\Delta t$ increases. This is likely the result of a selection bias: the spectral pairs in the lowest $\Delta t$ bin are more likely to have both observations made during the BOSS survey. For sources in the higher $\Delta t$ bins the probability is larger that the first of the spectra is from SDSS I/II. Sources that were in a relatively bright state during SDSS I/II were more likely to be included in DR7Q and therefore to be included in the Full Population sample. This makes it more likely to see transitions to dimmer states.  We conducted a similar analysis for $f_{2798}$ and included the results in Table~\ref{tab:dt_moments}. The continuum flux distributions show the same pattern as those of the line flux.\\
\indent The more interesting effects occur in the third and fourth moments of the distribution. The skewness indicates the distributions are skewed towards more positive values. This effect seems to decrease with $\Delta t$, however, particularly in the case of $\Delta f_{\mathrm{MgII}}$. A similar trend with $\Delta t$ can be seen in the kurtosis, which is a measure of the weight of the tails of a distribution. The two lower $\Delta t$ bins have quite prominent tails. This is also evident from the scatter in the top panel of Figure~\ref{fig:mgii_ldt_sdss}. The moment is significantly smaller for the higher bins, and more so for the line flux than for the continuum flux. The reduced number of strong outliers could indicate a reduced responsivity for objects in this bin. The data in Table~\ref{tab:dt_moments} indicate $|\Delta f_{2798}|$ and $\Delta t$ are therefore correlated parameters, making it difficult to see which of the two parameters most strongly affects the observed changes in the Mg~II line flux. However, we do note that large line flux changes are observed on all time scales. This is an indication that large continuum changes are an important driving force for changes in the line flux.
\begin{table*}
\caption{The sample moments of the $\Delta f_{\mathrm{MgII}}$ and $\Delta f_{2798}$ distributions for the $\Delta t$ subsamples shown in Figure~\ref{fig:mgii_ldt_sdss}. $\Delta t$ is in days in the restframe. There is an indication of change in the sample distributions over time. For the calculation of the moments the distributions have been filtered requiring $|\Delta f|<7$ (i.e. no flux changes greater than a factor 7 between epochs).}
\label{tab:dt_moments}
\noindent
\makebox[\linewidth]{\begin{tabular}{l | c c c | c c c}
\toprule
\toprule
& \multicolumn{3}{c|}{$\Delta f_{\mathrm{MgII}}$} & \multicolumn{3}{c}{$\Delta f_{2798}$}\\
& $0<\Delta t<500$ & $1000<\Delta t<1500$ & $2000<\Delta t<3500$ & $0<\Delta t<500$ & $1000<\Delta t<1500$ & $2000<\Delta t<3500$ \\
\hline
N & 6282 & 2472 & 502 & 6273 & 2475 & 499 \\
$\mu$ & 0.02 & -0.04 & -0.08 & 0.10 & -0.08 & -0.26 \\
$\sigma$ & 0.48 & 0.35 & 0.37 & 0.39 & 0.42 & 0.78 \\
skewness & -2.71 & -3.16 & 0.04 & -4.13 & -3.54 & -3.35 \\
kurtosis & 41.9 & 23.9 & 12.7 & 60.4 & 32.0 & 17.2 \\
\bottomrule
\bottomrule
\end{tabular}}
\end{table*}

\section{The Mg~II Line Profile}
\label{sec:profile}
\subsection{Changes in the Line Profile}
\label{sec:profile_quant}
\begin{figure}
\centering
\includegraphics[width=\columnwidth]{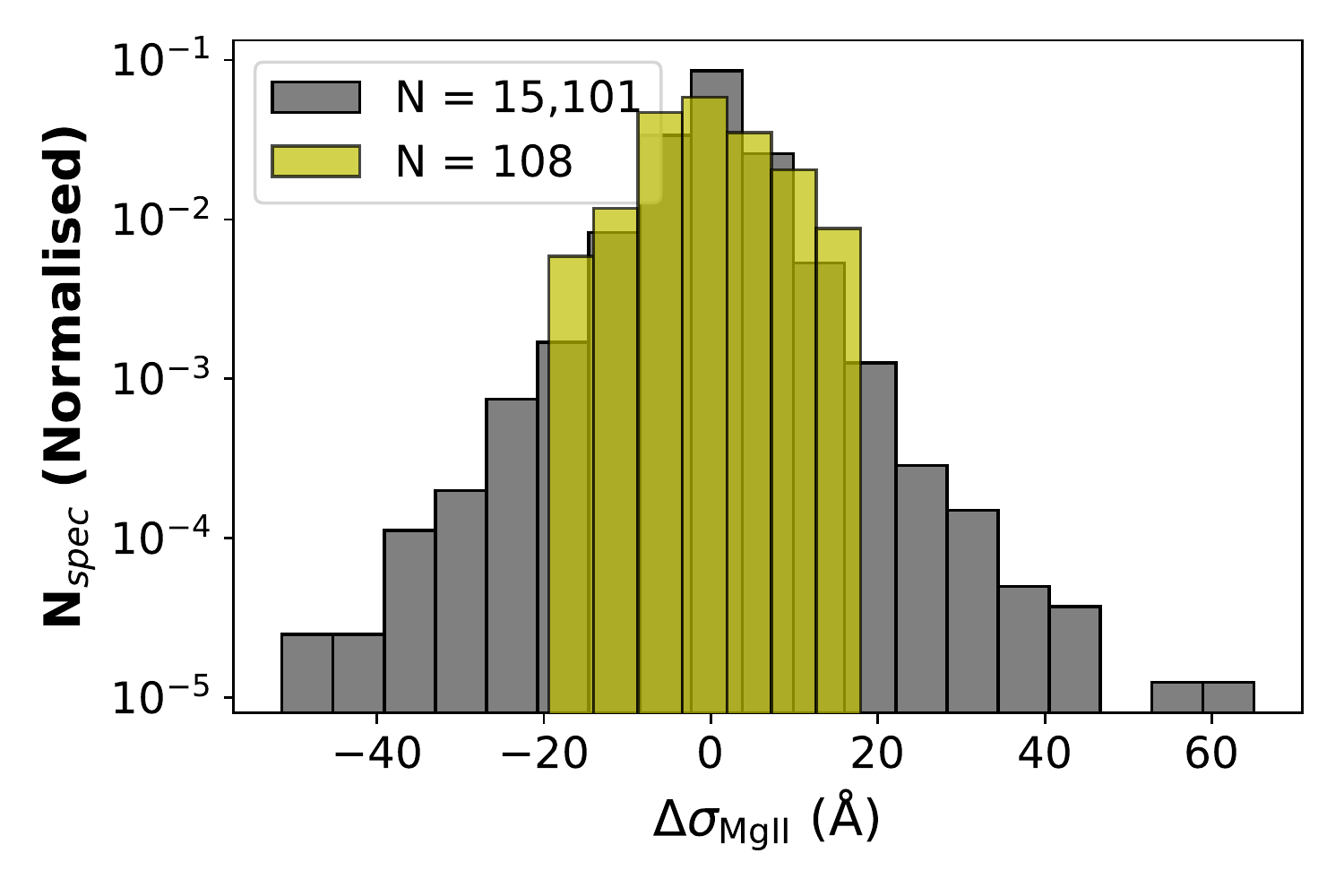}
\caption{The distribution of $\Delta\sigma_{\mathrm{MgII}}$ for the Full Population sample (\emph{grey}) and the Supervariable sample (\emph{yellow}). Each distribution has been normalised such that the surface area under the histogram equals one. N$_{spec}$ is the number of spectra in each bin. The distributions are strongly centred around objects with no observable change in line width.}
\label{fig:mgii_delsig_sdss}
\end{figure}

\begin{figure}
\includegraphics[width=\columnwidth]{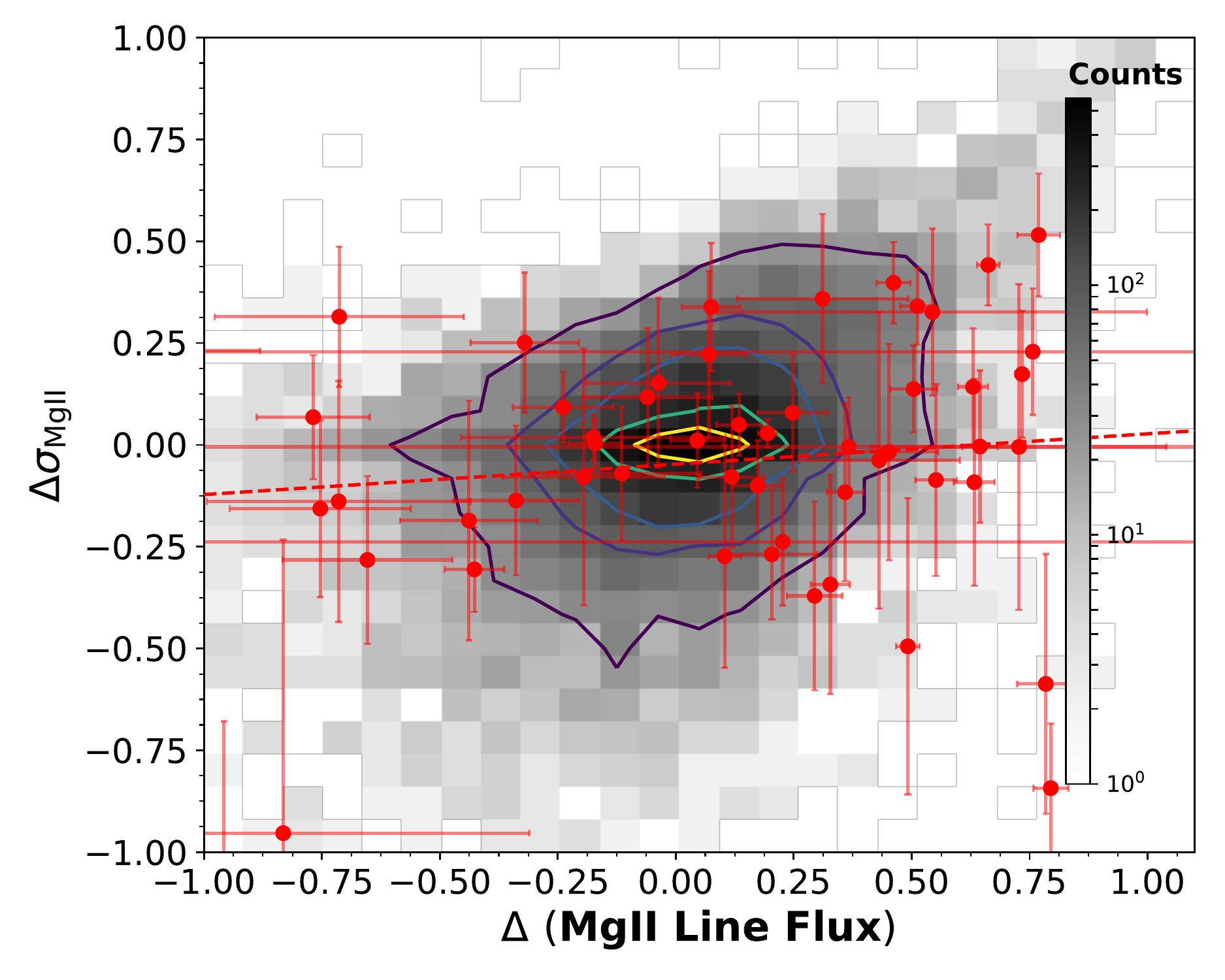}
\caption{The change in line width ($\Delta\sigma_{\mathrm{MgII}}$) plotted against the change in line flux. The values have been normalised to the preceding epoch and therefore represent a fractional change with respect to the older spectrum. The data for the Supervariable sample is shown in red, including the result of a linear regression of $\Delta\sigma_{\mathrm{MgII}}$ on $\Delta f_{\mathrm{MgII}}$. The Full Population sample is represented by the grey histogram. The contours indicate 5, 15, 25, 50 and 75\% of the peak value of the histogram. For most objects in the Full Population sample the two parameters do not seem to correspond at all. For large changes in line flux, associated with more extreme variability, a connection is possibly present.
}
\label{fig:mgii_sigvsline_both}
\end{figure}

\begin{figure}
\includegraphics[scale=.45]{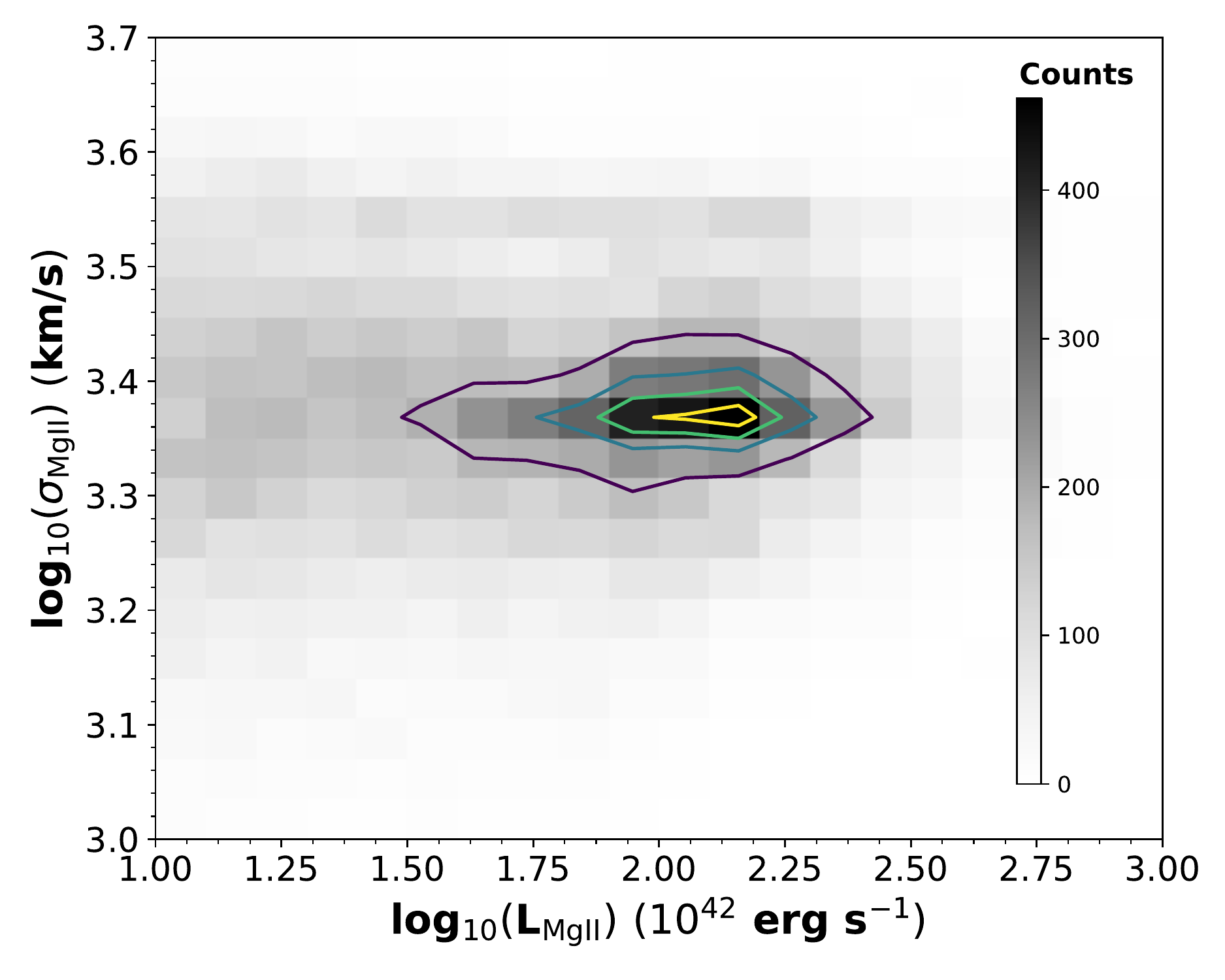}
\caption{The line width ($\sigma_{\mathrm{MgII}}$) plotted against the line luminosity for the Full Population sample. The contours represent 40, 60, 75, and 90\% of the peak numbers of counts per bin. A correlation is possibly present for lower luminosities, however most objects do not show such a correlation. A lack of correlation between line width and continuum luminosity would be in agreement with the findings in \citet{SHE11}.}
\label{fig:mgii_sigvsline_sdss}
\end{figure}

Line profiles provide insight into the kinematics of the line forming regions. The measure used to quantify the Mg~II profile is the width ($\sigma$) of the broadest Gaussian fit to the line. The widths are defined in the QSO restframe. Table~\ref{tab:var_results} lists the change in line width for the spectral pair with the largest change in continuum ($\Delta\sigma_{\mathrm{MgII}}$) for all targets in the Supervariable sample. The distribution of $\Delta\sigma_{\mathrm{MgII}}$ is shown in Figure~\ref{fig:mgii_delsig_sdss}. The distribution shows that most objects have relatively small changes in line width ($<10 {\rm\AA}$). However, there is also an extended tail towards larger changes. The large spread in the sample is another indication of the broad range of Mg~II variability among extremely variable quasars.\\
\indent The largest values of $\Delta\sigma_{\mathrm{MgII}}$ are $\sim$20{\AA} for the Supervariable sample. Under the assumption of Keplerian rotation of the BLR gas this corresponds to a change in rotational speed (along our line of sight) of approximately 2,100 km/s. For Mg~II the Keplerian speeds deduced from the FWHM are in the order 10$^{3.6}$ km/s \citep{SHE11}. The extremes in the observed changes in line width therefore imply large transitions.\\
\indent For the Full Population sample the distribution is strongly peaked around $\Delta\sigma_{\mathrm{MgII}}=0$, with a slight asymmetry towards decreasing line widths. Large changes in $\sigma$ are possible, but most objects exhibit only limited variability. This is likely the same effect as the observed concentration of flux changes around small values (Figure~\ref{fig:mgii_rm_hist_sdssvar}): most quasars in the sample have changed only slightly between epochs. For the Full Population sample we have visually inspected the spectral fits of 164 outliers in $|\Delta\sigma_{\mathrm{MgII}}|$ to exclude the possibility of issues with the fit or the data quality as a cause for the behaviour.\\
\indent Figure~\ref{fig:mgii_sigvsline_both} shows the change in line width plotted against the change in Mg~II flux, where both quantities have been normalised by dividing by the value from the first epoch of the pair. The two parameters represent a fractional change over time. The figure includes the result of a linear regression of $\Delta\sigma_{\mathrm{MgII}}$ on $\Delta f_{\mathrm{MgII}}$ for the Supervariable sample (\emph{red}). The slope of the line is $0.07 \pm 0.05$. Although the scatter in the data is strong, it is possible that there is a positive correlation between the magnitude \emph{and} direction of change of the line flux and the line width. For the Full Population sample a linear regression of $\Delta\sigma_{\mathrm{MgII}}$ on the change in line flux has a gradient of $0.09 \pm 0.01$.\\
\indent For the Full Population sample we show a comparison of $\sigma_{\mathrm{MgII}}$ with the fitted line luminosities in Figure~\ref{fig:mgii_sigvsline_sdss}. For the bulk of the spectra the line width appears unrelated to either the line or the continuum luminosity. This would be in agreement with results from \citet{SHE11}, who find no correlation between the FWHM of Mg~II and the 3000{\AA} luminosity. This is in contradiction with the expectation based on BLR breathing and could be an indication that the Mg~II line width is not entirely set by the gravitational potential or that the location of the line forming region does not vary strongly with the continuum flux.

\subsection{Skewed Profiles}
\label{sec:profile_skew}
\begin{figure}
\centering
\includegraphics[width=\columnwidth]{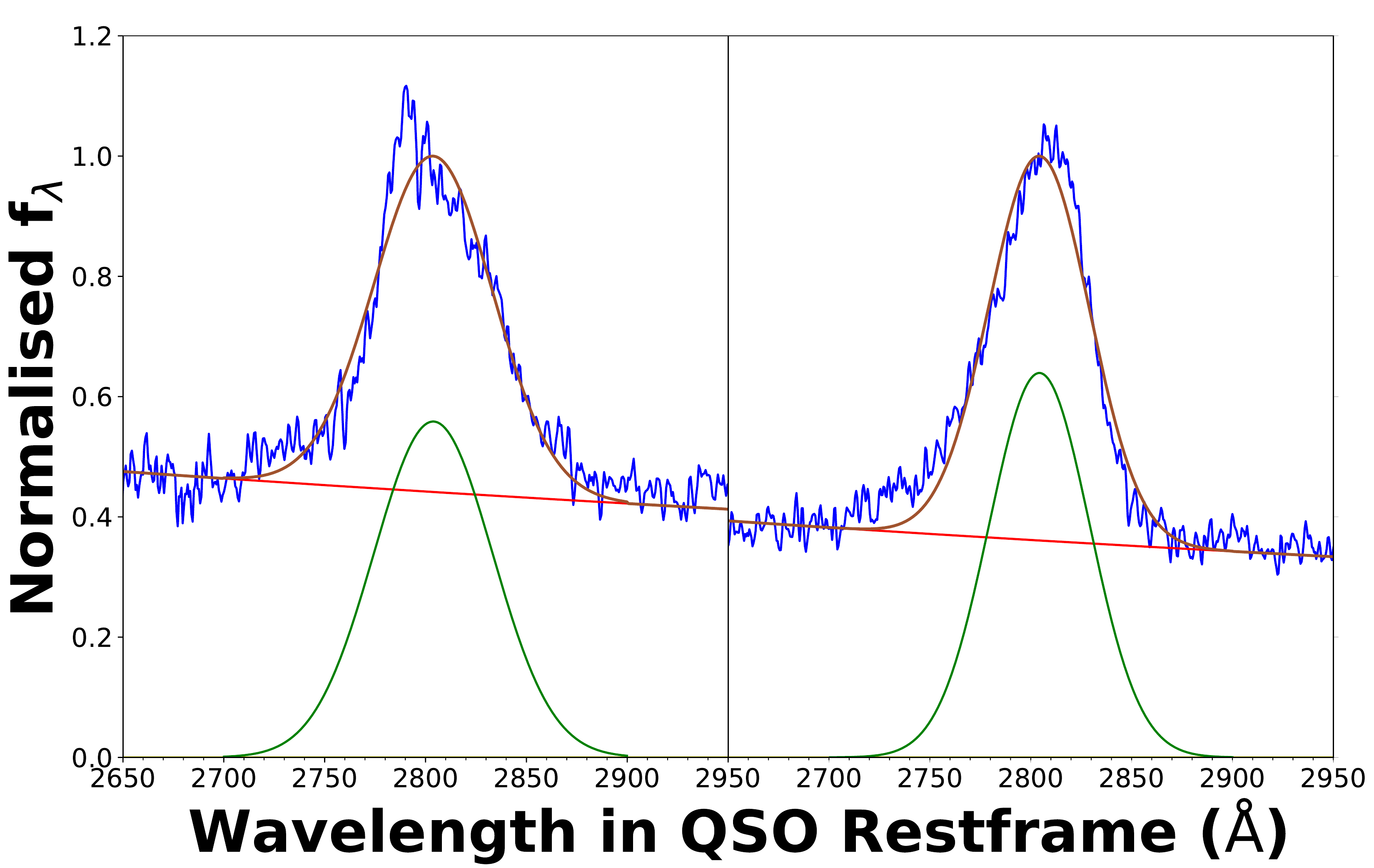}
\caption{Examples of asymmetry in the Mg~II line: $left$ J225240 observed at MJD 55500 (SDSS), $right$ J111348 observed at 57539 MJD (WHT). The fluxes have been normalised to the maximum flux of the fitted functions. In both objects the line appears skewed with respect to the symmetric Gaussian. The skewness of the lines is to different sides: to the blue for J225240 and to the red for J111348. A possible explanation for this effect is an offset between different line components.}
\label{fig:mgii_profile_ex}
\end{figure}

\begin{figure}
\centering
\includegraphics[scale=.4]{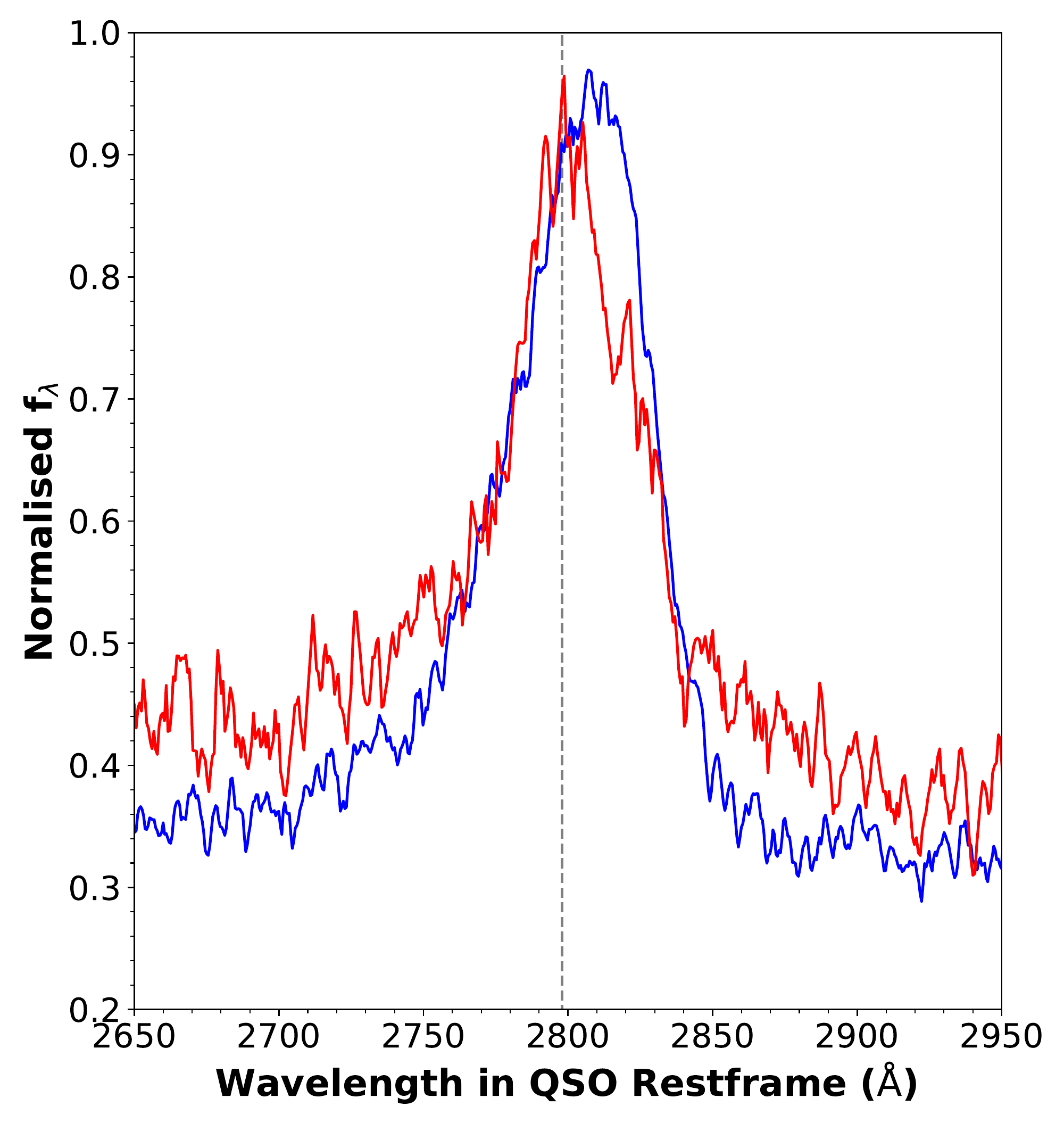}
\caption{The profile of the Mg~II$\lambda$2798 line in J111348 as observed on MJD 52438 ($red$) and 57539 ($blue$). The fluxes have been normalised to the maximum of the Mg~II line flux. Over the period between the two epochs the line flux dropped by a factor $\sim$4. In the more recent spectrum, associated with the lower emission state, the line has acquired a skewness to the red. The vertical line marks 2798{\AA} in the QSO restframe.}
\label{fig:mgii_profile_ch}
\end{figure}

There are five objects in the Supervariable sample that display a skewed line profile: J022556, J034144, J111348, J225240, and J234623. Of these J034144 and J111348 show a skewness to the red, the others to the blue. There are more spectra that can tentatively be identified as asymmetric, but the data quality does not allow for a definitive identification. Two examples of skewed line profiles in our sample are shown in Figure~\ref{fig:mgii_profile_ex}.\\
\indent Only one of the five objects, J111348, shows an indication of a variable line profile over time. Between the epochs, the maximum of the line, possibly associated with a distinct component, appears to shift from the centre to the red. The two epochs are shown in Figure~\ref{fig:mgii_profile_ch}. The change to a more asymmetric profile was associated with a large drop in flux, where the 2798{\AA} continuum dropped by a factor of approximately 4.

\section{Connection to Physical Parameters}
\label{sec:phys}
\begin{figure}
\centering
\includegraphics[width=\columnwidth,trim={.5cm .5cm .5cm .3cm},clip]{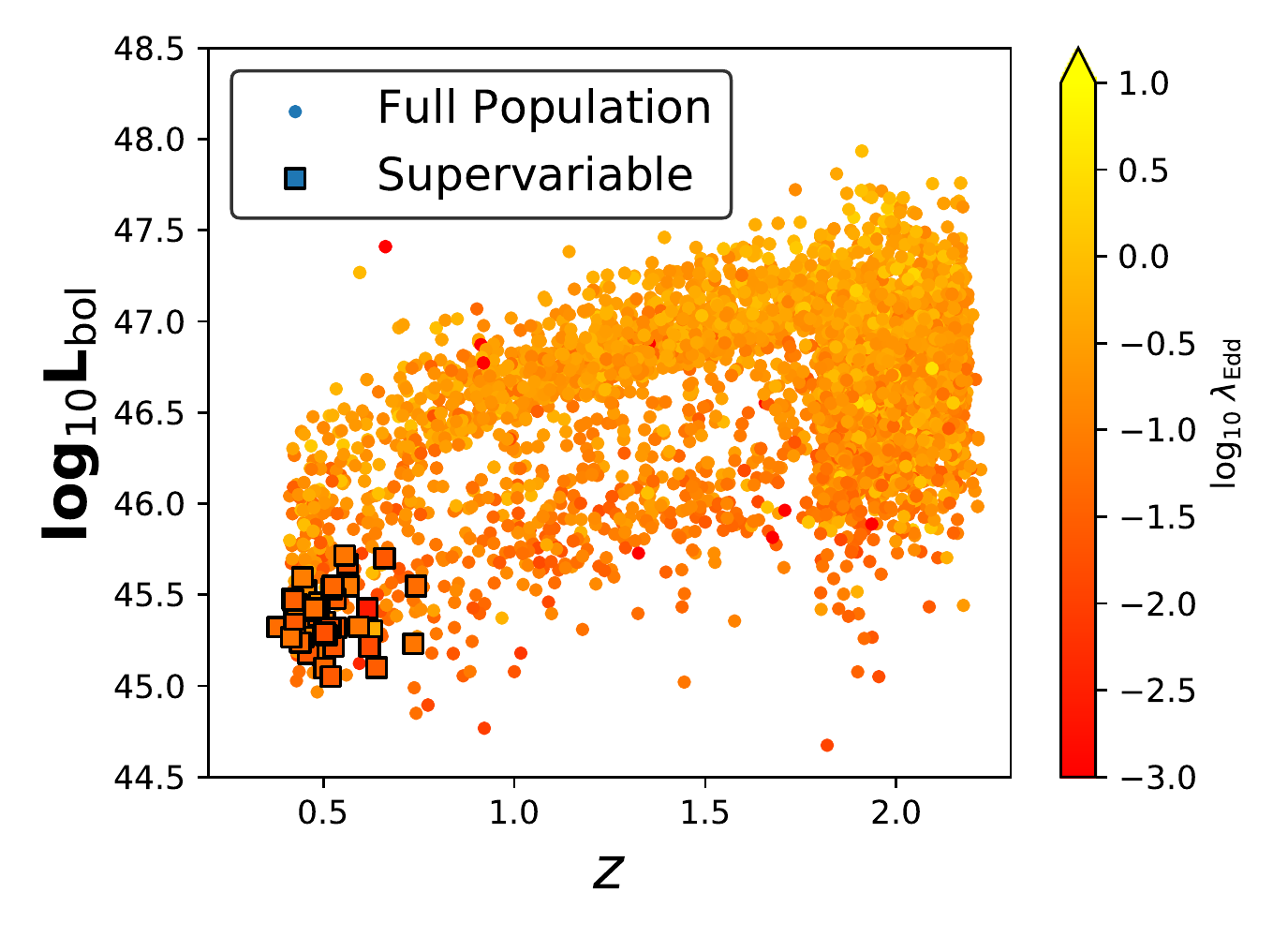}
\caption{The bolometric luminosity plotted against the redshift for the two QSO samples; colour indicates the Eddington ratio. The Supervariable sample is on average intrinsically dimmer as well as having a lower Eddington ratio. The requirements for the Supervariable sample (table~\ref{tab:clq_candidate_sel}) and the SDSS quasar criteria introduce a number of selection effects evident in the distributions of the samples.}
\label{fig:mgii_cat_both}
\end{figure}

\noindent To investigate whether the differences in spectral variability are connected to physical parameters, we will compare different subsets of the samples. The most important parameters to consider are the bolometric luminosity (L$_{\rm bol}$), the black hole mass (M$_{\rm BH}$), and the combination of these two values into the Eddington ratio ($\lambda_{\rm Edd}$). The Eddington ratio can be used as a measure of the accretion rate and EVQs appear to have relatively low Eddington rates on average (\citealt{RUM18}, MCL19). The influence of this parameter on the variability of Mg~II is therefore worth investigating. We use the catalogue provided by \citet{SHE11} for our values of L$_{\rm bol}$, M$_{\rm BH}$, and $\lambda_{\rm Edd}$, which means the quasars included in this part of the analysis are objects that were included in DR7Q. The number of objects included in this part of the analysis is 3,494. The black hole masses are the fiducial estimates, based on the single epoch virial method \citep{SHE11}.\\
\indent The combined distributions, for both samples, of log L$_{\rm bol}$, log $\lambda_{\rm Edd}$, and the redshift are shown in Figure~\ref{fig:mgii_cat_both}. A number of selection effects distinguishes the Supervariable sample from the general quasar population. The Eddington ratio for the strongly variable objects is on average lower than that of the Full Population sample. This is in agreement with the results presented in Figure 6 of MCL19 and the EVQ sample discussed by \citet{RUM18}. However, the strong selection biases that separate the samples imply that a direct comparison of the two samples should be treated with caution.

\subsection{Data Subsets Based on Normalised Flux}
\label{sec:phys_flux}
\begin{figure*}
\centering
\includegraphics[scale=0.6,trim={1cm .5cm .5cm .4cm},clip]{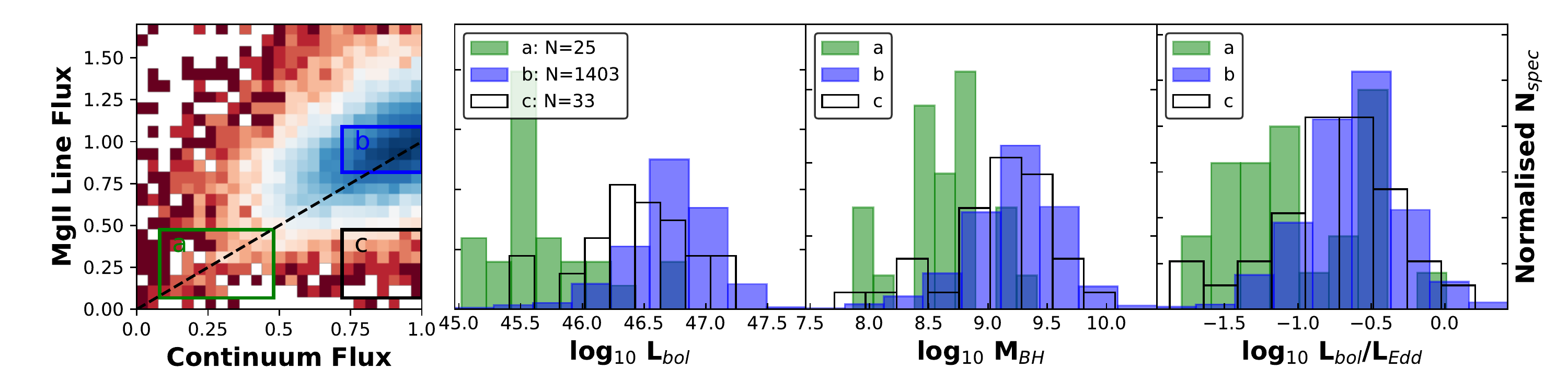}
\caption{Quasar properties for the subsets of the Full Population sample, normalised to the epoch of highest continuum emission. \emph{left}: The selection in the normalised flux space. A 1:1 ratio of line and continuum flux is indicated by the dashed line. \emph{right}: Three histograms displaying (from left to right) the normalised distribution of the bolometric luminosity, the fiducial black hole mass, and the Eddington ratio. The values for the distributions are from \citet{SHE11}.}
\label{fig:mgii_cats7_max}
\end{figure*}

\noindent By making selections in the normalised flux space, it is possible to create subsets of the samples based on spectral variability. Figure~\ref{fig:mgii_cats7_max} illustrates a comparison of physical parameters associated with different regimes of spectral variability. We consider the $\lambda_{\rm Edd}$ distribution in more detail in Table~\ref{tab:ledd_loc}.\\
\indent Figure~\ref{fig:mgii_cats7_max} is based on the normalisation to the minimum continuum epoch. The three sections have been selected to represent different types of responsivity. The sections have been made as follows: section $a$ targets highly variable and highly responsive objects, section $b$ selects a general quiescent sample, and section $c$ targets objects where the line is relatively more variable than the continuum. Section $c$ overlaps with $b$ in the range of continuum fluxes, and with $a$ in the range of line fluxes.\\
\indent The distributions in the right panels of Figure~\ref{fig:mgii_cats7_max} show that the subset $a$ is a clearly distinct selection of quasars from those in $b$ and $c$. The most variable objects ($a$) are characterised by lower luminosities, lower masses, and lower Eddington ratios. A correlation between continuum variability and Eddington ratio is therefore evident. Considering the Mg~II line variability, the $\lambda_{\rm Edd}$ distributions of the objects in $b$ and $c$ indicate that these populations are quite similar, suggesting strong Mg~II variability can occur in both high and low Eddington systems. However, the averages and means for the distributions (Table~\ref{tab:ledd_loc}) do show an offset: the objects with greater line variability have on average somewhat smaller Eddington ratios.  We find an average of the log$_{10}\lambda_{\rm Edd}$ distribution of -1.02 for section $a$ and -0.75 for section $b$, compared to -0.62 for section $c$. The connection of Mg~II flux variability with the Eddington ratio is therefore noticeably weaker than the connection between $\lambda_{\rm Edd}$ and the continuum flux.\\
\indent For the Supervariable sample we make the same selections in flux-flux parameter space. We characterise the distributions of log$_{10}\lambda_{\rm Edd}$, calculating the median, sample average and standard deviation of the data subsets in Table~\ref{tab:ledd_loc}. For the Supervariable sample the comparison shows that the dependence of the magnitude of relative continuum changes on the Eddington ratio also applies. The average value of log$_{10}\lambda_{\rm Edd}$ for the highest continuum variability subset $a$ is -1.48 and for $b$ log$_{10}\lambda_{\rm Edd}$=-0.90. The number of objects in the subsets of the Supervariable sample is low, making a reliable comparison difficult. No objects matched the requirements for subset $c$, most likely because the Supervariable sample was selected based on strong continuum variability.
\begin{table}
\centering
\caption{Summary statistics of the distributions of log$_{10}\lambda_{\mathrm{Edd}}$ distributions for the data subsets defined in Figure~\ref{fig:mgii_cats7_max}. The results for the Full Population and Supervariable sample are included. The sections in the normalised flux space that define the subset were the same for each sample. The columns represent, from left to right, the sample size, median, sample average, and standard deviation.}
\label{tab:ledd_loc}
\begin{tabular}{l| c c c c | c c c c}
\toprule
\toprule
& \multicolumn{4}{c|}{Full Population} & \multicolumn{4}{c}{Supervariable}\\
& N & $\widetilde{\mathrm{X}}$ & $\bar{\mathrm{X}}$ & $\sigma$ & N & $\widetilde{\mathrm{X}}$ & $\bar{\mathrm{X}}$ & $\sigma$\\
\hline
a & 25 & -1.14 & -1.02 & 0.45 & 24 & -1.23 & -1.48 & 0.62\\
b & 1403 & -0.61 & -0.62 & 0.33 & 2 & $-$ & -0.90 & 0.13\\
c & 33 & -0.73 & -0.75 & 0.44 & 0 & $-$ & $-$ & $-$\\
\bottomrule
\bottomrule
\end{tabular}
\end{table}

\subsection{Considering the Responsivity Measure}
\begin{figure}
\centering
\includegraphics[width=\columnwidth]{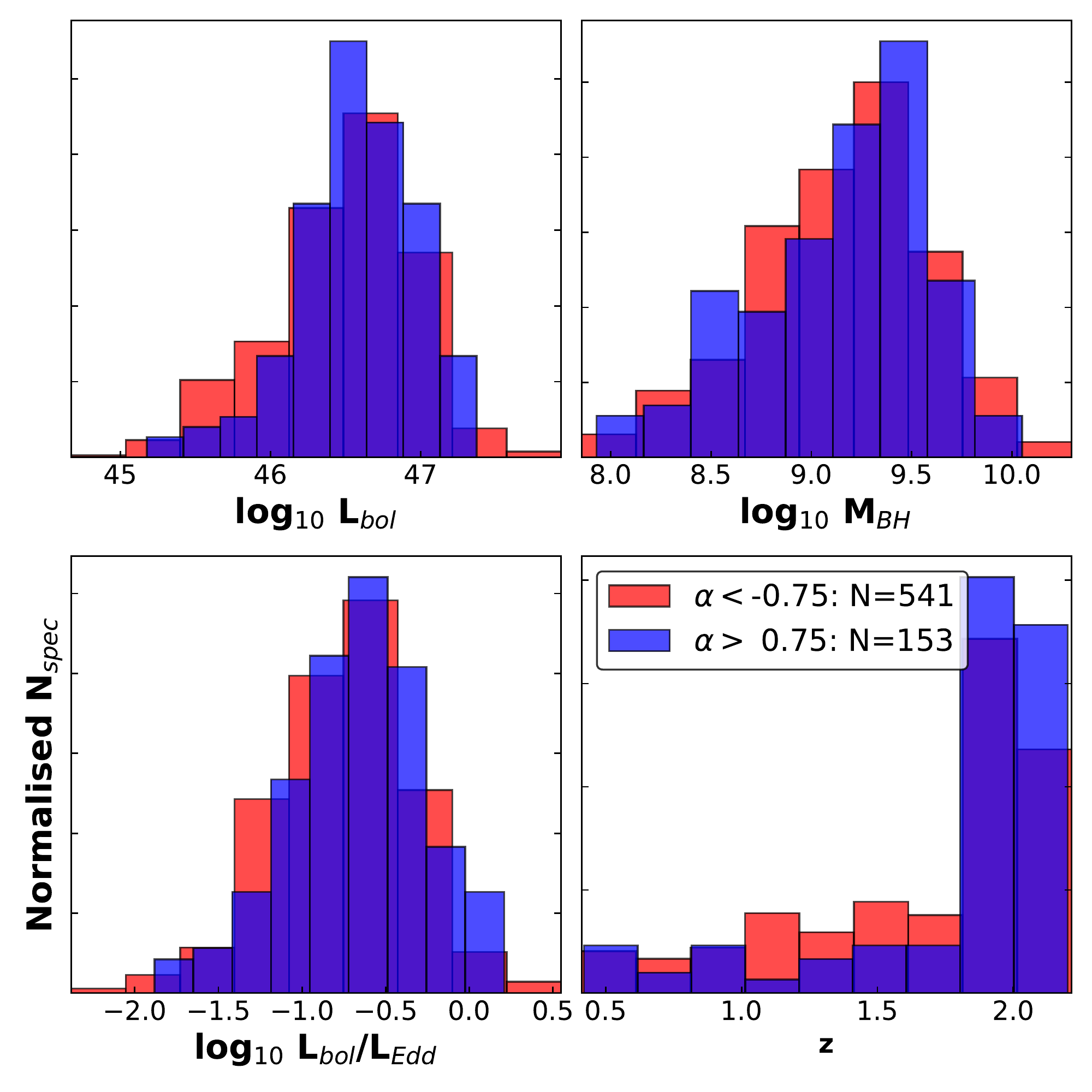}
\caption{A comparison of the distribution of AGN properties among subsections of the Full Population sample based on the responsivity measure. The properties are: bolometric luminosity (\emph{top left}), black hole mass (\emph{top right}), Eddington ratio (\emph{bottom left}), and redshift (\emph{bottom right}). No clear distinction is visible between the subsets, despite a cut based on a relatively large value of $|\alpha_{rm}|$.}
\label{fig:mgii_cats7}
\end{figure}

\noindent We now consider an analysis of sample subsets based on $\alpha_{rm}$. The first selection simply splits the samples between positive and negative $\alpha_{rm}$, i.e. the over and under responsive subsets respectively. A second selection focuses on the more extreme cases by requiring $|\alpha_{rm}|>0.75$. This selection will function as a check on the first results: if there is a distinction in physical parameters based on $\alpha_{rm}$, the distinction should be more prominent among the outliers.\\
\indent The results for the selection $|\alpha_{rm}|>0.75$ are shown in Figure~\ref{fig:mgii_cats7}. For none of the four parameters considered (L$_{bol}$, M$_{BH}$, $\lambda_{\rm Edd}$, and $z$) do the distributions of the two sets of parameters differ visibly. For the distributions of $\lambda_{\rm Edd}$ summary statistics are listed in Table ~\ref{tab:ledd_rm}. However, the medians and averages in Table~\ref{tab:ledd_rm} do indicate a distinction between the two $|\alpha_{rm}|>0.75$ samples: the over-responsive objects have a slightly higher $\lambda_{\rm Edd}$. For the Supervariable sample a similarly small distinction based on $\alpha_{rm}$ can be detected (Table~\ref{tab:ledd_rm}).\\
\indent A lower value of the Eddington ratio for objects with a negative responsivity measure would be in agreement with the results based on the subsets in flux. The correlation between the continuum flux and $\lambda_{\rm Edd}$ is stronger than the correlation of the line flux and $\lambda_{\rm Edd}$. Therefore objects in which the continuum has relatively changed the most ($\alpha_{rm}<0$) can be expected to have lower Eddington ratios. The difference between the two distributions is slight, making a firmer conclusion difficult.
\begin{table*}
\centering
\caption{Summary statistics for the distributions of the Eddington ratio in subsets of the Full Population and Supervariable samples based on $\alpha_{rm}$. The statistics are, from left to right: the sample size, median, sample average, and standard deviation. The same statistics for the samples in their entirety are listed in the bottom row. J081916 was not part of DR7, reducing the total size of the Supervariable sample from 43 to 42.}
\label{tab:ledd_rm}
\begin{tabular}{l | c c c c | c c c c }
\toprule
\toprule
& \multicolumn{4}{c|}{Full Population} & \multicolumn{4}{c}{Supervariable}\\
Sample & N & Median & $\bar{\mathrm{X}}$ & $\sigma$ & N & Median & $\bar{\mathrm{X}}$ & $\sigma$\\
\hline
$\alpha_{rm}<0$ & 5374 & -0.64 & -0.65 & 0.37 & 35 & -1.38 & -1.40 & 0.48\\
$\alpha_{rm}>0$ & 1979 & -0.63 & -0.65 & 0.36 & 7 & -1.29 & -1.33 & 0.31\\
\hline
$\alpha_{rm}<0.75$ & 541 & -0.72 & -0.76 & 0.42 & 18 & -1.34 & -1.38 & 0.28\\
$\alpha_{rm}>0.75$ & 153 & -0.65 & -0.67 & 0.41 & 2 & $-$ & -1.16 & 0.00\\
\hline
Full Sample & 7407 & -0.63 & -0.65 & 0.36 & 42 & -1.28 & -1.36 & 0.47\\
\bottomrule
\bottomrule
\end{tabular}
\end{table*}

\section{Interpretation}
\label{sec:interpretation}
\subsection{Flux Variability}
\noindent The broad range in Mg~II line flux variability noted in both samples agrees with previous studies. Differences from object to object can be seen among RM studies \citep[e.g.][]{CLA91,HRY14,TRE07,CAC15} and are a likely contribution to the strong scatter found in Mg~II variability in larger samples of quasars \citep{KOK14,SUN15,ZHU17}, as well as in the EVQ sample presented by \citet{YAN19}. Even for very large variations in the (ionising) continuum, the response can differ considerably. As was evident from the example of J022556, the behaviour does not only differ among objects, but evolves for individual quasars over time. The range in behaviours illustrated in Figure~\ref{fig:mgii_var_behaviour} appears representative of the EVQ population, when compared to other objects in the literature \citep{ROSS18,YAN19,KYN19,GUO19a,DEX19b}: strong changes in the optical or UV continuum can be followed by dramatic changes in the Mg~II flux, but this is not always the case .\\
\indent The primary behaviour observed in the Mg~II response is that the line tracks the continuum. This is true for both the Supervariable and the Full Population sample. The correlation between line and continuum has been found in other studies \citep{SUN15,ZHU17,YAN19}, although not all authors agree on the strength of the response \citep{KOK14}. Although a more complex flux response (e.g. a saturation effect, Figure~\ref{fig:mgii_linfit_min}) is certainly possible, the approximation to first order is the clearest effect in the data. To compare the gradient in our sample with that presented in \citet{YAN19} we calculate $\Delta$log$_{10}$L$_{\mathrm{MgII}}$/$\Delta$log$_{10}$L$_{2798}$. We find $\Delta$log$_{10}$L$_{\mathrm{MgII}}=(0.58\pm 0.05)\Delta$log$_{10}$L$_{2798}$ for the Supervariable sample, and $0.35\pm 0.01$ for the Full Population sample. \citet{YAN19} find a value of $0.47\pm 0.05$ for $\Delta$log$_{10}$L$_{\mathrm{MgII}}$/$\Delta$log$_{10}$L$_{3000}$ in an EVQ sample including SDSS quasars. The results are similar and, furthermore, emphasise the stronger overall response observed in the Supervariable sample.\\
\indent The timescales for strong Mg~II variability are $\sim$years (\citet{ROSS18,GUO19a}, but see \citet{HRY14}). For the objects in the Supervariable sample as well as for a large fraction of the objects in the Full Population sample (see Figure~\ref{fig:mgii_dt_hist}) the timescales are of the same order of magnitude. On shorter timescales the linear correlation between line and continuum is not as clear, although it can be detected for a fraction of the population \citep{SHE16a}. Interestingly, a timescale of years corresponds with a dynamical timescale ($\tau_{dyn}$) in the BLR. The fact that a consistent positive correlation (albeit with a large scatter in the amplitude of response) is clearer on longer timescales, could indicate that the ability of the Mg~II line to reprocess the ionising flux represents a physical change in the BLR.\\
\indent A difficulty in linking the evolution of the line response over time to $\tau_{dyn}$ is that a larger $\Delta t$ is on average related to a larger change in continuum. The entanglement of $\Delta f_{\mathrm{MgII}}$ and $\Delta f_{2798}$ makes a firm conclusion about the effect problematic. The moments of the distributions of $\Delta f_{\mathrm{MgII}}$ and $\Delta f_{2798}$ in different $\Delta t$ bins, listed in Table~\ref{tab:dt_moments}, show that the distributions evolve in a similar way. Disentangling the two parameters in this manner therefore does not appear possible. An indicator that $\Delta t$ is of significant importance is that the gradient of the responsivity is larger in the Full Population sample than in the Supervariable sample (Table~\ref{tab:test_varsdss}). It should be stressed that this response is to \emph{fractional changes}: the Full Population sample is dominated by objects with only small changes in both continuum and Mg~II. For the most dramatic changes in Mg~II flux a large continuum flare or drop might still be required.

\noindent An established population effect for quasars is the anti-correlation of continuum variability with the Eddington ratio \citep{MCL12,RUM18,GRAH19}. Objects with larger Eddington ratios, linked to higher accretion rates, on average show a smaller amplitude in variability. This effect is evident in both samples, as can be seen in Figure~\ref{fig:mgii_cats7_max} and Table~\ref{tab:ledd_loc}. The data also show a tentative correlation between the change in Mg~II line flux and $\lambda_{\rm Edd}$. As the line flux tracks the continuum, a correspondence between the line flux and $\lambda_{\rm Edd}$ can be expected, based on the correlation between the continuum variability and this parameter. A similar relation is reported by \citet{DON09}, who find that the EW of Mg~II has an inverse correlation with $\lambda_{\rm Edd}$.\\
\indent Both \citet{SUN15} and \citet{ZHU17} find that the response of the Mg~II line depends on L$_{bol}$. \citet{SUN15} report a decrease in the line variability with increasing L$_{bol}$. \citet{ZHU17} find that more luminous objects have a shallower slope in the linear response. The connection between L$_{bol}$ and the Mg~II line change is not clear in the data for the Supervariable sample, although the effect $is$ visible for the continuum variability (Figure~\ref{fig:mgii_cats7_max}). The decrease of the continuum variability with bolometric luminosity is also found in DRW studies \citep{MCL10}.

\noindent The connection between H$\beta$ and Mg~II variability can provide information about the BLR processes that govern the emission. The line forming regions for these two species are thought to be in close proximity \citep{WOO08,SHE12,KOR04}. The variability of the two lines in relation to each other can show a wide range, as seen in Table~\ref{tab:var_avgtime}. A class of objects in which Mg~II and H$\beta$ vary independently would agree with the BOSS objects presented in \citet{ROI13}. The authors of this study reveal the presence of a class of objects in the BOSS sample for which Mg~II is significantly stronger than the Balmer emission lines, but which lack signs of reddening. \citet{KOK14} argue that Mg~II is intrinsically less variable than H$\beta$ in the entire quasar population. Although this is in agreement with RM studies, the research presented here indicates that the Mg~II behaviour is more complex than a suppressed response compared to the Balmer lines. The observed pattern of behaviour agrees with the CL `sequence' presented in \citet{GUO19b}, although we would argue that the marked difference in possible responses among the AGN population suggests that the response is not determined by changes in the luminosity alone. Rather, a range of conditions in the BLR likely conspire to create a specific responsivity \citep{HOM20}.\\
\indent Several causes for the difference between H$\beta$ and Mg~II are discussed in the literature. The first possibility is that the material emitting H$\beta$ has a different geometric distribution than the Mg~II emitting material \citep{SUN15}. The `BLR' that forms H$\beta$ would then be physically distinct from the one that forms Mg~II. For a more extended Mg~II region, any change in incident flux would be geometrically diluted, illuminating different radii at different times. This could explain the lack of effect in some of the RM studies, although it is more difficult to explain the strong instantaneous response measured by e.g. \citet{HRY14} in this scenario.\\
\indent Another issue with an interpretation based on differing BLR geometries is the physical conditions in which the two lines are formed: the Mg~II population is likely present in the cooler, self-shielded `back' of the cloud \citep[e.g.][]{BAS14a}. This region largely overlaps with the parts of the cloud where H is neutral. As the collisional excitation of Mg~II depends on the the electron density ($n_e$) and the temperature \citep{GUO19b}, the formation region of the 2798{\AA} line is likely associated with the transition region between the HII and HI regions of BLR clouds, where H$\beta$ is also formed \citep{MCA72,NET80,COL86,KRO}. If the line forming regions of H$\beta$ and Mg~II are so intimately linked, a very different geometry for the two appears less probable.\\
\indent A second possible explanation is based on the difference in optical depths between the two lines \citep{KOR04,SUN15,YAN19}. Mg~II is a resonance line, whereas H$\beta$ is not. Mg~II can therefore be expected to have a greater optical depth, increasing the number of absorptions and reemissions of any line photon before it escapes the cloud. This could also dilute the response to a change in incident flux, spreading it out over time. A possible problem with this interpretation is that although Mg~II is a resonance line, the population of ground state atoms is itself subject to a collisional equilibrium which could suppress the number of absorbing atoms. Photoionisation modelling does indicate the optical depth effect could be significant \citep{KOR04,BAS14a}. If optical depth for the lines is a critical factor, the observed evolution with $\lambda_{\rm Edd}$ could be the result of a changing ionisation structure in the cloud, as a result of a changing shape of the SED. A difference in temperature or $n_e$ could affect the collisional excitation of Mg~II in BLR clouds.\\

\subsection{Line Profile}
\noindent The Supervariable sample shows a range of line profiles. If the skewness seen in some spectra results from an offset between broad and narrow components, a radial outflow or inflow would be a plausible explanation. Another option is to associate the different line components with a blueshifted and a redshifted component. These components could be the result of rotation of the BLR or possibly even the outer regions of the disc, where the line could be formed. Such a scenario would be akin to the interpretation for quasars with double peaked emission lines \citep[e.g.][]{ERA09}. It is impossible to discern the different dynamic components in either of the two samples, however.\\
\indent To quantify the behaviour of the line profile, we have made use of the line width ($\sigma$) of the broadest Gaussian fit to the line. For both the Supervariable and the Full Population sample $\sigma$ lacks a correlation with the luminosity. We compare our results to those of \citet{YAN19}. We find a slope of $\Delta$log$_{10}$FWHM$_{\mathrm{MgII}}$/$\Delta$log$_{10}$L$_{2798}$ of $0.013\pm 0.035$ for the Supervariable sample and $0.008\pm 0.066$ for the Full Population sample, compared to $\Delta$log$_{10}$FWHM$_{\mathrm{MgII}}=0.012\pm 0.012 \Delta$log$_{10}$L$_{3000}$. These values are in agreement with other results \citep{WOO08,SHE11} and indicate that the Mg~II line does not `breathe'.\\
\indent The lack of breathing presents another marked difference with the behaviour of H$\beta$ \citep{BEN13}. A possible explanation is that the Mg~II emission region is confined to a smaller set of radii \citep{YAN19}, perhaps as the result of the small geometrical depth of the Mg~II line emitting region in each cloud.

\section{Conclusions}
\label{sec:conclusion}
\noindent The Mg~II line exhibits a complex behaviour. The analysis presented in this work falls into two broad categories. The first is flux responsivity: the changing levels of the line flux under a varying continuum provide information about the incident ionising flux and the physical conditions in the line forming region. The second category is the line profile: changes in line width and skewness can provide information about the kinematics of the gas creating the Mg~II emission. Below is a brief summary of the results.
\subsection{Flux Responsivity}
\begin{enumerate}[label=\textbf{\roman*)}]
\item There is a broad range of behaviour in Mg~II variability. In some objects the line changes more than the continuum, whereas in others the line shows barely any response at all. Based on the responsivity measure, it is clear that the objects where the line varies less than the continuum form the largest part of the quasar population.
\item $|\Delta L_{\mathrm{MgII}}|$ and $|\Delta\lambda L_{2798}|$ show a linear correlation, visible for the Supervariable sample (Figure~\ref{fig:mgii_lum_var}). The relation holds over five orders of magnitude.
\item The connection between Mg~II variability and CLQ behaviour is not one-to-one (Table~\ref{tab:var_avgtime}). Considering H$\beta$ and Mg~II in the same object, the Supervariable sample contains objects where either, both and neither of the lines vary. Joint and separate Mg~II and H$\beta$ variability occur in approximately equal measure.
\item The normalised fluxes show a linear correlation between the Mg~II line and the 2798{\AA} continuum. This holds for both normalisations, to the minimum and the maximum state (Figures~\ref{fig:mgii_linfit_max} and \ref{fig:mgii_linfit_min}; Table~\ref{tab:test_varsdss}). The scatter in the data is considerable. Although the scatter is partially the result of measurement and fitting uncertainties, it is indicative of the range in Mg~II variability.
\item There is clear evidence of a changing responsivity over time in the spectra of J022556 (Figure~\ref{fig:mgii_022556}), for which relatively high cadence observations are available. The response of Mg~II not only changes from object to object, but also from epoch to epoch.
\item A change in the linear responsivity of Mg~II with time can be observed in the Full Population sample. The responsivity appears to decrease for longer elapsed times between spectral epochs (Figure~\ref{fig:mgii_dtbin_sdss}). When modelled as a simple linear response, the change in gradient is modest, but steady over the $\Delta t$ bins (Table~\ref{tab:dt_slopes}).
\item The distributions of $\Delta f_{\mathrm{MgII}}$ and $\Delta f_{2798}$ change with $\Delta t$ (Figure~\ref{fig:mgii_ldt_sdss} and Table~\ref{tab:dt_moments}). The averages of the distributions become more negative over time and the skewness increases. The change in the average is likely the result of a selection bias (see the discussion in section~\ref{sec:mgii_sdss_dt}).
\item The Eddington ratio is connected to the magnitude of changes in the continuum. Objects with larger continuum changes on average have a lower $\lambda_{\rm Edd}$. The difference is visible in the offset between the $\lambda_{\rm Edd}$ distributions of the Supervariable sample and the Full Population sample (Figure~\ref{fig:mgii_cats7_max} and \ref{fig:mgii_cats7} and Tables~\ref{tab:ledd_loc} and \ref{tab:ledd_rm}). For changes in the line flux this connection possibly also exists, but is not as strong as for the continuum.
\end{enumerate}

\subsection{Kinematics}
\begin{enumerate}[label=\textbf{\roman*)}]
\item Some of the objects in the Supervariable sample show a clear asymmetry in the Mg~II profile. This asymmetry manifests itself as a skewness of the lines. The skewness can be both to the blue and to the red of the line centre (Figure~\ref{fig:mgii_profile_ex}).
\item Asymmetry of the Mg~II line can evolve over time, as observed for J111348 (Figure~\ref{fig:mgii_profile_ch}). The evolution of the line profile could be due to a change in the relative strength of different line components. These could be the broad and narrow components, but it is also possible they represent a more dynamic structure.
\item The Mg~II line width lacks a correlation with either the 2798{\AA} continuum or the line flux (Figure~\ref{fig:mgii_sigvsline_sdss}). There is some tentative evidence of a correspondence for the lowest luminosities, however the bulk of the distribution shows a flat response. The Mg~II shows no clear signs on breathing. This is in agreement with previous results \citep[e.g.][]{SHE11}.
\end{enumerate}

\section*{Acknowledgements}
D.H. was supported by a Principal's Career Development PhD Scholarship from the University of Edinburgh.
N.P.R. acknowledges support from the STFC and the Ernest Rutherford Fellowship scheme. 

\bibliographystyle{mnras}
\bibliography{References}

\begin{thebibliography}{}
\makeatletter
\relax
\def\mn@urlcharsother{\let\do\@makeother \do\$\do\&\do\#\do\^\do\_\do\%\do\~}
\def\mn@doi{\begingroup\mn@urlcharsother \@ifnextchar [ {\mn@doi@}
  {\mn@doi@[]}}
\def\mn@doi@[#1]#2{\def\@tempa{#1}\ifx\@tempa\@empty \href
  {http://dx.doi.org/#2} {doi:#2}\else \href {http://dx.doi.org/#2} {#1}\fi
  \endgroup}
\def\mn@eprint#1#2{\mn@eprint@#1:#2::\@nil}
\def\mn@eprint@arXiv#1{\href {http://arxiv.org/abs/#1} {{\tt arXiv:#1}}}
\def\mn@eprint@dblp#1{\href {http://dblp.uni-trier.de/rec/bibtex/#1.xml}
  {dblp:#1}}
\def\mn@eprint@#1:#2:#3:#4\@nil{\def\@tempa {#1}\def\@tempb {#2}\def\@tempc
  {#3}\ifx \@tempc \@empty \let \@tempc \@tempb \let \@tempb \@tempa \fi \ifx
  \@tempb \@empty \def\@tempb {arXiv}\fi \@ifundefined
  {mn@eprint@\@tempb}{\@tempb:\@tempc}{\expandafter \expandafter \csname
  mn@eprint@\@tempb\endcsname \expandafter{\@tempc}}}

\bibitem[\protect\citeauthoryear{{Baskin}, {Laor}  \& {Stern}}{{Baskin}
  et~al.}{2014}]{BAS14a}
{Baskin} A.,  {Laor} A.,   {Stern} J.,  2014, \mn@doi [\mnras]
  {10.1093/mnras/stt2230}, \href
  {https://ui.adsabs.harvard.edu/abs/2014MNRAS.438..604B} {438, 604}

\bibitem[\protect\citeauthoryear{{Bentz} et~al.,}{{Bentz} et~al.}{2013}]{BEN13}
{Bentz} M.~C.,  et~al., 2013, \mn@doi [\apj] {10.1088/0004-637X/767/2/149},
  \href {https://ui.adsabs.harvard.edu/abs/2013ApJ...767..149B} {767, 149}

\bibitem[\protect\citeauthoryear{{Boller}, {Balestra}  \&
  {Kollatschny}}{{Boller} et~al.}{2007}]{BOL07}
{Boller} T.,  {Balestra} I.,   {Kollatschny} W.,  2007, \mn@doi [\aap]
  {10.1051/0004-6361:20066343}, \href
  {http://adsabs.harvard.edu/abs/2007A%26A...465...87B} {465, 87}

\bibitem[\protect\citeauthoryear{Bolton et~al.,}{Bolton et~al.}{2012}]{SDSSIII}
Bolton A.~S.,  et~al., 2012, \aj, 144, 144

\bibitem[\protect\citeauthoryear{{Cackett}, {G{\"u}ltekin}, {Bentz},
  {Fausnaugh}, {Peterson}, {Troyer}  \& {Vestergaard}}{{Cackett}
  et~al.}{2015}]{CAC15}
{Cackett} E.~M.,  {G{\"u}ltekin} K.,  {Bentz} M.~C.,  {Fausnaugh} M.~M.,
  {Peterson} B.~M.,  {Troyer} J.,   {Vestergaard} M.,  2015, \mn@doi [\apj]
  {10.1088/0004-637X/810/2/86}, \href
  {https://ui.adsabs.harvard.edu/abs/2015ApJ...810...86C} {810, 86}

\bibitem[\protect\citeauthoryear{{Clavel} et~al.,}{{Clavel}
  et~al.}{1991}]{CLA91}
{Clavel} J.,  et~al., 1991, \mn@doi [\apj] {10.1086/169540}, \href
  {https://ui.adsabs.harvard.edu/abs/1991ApJ...366...64C} {366, 64}

\bibitem[\protect\citeauthoryear{{Collin-Souffrin}, {Dumnont}, {Joly}  \&
  {Pequignot}}{{Collin-Souffrin} et~al.}{1986}]{COL86}
{Collin-Souffrin} S.,  {Dumnont} S.,  {Joly} M.,   {Pequignot} D.,  1986, \aap,
  \href {https://ui.adsabs.harvard.edu/abs/1986A&A...166...27C} {166, 27}

\bibitem[\protect\citeauthoryear{{Dexter} et~al.,}{{Dexter}
  et~al.}{2019}]{DEX19b}
{Dexter} J.,  et~al., 2019, \mn@doi [\apj] {10.3847/1538-4357/ab4354}, \href
  {https://ui.adsabs.harvard.edu/abs/2019ApJ...885...44D} {885, 44}

\bibitem[\protect\citeauthoryear{{Dong}, {Wang}, {Wang}, {Fan}, {Wang}, {Zhou}
  \& {Yuan}}{{Dong} et~al.}{2009}]{DON09}
{Dong} X.-B.,  {Wang} T.-G.,  {Wang} J.-G.,  {Fan} X.,  {Wang} H.,  {Zhou} H.,
   {Yuan} W.,  2009, \mn@doi [\apjl] {10.1088/0004-637X/703/1/L1}, \href
  {https://ui.adsabs.harvard.edu/abs/2009ApJ...703L...1D} {703, L1}

\bibitem[\protect\citeauthoryear{{Drake} et~al.,}{{Drake} et~al.}{2009}]{DRA09}
{Drake} A.~J.,  et~al., 2009, \mn@doi [\apj] {10.1088/0004-637X/696/1/870},
  \href {https://ui.adsabs.harvard.edu/abs/2009ApJ...696..870D} {696, 870}

\bibitem[\protect\citeauthoryear{{Eracleous}, {Lewis}  \& {Flohic}}{{Eracleous}
  et~al.}{2009}]{ERA09}
{Eracleous} M.,  {Lewis} K.~T.,   {Flohic} H.~M.~L.~G.,  2009, \mn@doi [\nar]
  {10.1016/j.newar.2009.07.005}, \href
  {https://ui.adsabs.harvard.edu/abs/2009NewAR..53..133E} {53, 133}

\bibitem[\protect\citeauthoryear{{Graham} et~al.,}{{Graham}
  et~al.}{2020}]{GRAH19}
{Graham} M.~J.,  et~al., 2020, \mn@doi [\mnras] {10.1093/mnras/stz3244}, \href
  {https://ui.adsabs.harvard.edu/abs/2020MNRAS.491.4925G} {491, 4925}

\bibitem[\protect\citeauthoryear{{Gunn} et~al.,}{{Gunn} et~al.}{2006}]{GUN}
{Gunn} J.~E.,  et~al., 2006, \mn@doi [\aj] {10.1086/500975}, \href
  {http://adsabs.harvard.edu/abs/2006AJ....131.2332G} {131, 2332}

\bibitem[\protect\citeauthoryear{{Guo}, {Sun}, {Liu}, {Wang}, {Kong}, {Wang},
  {Sheng}  \& {He}}{{Guo} et~al.}{2019}]{GUO19a}
{Guo} H.,  {Sun} M.,  {Liu} X.,  {Wang} T.,  {Kong} M.,  {Wang} S.,  {Sheng}
  Z.,   {He} Z.,  2019, \mn@doi [\apjl] {10.3847/2041-8213/ab4138}, \href
  {https://ui.adsabs.harvard.edu/abs/2019ApJ...883L..44G} {883, L44}

\bibitem[\protect\citeauthoryear{{Guo} et~al.,}{{Guo} et~al.}{2020}]{GUO19b}
{Guo} H.,  et~al., 2020, \mn@doi [\apj] {10.3847/1538-4357/ab5db0}, \href
  {https://ui.adsabs.harvard.edu/abs/2020ApJ...888...58G} {888, 58}

\bibitem[\protect\citeauthoryear{{Hamann}, {Herbst}, {Paris}  \&
  {Capellupo}}{{Hamann} et~al.}{2019}]{HAM19}
{Hamann} F.,  {Herbst} H.,  {Paris} I.,   {Capellupo} D.,  2019, \mn@doi
  [\mnras] {10.1093/mnras/sty2900}, \href
  {https://ui.adsabs.harvard.edu/abs/2019MNRAS.483.1808H} {483, 1808}

\bibitem[\protect\citeauthoryear{{Harris} et~al.,}{{Harris} et~al.}{2016}]{HAR}
{Harris} D.~W.,  et~al., 2016, \mn@doi [\aj] {10.3847/0004-6256/151/6/155},
  \href {http://adsabs.harvard.edu/abs/2016AJ....151..155H} {151, 155}

\bibitem[\protect\citeauthoryear{{Heinis} et~al.,}{{Heinis} et~al.}{2016}]{HEI}
{Heinis} S.,  et~al., 2016, \mn@doi [\apj] {10.3847/0004-637X/826/1/62}, \href
  {https://ui.adsabs.harvard.edu/abs/2016ApJ...826...62H} {826, 62}

\bibitem[\protect\citeauthoryear{{Homan} et~al.,}{{Homan} et~al.}{2020}]{HOM20}
{Homan} D.,  et~al., 2020, Markarian 110 BLR Response, in preparation

\bibitem[\protect\citeauthoryear{{Hryniewicz}, {Czerny}, {Pych}, {Udalski},
  {Krupa}, {{\'S}wi{\c{e}}to{\'n}}  \& {Kaluzny}}{{Hryniewicz}
  et~al.}{2014}]{HRY14}
{Hryniewicz} K.,  {Czerny} B.,  {Pych} W.,  {Udalski} A.,  {Krupa} M.,
  {{\'S}wi{\c{e}}to{\'n}} A.,   {Kaluzny} J.,  2014, \mn@doi [\aap]
  {10.1051/0004-6361/201322487}, \href
  {https://ui.adsabs.harvard.edu/abs/2014A&A...562A..34H} {562, A34}

\bibitem[\protect\citeauthoryear{{Khachikian} \& {Weedman}}{{Khachikian} \&
  {Weedman}}{1971}]{KHA71}
{Khachikian} E.~E.,  {Weedman} D.~W.,  1971, Astrofizika, 7, 389

\bibitem[\protect\citeauthoryear{{Kokubo}, {Morokuma}, {Minezaki}, {Doi},
  {Kawaguchi}, {Sameshima}  \& {Koshida}}{{Kokubo} et~al.}{2014}]{KOK14}
{Kokubo} M.,  {Morokuma} T.,  {Minezaki} T.,  {Doi} M.,  {Kawaguchi} T.,
  {Sameshima} H.,   {Koshida} S.,  2014, \mn@doi [\apj]
  {10.1088/0004-637X/783/1/46}, \href
  {https://ui.adsabs.harvard.edu/abs/2014ApJ...783...46K} {783, 46}

\bibitem[\protect\citeauthoryear{{Korista} \& {Goad}}{{Korista} \&
  {Goad}}{2004}]{KOR04}
{Korista} K.~T.,  {Goad} M.~R.,  2004, \mn@doi [\apj] {10.1086/383193}, \href
  {https://ui.adsabs.harvard.edu/abs/2004ApJ...606..749K} {606, 749}

\bibitem[\protect\citeauthoryear{Krolik}{Krolik}{1999}]{KRO}
Krolik J.,  1999, Active Galactic Nuclei. From the Central Black Hole to the
  Galactic Environment.
Princeton

\bibitem[\protect\citeauthoryear{{Kynoch}, {Ward}, {Lawrence}, {Bruce}, {Landt}
   \& {MacLeod}}{{Kynoch} et~al.}{2019}]{KYN19}
{Kynoch} D.,  {Ward} M.~J.,  {Lawrence} A.,  {Bruce} A.~G.,  {Landt} H.,
  {MacLeod} C.~L.,  2019, \mn@doi [\mnras] {10.1093/mnras/stz517}, \href
  {https://ui.adsabs.harvard.edu/abs/2019MNRAS.485.2573K} {485, 2573}

\bibitem[\protect\citeauthoryear{LaMassa, Cales, Moran, Myers, Richards,
  Eracleous, Heckman  \& Gallo}{LaMassa et~al.}{2015}]{LAM15}
LaMassa S.,  Cales S.,  Moran E.,  Myers A.,  Richards G.,  Eracleous M.,
  Heckman T.,   Gallo L.~Urry C.,  2015, \apj, 800, 144

\bibitem[\protect\citeauthoryear{{Lawrence} et~al.,}{{Lawrence}
  et~al.}{2016}]{LAW16}
{Lawrence} A.,  et~al., 2016, \mn@doi [\mnras] {10.1093/mnras/stw1963}, \href
  {https://ui.adsabs.harvard.edu/abs/2016MNRAS.463..296L} {463, 296}

\bibitem[\protect\citeauthoryear{{MacAlpine}}{{MacAlpine}}{1972}]{MCA72}
{MacAlpine} G.~M.,  1972, \mn@doi [\apj] {10.1086/151536}, \href
  {https://ui.adsabs.harvard.edu/abs/1972ApJ...175...11M} {175, 11}

\bibitem[\protect\citeauthoryear{MacLeod et~al.,}{MacLeod et~al.}{2010}]{MCL10}
MacLeod C.,  et~al., 2010, The Astrophysical Journal, 721, 1014

\bibitem[\protect\citeauthoryear{MacLeod et~al.,}{MacLeod et~al.}{2012}]{MCL12}
MacLeod C.,  et~al., 2012, \apj, 753, 106

\bibitem[\protect\citeauthoryear{MacLeod, Ross, Lawrence, Goad  \&
  Horne}{MacLeod et~al.}{2016}]{MCL16}
MacLeod C.,  Ross N.,  Lawrence A.,  Goad M.,   Horne K.,  2016, \mnras, 457,
  389

\bibitem[\protect\citeauthoryear{{MacLeod} et~al.,}{{MacLeod}
  et~al.}{2019}]{MCL19}
{MacLeod} C.~L.,  et~al., 2019, \mn@doi [\apj] {10.3847/1538-4357/ab05e2},
  \href {https://ui.adsabs.harvard.edu/abs/2019ApJ...874....8M} {874, 8}

\bibitem[\protect\citeauthoryear{{Margala}, {Kirkby}, {Dawson}, {Bailey},
  {Blanton}  \& {Schneider}}{{Margala} et~al.}{2016}]{MAR}
{Margala} D.,  {Kirkby} D.,  {Dawson} K.,  {Bailey} S.,  {Blanton} M.,
  {Schneider} D.~P.,  2016, \mn@doi [The Astrophysical Journal]
  {10.3847/0004-637X/831/2/157}, \href
  {http://adsabs.harvard.edu/abs/2016ApJ...831..157M} {831, 157}

\bibitem[\protect\citeauthoryear{{McLure} \& {Dunlop}}{{McLure} \&
  {Dunlop}}{2004}]{MCLU04}
{McLure} R.~J.,  {Dunlop} J.~S.,  2004, \mn@doi [\mnras]
  {10.1111/j.1365-2966.2004.08034.x}, \href
  {https://ui.adsabs.harvard.edu/abs/2004MNRAS.352.1390M} {352, 1390}

\bibitem[\protect\citeauthoryear{{Metzroth}, {Onken}  \& {Peterson}}{{Metzroth}
  et~al.}{2006}]{MET06}
{Metzroth} K.~G.,  {Onken} C.~A.,   {Peterson} B.~M.,  2006, \mn@doi [\apj]
  {10.1086/505525}, \href
  {https://ui.adsabs.harvard.edu/abs/2006ApJ...647..901M} {647, 901}

\bibitem[\protect\citeauthoryear{{Netzer}}{{Netzer}}{1980}]{NET80}
{Netzer} H.,  1980, \mn@doi [\apj] {10.1086/157757}, \href
  {https://ui.adsabs.harvard.edu/abs/1980ApJ...236..406N} {236, 406}

\bibitem[\protect\citeauthoryear{Osmer \& Shields}{Osmer \&
  Shields}{1999}]{OSM99}
Osmer P.,  Shields J.,  1999, in Ferland G.,  Baldwin J.,  eds,  ASP Conference
  Series Vol. 162, Quasars and Cosmology. p.~235

\bibitem[\protect\citeauthoryear{Osterbrock}{Osterbrock}{1981}]{OST81}
Osterbrock D.,  1981, The Astrophysical Journal, 249, 462

\bibitem[\protect\citeauthoryear{{P\^{a}ris} et~al.,}{{P\^{a}ris}
  et~al.}{2014}]{PAR14}
{P\^{a}ris} I.,  et~al., 2014, \aap, 563, A54

\bibitem[\protect\citeauthoryear{{Peterson}}{{Peterson}}{2008}]{PET08}
{Peterson} B.~M.,  2008, \mn@doi [\nar] {10.1016/j.newar.2008.06.005}, \href
  {https://ui.adsabs.harvard.edu/abs/2008NewAR..52..240P} {52, 240}

\bibitem[\protect\citeauthoryear{{Raki{\'c}}, {La Mura}, {Ili{\'c}},
  {Shapovalova}, {Kollatschny}, {Rafanelli}  \& {Popovi{\'c}}}{{Raki{\'c}}
  et~al.}{2017}]{RAK17}
{Raki{\'c}} N.,  {La Mura} G.,  {Ili{\'c}} D.,  {Shapovalova} A.~I.,
  {Kollatschny} W.,  {Rafanelli} P.,   {Popovi{\'c}} L.~{\v{C}}.,  2017,
  \mn@doi [\aap] {10.1051/0004-6361/201630085}, \href
  {https://ui.adsabs.harvard.edu/abs/2017A&A...603A..49R} {603, A49}

\bibitem[\protect\citeauthoryear{{Reichert} et~al.,}{{Reichert}
  et~al.}{1994}]{REI94}
{Reichert} G.~A.,  et~al., 1994, \mn@doi [\apj] {10.1086/174007}, \href
  {https://ui.adsabs.harvard.edu/abs/1994ApJ...425..582R} {425, 582}

\bibitem[\protect\citeauthoryear{{Roig}, {Blanton}  \& {Ross}}{{Roig}
  et~al.}{2014}]{ROI13}
{Roig} B.,  {Blanton} M.~R.,   {Ross} N.~P.,  2014, \apj, \href
  {https://ui.adsabs.harvard.edu/abs/2013arXiv1312.1249R} {781, 72}

\bibitem[\protect\citeauthoryear{{Ross} et~al.,}{{Ross} et~al.}{2018}]{ROSS18}
{Ross} N.~P.,  et~al., 2018, \mn@doi [\mnras] {10.1093/mnras/sty2002}, \href
  {https://ui.adsabs.harvard.edu/abs/2018MNRAS.480.4468R} {480, 4468}

\bibitem[\protect\citeauthoryear{{Rumbaugh} et~al.,}{{Rumbaugh}
  et~al.}{2018}]{RUM18}
{Rumbaugh} N.,  et~al., 2018, \mn@doi [\apj] {10.3847/1538-4357/aaa9b6}, \href
  {https://ui.adsabs.harvard.edu/abs/2018ApJ...854..160R} {854, 160}

\bibitem[\protect\citeauthoryear{Schneider et~al.,}{Schneider
  et~al.}{2010}]{SCH10}
Schneider D.~P.,  et~al., 2010, \aj, 139, 2360

\bibitem[\protect\citeauthoryear{{Shappee} et~al.,}{{Shappee}
  et~al.}{2014}]{SHAPP14}
{Shappee} B.~J.,  et~al., 2014, \mn@doi [\apj] {10.1088/0004-637X/788/1/48},
  \href {https://ui.adsabs.harvard.edu/abs/2014ApJ...788...48S} {788, 48}

\bibitem[\protect\citeauthoryear{{Shen} \& {Liu}}{{Shen} \&
  {Liu}}{2012}]{SHE12}
{Shen} Y.,  {Liu} X.,  2012, \mn@doi [\apj] {10.1088/0004-637X/753/2/125},
  \href {https://ui.adsabs.harvard.edu/abs/2012ApJ...753..125S} {753, 125}

\bibitem[\protect\citeauthoryear{{Shen} et~al.,}{{Shen} et~al.}{2011}]{SHE11}
{Shen} Y.,  et~al., 2011, The Astrophysical Journal Supplement, 194, 45

\bibitem[\protect\citeauthoryear{{Shen} et~al.,}{{Shen} et~al.}{2016a}]{SHE16b}
{Shen} Y.,  et~al., 2016a, \mn@doi [\apj] {10.3847/0004-637X/818/1/30}, \href
  {https://ui.adsabs.harvard.edu/abs/2016ApJ...818...30S} {818, 30}

\bibitem[\protect\citeauthoryear{{Shen} et~al.,}{{Shen} et~al.}{2016b}]{SHE16a}
{Shen} Y.,  et~al., 2016b, \mn@doi [\apj] {10.3847/0004-637X/831/1/7}, \href
  {https://ui.adsabs.harvard.edu/abs/2016ApJ...831....7S} {831, 7}

\bibitem[\protect\citeauthoryear{{Stoughton} et~al.,}{{Stoughton}
  et~al.}{2002}]{SDSSI}
{Stoughton} C.,  et~al., 2002, \aj, 123, 485

\bibitem[\protect\citeauthoryear{Sun et~al.,}{Sun et~al.}{2015}]{SUN15}
Sun M.,  et~al., 2015, \apj, 811, 42

\bibitem[\protect\citeauthoryear{{Tohline} \& {Osterbrock}}{{Tohline} \&
  {Osterbrock}}{1976}]{TOH76}
{Tohline} J.~E.,  {Osterbrock} D.~E.,  1976, \mn@doi [\apjl] {10.1086/182317},
  \href {https://ui.adsabs.harvard.edu/abs/1976ApJ...210L.117T} {210, L117}

\bibitem[\protect\citeauthoryear{{Trevese}, {Paris}, {Stirpe}, {Vagnetti}  \&
  {Zitelli}}{{Trevese} et~al.}{2007}]{TRE07}
{Trevese} D.,  {Paris} D.,  {Stirpe} G.~M.,  {Vagnetti} F.,   {Zitelli} V.,
  2007, \mn@doi [\aap] {10.1051/0004-6361:20077237}, \href
  {https://ui.adsabs.harvard.edu/abs/2007A&A...470..491T} {470, 491}

\bibitem[\protect\citeauthoryear{{Vestergaard} \& {Wilkes}}{{Vestergaard} \&
  {Wilkes}}{2001}]{VES}
{Vestergaard} M.,  {Wilkes} B.~J.,  2001, \mn@doi [the Astrophysical Journal
  Supplement] {10.1086/320357}, \href
  {http://adsabs.harvard.edu/abs/2001ApJS..134....1V} {134, 1}

\bibitem[\protect\citeauthoryear{{Wang} et~al.,}{{Wang} et~al.}{2019}]{WAN19}
{Wang} S.,  et~al., 2019, \mn@doi [\apj] {10.3847/1538-4357/ab322b}, \href
  {https://ui.adsabs.harvard.edu/abs/2019ApJ...882....4W} {882, 4}

\bibitem[\protect\citeauthoryear{{Welsh}, {Wheatley}  \& {Neil}}{{Welsh}
  et~al.}{2011}]{WEL11}
{Welsh} B.~Y.,  {Wheatley} J.~M.,   {Neil} J.~D.,  2011, \mn@doi [\aap]
  {10.1051/0004-6361/201015865}, \href
  {https://ui.adsabs.harvard.edu/abs/2011A&A...527A..15W} {527, A15}

\bibitem[\protect\citeauthoryear{{Woo}}{{Woo}}{2008}]{WOO08}
{Woo} J.-H.,  2008, \mn@doi [\aj] {10.1088/0004-6256/135/5/1849}, \href
  {https://ui.adsabs.harvard.edu/abs/2008AJ....135.1849W} {135, 1849}

\bibitem[\protect\citeauthoryear{{Yang} et~al.,}{{Yang} et~al.}{2020}]{YAN19}
{Yang} Q.,  et~al., 2020, \mn@doi [\mnras] {10.1093/mnras/staa645}, \href
  {https://ui.adsabs.harvard.edu/abs/2020MNRAS.493.5773Y} {493, 5773}

\bibitem[\protect\citeauthoryear{{Zhu}, {Sun}  \& {Wang}}{{Zhu}
  et~al.}{2017}]{ZHU17}
{Zhu} D.,  {Sun} M.,   {Wang} T.,  2017, \mn@doi [\apj]
  {10.3847/1538-4357/aa76e7}, \href
  {http://adsabs.harvard.edu/abs/2017ApJ...843...30Z} {843, 30}

\makeatother
\end{thebibliography}

\newpage
\appendix
\section{Statistical Tests}
\label{app:tests}
\subsubsection*{$\chi^2$ Tests}
We use $\chi^2$ as the goodness-of-fit parameter when fitting responsivity functions to the normalised flux data. The one-component and two-component responsivity functions are nested functions. In the case of nested functions, the difference between the two $\chi^2$ ($\Delta\chi^2$) of the fits is distributed according to an $M$ d.o.f. $\chi^2$ distribution, where $M$ is the difference in number of parameters between the functions.\\
\indent In the case of our study M=1, therefore $P(\chi^2_1 > \Delta\chi^2_f) = 1 - \mathrm{CDF}(\Delta\chi^2_f)$, CDF is the cumulative distribution function associated with $\chi^2_1$. If this $p$-value is small, this indicates that the two-component fit is the preferred model.
\subsubsection*{Sequence Test}
The `sequence test' is intended to check the distribution of the data points around the fitted responsivity functions. If the function truly represents the behaviour of the fluxes, the scatter of the data points should be random, as it would result only from the uncertainties in the measurements. This can be pictured by considering the residuals (see Figure~\ref{fig:mgii_linfit_max}). Any data point is just as likely to be above the line of the function as it is to be below it.\\
\indent The sequence test is performed on the residuals of the fit, and counts the number of data points that have the same sign ($\pm$) as the preceding data point. The test statistic is labelled $S$, and represents the number of data points that have the same sign as the preceding point. The sampling distribution for this statistic is the binomial distribution for $N$-1 trials, where $N$ is the sample size. For a perfect fit each data point (after the first one) has a $\frac{1}{2}$ probability of lying on the same side of the response function. $S$ is normalised by dividing by $N$-1. The expectation value for $S$ is therefore $\frac{1}{2}$ and the variance $\frac{1}{4}$. Using the sample distribution a p-value can be calculated for a given $S$ statistic ($p_s$). This value represents the probability that the S statistic is not drawn from the binomial distribution. A low value of $p_s$ suggests a good fit. We verified the robustness of our test with a bootstrap MC (N=$10^4$) simulation, using the spectra of the Supervariable sample.
\subsubsection*{Responsivity Measure: $\alpha_{rm}$}
The responsivity measure is constructed using the normalisation to the minimum continuum state. This metric is based on the perpendicular distance in the normalised flux plane from the line that would indicate a 1:1 correspondence between line and continuum flux.\\
\indent Different types of behaviour can be associated with different values of $\alpha_{rm}$: the tracking of the line by the continuum will result in $\alpha_{rm}$$\sim$$0$, as will a stable line and continuum; a positive responsivity measure is associated with a larger line flux change than continuum flux change; a negative measure corresponds to objects with a continuum flux change that outdoes the line flux change.
\subsubsection*{Test for Normal Distribution}
\indent There are several statistical tests available to test for normalcy of a distribution. For the purposes of our analysis we have tested the Lilliefors and Shapiro-Wilk tests on their suitability. The Kolmogorov-Smirnov test was not directly applicable, as the normalisation of the data requires using the sample average and sample variance. The Lilliefors test is an adaptation of the KS test, designed to account for the fact that the normalisation uses properties derived from the sample distribution.\\
\indent Our testing consisted of a bootstrap MC (N=$10^4$) on the $\Delta f_{\mathrm{MgII}}$ of the Full Population Sample as well as on a test set of data drawn from a normal distribution. The Lilliefors test proved the most robust. We have applied the Lilliefors test as implemented in the \texttt{statsmodels} package\footnote{https://www.statsmodels.org/dev/}.

%%%%%%%%%%%%%%%%%%%%%%%%%%%%%%%%%%%%%%%%%%%%%%%%%%

% Don't change these lines
\bsp	% typesetting comment
\label{lastpage}
\end{document}